\documentclass{article}
\pdfoutput=1

\usepackage{fullpage}
\usepackage{amsmath}
\usepackage{paralist}
\usepackage{booktabs}
\usepackage{graphicx}
\usepackage{siunitx}
\usepackage{url}
\usepackage{hyperref}
\graphicspath{{figures/}}
\usepackage[round]{natbib}

\usepackage{etoolbox}

\hypersetup{
  pdfauthor={Barbu \emph{et al}},
  pdftitle={The Compositional Nature of Event Representations in the Human Brain},
  colorlinks,
  linkcolor={blue},
  citecolor={blue},
  urlcolor={blue}
}

\newcommand{\eg}{\emph{e.g.},}

\newcommand{\vs}{\emph{vs.}}

\newcommand{\Cf}{\emph{Cf.},}

\usepackage[usenames,dvipsnames,svgnames,table]{xcolor}
\newcommand{\actor}{{\color{BlueViolet}actor}}
\newcommand{\Verb}{{\color{Goldenrod}verb}}
\newcommand{\object}{{\color{DarkRed}object}}
\newcommand{\direction}{{\color{Blue}direction}}
\newcommand{\location}{{\color{SeaGreen}location}}

\title{The Compositional Nature of Event Representations\\in the Human Brain}

\newcommand{\nam}[3]{\href{mailto:#2}{#1}$^{\ref*{item:#3}}$}

\author{
  \nam{Andrei Barbu}{andrei@0xab.com}{mit} \quad
  \nam{N. Siddharth}{siddharth@iffsid.com}{stanford} \quad
  \nam{Caiming Xiong}{caimingxiong@ucla.edu}{ucla} \quad
  \nam{Jason J. Corso}{jcorso@buffalo.edu}{michigan} \\
  \nam{Christiane D. Fellbaum}{fellbaum@princeton.edu}{princeton} \quad
  \nam{Catherine Hanson}{cat@psychology.rutgers.edu}{rutgers} \quad
  \nam{Stephen Jos\'{e} Hanson}{jose@psychology.rutgers.edu}{rutgers} \\
  \nam{S\'{e}bastien H\'{e}lie}{shelie@purdue.edu}{psych} \quad
  \nam{Evguenia Malaia}{malaia@uta.edu}{uta} \quad
  \nam{Barak A. Pearlmutter}{barak@pearlmutter.net}{nuim} \\
  \nam{Jeffrey Mark Siskind}{qobi@purdue.edu}{ece} \quad
  \nam{Thomas Michael Talavage}{tmt@purdue.edu}{ece} \quad
  \nam{Ronnie B. Wilbur}{wilbur@purdue.edu}{ling}
}

\begin{document}

\maketitle

\thispagestyle{empty}

\footnotetext[1]{Computer Science and Artificial Intelligence Laboratory, MIT, Cambridge
   MA 02139\label{item:mit}}
\footnotetext[2]{Department of Psychology, Jordan Hall, Building 01-420, Stanford
   University, 450 Serra Mall, Stanford CA 94305\label{item:stanford}}
\footnotetext[3]{Statistics, University of California at Los Angeles, Los
   Angeles CA 90095\label{item:ucla}}
\footnotetext[4]{Electrical Engineering and Computer Science, University of
  Michigan, Ann Arbor MI 48109\label{item:michigan}}
\footnotetext[5]{Department of Computer Science, Princeton University, Princeton NJ 08540\label{item:princeton}}
\footnotetext[6]{Department of Psychology and Rutgers Brain Imaging Center, Rutgers
   University, Newark NJ 07102\label{item:rutgers}}
\footnotetext[7]{Psychological Sciences, Purdue University, West Lafayette IN 47907\label{item:psych}}
\footnotetext[8]{Southwest Center for Mind, Brain, and Education, University of Texas at
   Arlington, Arlington TX 76019\label{item:uta}}
\footnotetext[9]{Hamilton Institute and Dept.\ of Computer Science, National
   University of Ireland Maynooth, Co.\ Kildare, Ireland\label{item:nuim}}
\footnotetext[10]{School of Electrical and Computer Engineering, Purdue University, West
   Lafayette IN 47907\label{item:ece}}
\footnotetext[11]{Dept.\ of Speech, Language, and Hearing Sciences and Linguistics
   Program, Purdue University, West Lafayette IN 47907\label{item:ling}}

\begin{abstract}\noindent
  How does the human brain represent simple compositions of constituents:
  actors, verbs, objects, directions, and locations?
  Subjects viewed videos during neuroimaging (fMRI) sessions from which
  sentential descriptions of those videos were identified by decoding the brain
  representations based only on their fMRI activation patterns.
  Constituents (\eg\ \emph{fold} and \emph{shirt}) were independently
  decoded from a single presentation.
  Independent constituent classification was then compared to joint
  classification of aggregate concepts (\eg\ \emph{fold}-\emph{shirt});
  results were similar as measured by accuracy and correlation.
  The brain regions used for independent constituent classification are
  largely disjoint and largely cover those used for joint classification.
  This allows recovery of sentential descriptions of stimulus
  videos by composing the results of the independent constituent classifiers.
  Furthermore, classifiers trained on the \emph{words} one set of
  subjects think of when watching a video can recognise \emph{sentences} a
  \emph{different} subject thinks of when watching a \emph{different} video.
\end{abstract}

\paragraph{Nonspecialist Summary}

When people see John folding a chair, they readily perceive that \emph{John}
performed an action \emph{fold} with a \emph{chair}, breaking down the
aggregate event into individual components.
We investigate if such compositional perception is reflected in the brain---Can
one identify component and aggregate events that someone saw, by looking at
their brain activity?
Do brain regions that activate when identifying aggregate events (John folding)
relate to those that activate when identifying individual components (John,
folding)?
Do different people exhibit similar representations?
Our findings indicate affirmative outcomes for all the above and that the
representations involved are indeed compositional.

\paragraph{Scientific Summary}

Our work investigates the neural basis for compositionality of event
representations.
We explore how the brain represents compositions of constituents such as
actors, verbs, objects, directions, and locations.
While the constituents and their compositions are themselves linguistic in
nature, the stimuli are purely visual, depicting complex activity.
This enables exploration of the linguistic basis of visual perception through
neuroimaging.
%
This work straddles a wide variety of disciplines:
cognitive science, vision, linguistics, computational neuroscience, and machine
learning.

\section{Introduction}

The compositional nature of thought is taken for granted by many in the
cognitive-science community.
%
%
The representations commonly employed compose aggregated concepts from
constituent parts
\citep{miller1976}.\footnote{\Cf\ \citet{jackendoff1983},
  \eg\ (10.10a--j) p.~192, \citet{pinker1989}, \eg\ (5.46) and~(5.47) p.~218,
  and \citet{kosslyn1996}, \textparagraph~2, p.~6.}
This has been articulated by Jackendoff as the \emph{Cognitive Constraint} and
the \emph{Conceptual Structure
  Hypothesis}.\footnote{\Cf\ \citet{jackendoff1983}, pp.~16--22 including~(1.3)
  and~(1.4).}
Humans need not employ compositional representations; indeed, many argue that
such representations may be ill suited as models of human
cognition \citep{brooks1991}.
This is because concepts like \emph{verb} or even \emph{object} are human
constructs; there is hence debate as to how they arise from
percepts \citep{smith1996}.
Recent advances in brain-imaging techniques enable exploration of the
compositional nature of brain activity.
To that end, subjects underwent functional magnetic resonance imaging (fMRI)
while exposed to stimuli that evoke complex brain activity which was decoded
piece by piece.
The video stimuli depicted events composed of an \textbf{actor}, a
\textbf{verb}, an \textbf{object}, and a \textbf{direction} of motion or a
\textbf{location} of the event in the field of view.
Instances of these constituents could form ordinary sentences like
\emph{Dan carried the tortilla leftward}.
Machine-learned classifiers were used to decode complex brain activity into its
constituent parts
The classifiers used a subset of voxels in the whole-brain scan that was
determined automatically by the machine-learning methods to be maximally
effective.
The study further demonstrates that:
%
%
\begin{compactitem}
\item Accuracy of classification by classifiers trained independently on the
  constituents is largely the same as that of classifiers trained jointly on
  constituent pairs and triples.
\item The brain regions employed by the per-constituent classifiers are largely
  pairwise disjoint.
\item The brain regions employed by the joint classifiers largely consist of
  the unions of the brain regions employed by the component constituent
  classifiers.
\end{compactitem}
This provides evidence for the neural basis of the compositionality of event
representations.
We know of no other work that demonstrates this neural basis by simultaneously
decoding the brain activity for all of these constituents from the same video
stimulus.

Compositionality can refer to at least two different notions.
It can refer to the result of a composition.
For example, $\text{2}+\text{3}=\text{5}$ composes~5 out of~2 and~3.
It is impossible to reconstruct~2 and~3 from the result~5.
It can also refer to a specification of the structure of the composition.
For example, `$\text{2}+\text{3}$'.
From such it is possible to extract `2' and `3'.
The same issue arises with semantics.
\emph{John} combines with \emph{walked} to yield \emph{John walked}.
The result of such a composition could be some nondecomposable representation.
Yet the structural specification of such a composition could be decomposed into
its constituents.
(There are also, of course, idiomatic expressions whose meanings are not derived
compositionally from their constituents.
Such is not the topic of study here.)

Does the brain employ such decomposable representations?
It is conceivable that representations are decomposable at some processing
stages but not others.
When seeing John walk, neural activity might encode regions in the field of
view that reflect the aggregate percept of John walking.
Moreover, that aggregate percept might be spread across neural activity in
space and/or time.
A behavioural response to seeing John walk, such as walking towards him, also
might reflect the aggregate percept, not the percepts of John alone or walk
alone, because the percept of John sitting or of Mary walking might evoke
different responses.
Motor response might reflect an aggregate percept which might also be spread
across neural activity in space and/or time.
Thus there appear to be at least some processing stages, particularly at the
inputs and outputs, where representations might not be decomposable.
The question is whether there exist other intermediate processing stages that
are.
This question is investigated.

Language itself is compositional.
Sentences are composed of words.
It seems likely that when a percept involves language, either auditory or
visual (orthographic, signed), the neural representation of that percept would
be decomposable, at least at the input.
It also seems likely that when a behavioural response involves language, oral or
visual (written, signed), the neural representation of that motor response
would be decomposable, at least at the output.
It would be surprising, however, if a purely visual task that involved no
linguistic stimuli or response would evoke decomposable brain activity.
The experiment design investigates just that.

A requirement for decomposable representations is a degree of independence of
the constituent representations.
It is not possible to recover~2 and~3 from~5 because extra information
enters into the process $\text{2}+\text{3}=\text{5}$, namely addition.
It would only be possible to recover `2' and `3' from `$\text{2}+\text{3}$'
if their representations were independent.
Just as decomposability need not be black and white---there may be both
decomposable and nondecomposable representations employed in different brain
regions---independence also need not be black and white---degree of
independence may vary.
Structural decomposability of aggregate compositional percepts is investigated
by measuring and demonstrating a high degree of independence of the
constituents.


Recent work on decoding brain activity has recovered object class from
nouns presented as image, video, aural, and orthographic
stimuli \citep{puce1996, hanson2004, miyawaki2008, hanson2009, just2010,
  connolly2012, pereira2012, huth2012}.
Similar work on verbs has primarily been concerned with identifying active
brain
regions \citep{kable2006, kemmerer2008, kemmerer2010, huth2012, coello2012}.
Other recent work has demonstrated the ability to decode the actor of an event
using personality traits \citep{hassabis2013}.
These past successes suggest that one can investigate this hypothesis using
significant extensions of these prior methods combined with several novel
analyses.

Compositionality is investigated as applied to sentence structure---objects
fill argument positions in predicates that combine to form sentential
meaning---and decompose such into independent constituents.
Recent work has identified brain regions correlated with compositionality that
may not be decomposable using a task called \emph{complement
  coercion} \citep{pylkkanen2011}.
Subjects were presented with sentences whose meanings were partly implied
rather than fully expressed overtly through surface constituents.
For example, the sentence \emph{The boy finished the pizza} is understood as
meaning that the pizza was eaten, even though the verb \emph{eat} does not
appear anywhere in the sentence \citep{pustejovsky1995}.
The presence of \emph{pizza}, belonging to the category \emph{food}, coerces
the interpretation of \emph{finish} as \emph{finish eating}.
By contrast, \emph{He finished the newspaper} induces the interpretation
\emph{finish reading}.
Because syntactic structure in this prior experiment was held constant, the
assumption was that coercion reflects incorporation of extra information in the
result that is absent in the constituents.
Brain activity measured using magnetoencephalography (MEG) showed activity
related to coercion in the anterior midline field.
This result suggests that there may be some regions that do not exhibit
decomposable brain activity but does not rule out the possibility that there
are other regions that do.

It is conceivable that different subjects represent such compositional
information differently, perhaps in different brain regions.
Evidence is presented for why this might not be the case by demonstrating
cross-subject train and test: training classifiers on a set of subjects
watching one set of videos and testing on a \emph{different} subject watching
a \emph{different} set of videos.

\section{Experiment Design and Analysis}

The existence of brain regions that exhibit decomposable brain activity was
hypothesised and an experiment was conducted to evaluate this hypothesis by
demonstrating the ability to decode the brain activity evoked by a complex
visual stimulus into a sentence that describes that stimulus by independently
decoding the constituent words.
During neuroimaging (fMRI), subjects were shown videos that depicted events
that can be described by sentences of the form: the \textbf{actor}
\textbf{verb} the \textbf{object} \textbf{direction}/\textbf{location},
\eg\ \emph{Siddharth folded the chair on the right}.
They were asked to think about the sentence depicted by each video but were not
required to provide a specific behavioural response.

The videos were nonlinguistic; they showed one of four \textbf{actor}s
performing one of three \textbf{verb}s on one of three \textbf{object}s in one
of two \textbf{direction}s or \textbf{location}s.
The videos were also combinatorial in nature; any actor could perform any verb
on any object in any direction or location.
The following questions were asked:
\begin{compactitem}
  \item Can one recover these individual constituents from the aggregate
    stimulus?
  \item Can one recover combinations of these from the aggregate stimulus?
  \item How does accuracy, when doing so, depend on whether the classifiers were
    trained only on the individual constituents or jointly on the combined
    concepts?
  \item Do classifiers trained jointly on the combined concepts use different
    brain regions than those trained on the individual constituents?
  \item Do such stimuli evoke different neural activity patterns in different
    brain regions in different subjects?
\end{compactitem}

The combinatorial nature of the stimuli facilitates investigating these
questions by allowing one to train classifiers for the independent
constituents that occur in each stimulus:
\begin{compactdesc}
\item[\hspace*{2em}actor] one-out-of-four actor identity
\item[\hspace*{2em}verb] one-out-of-three verb (\emph{carry}, \emph{fold}, and
  \emph{leave})
\item[\hspace*{2em}object] one-out-of-three noun (\emph{chair}, \emph{shirt},
  and \emph{tortilla})
\item[\hspace*{2em}direction] one-out-of-two direction of motion for
  \emph{carry} and \emph{leave} (\emph{leftward} \vs\ \emph{rightward})
\item[\hspace*{2em}location] one-out-of-two location in the field of view for
  \emph{fold} (\emph{on the left} \vs\ \emph{on the right})
\end{compactdesc}
This design further facilitates the investigation by also allowing one to
train classifiers for combinations of the constituents: pairs
(\textbf{actor}-\textbf{verb}, \textbf{actor}-\textbf{object},
\textbf{actor}-\textbf{direction}, \textbf{actor}-\textbf{location},
\textbf{verb}-\textbf{object}, \textbf{verb}-\textbf{direction},
\textbf{object}-\textbf{direction}, and \textbf{object}-\textbf{location}),
triples (\textbf{actor}-\textbf{verb}-\textbf{object},
\textbf{actor}-\textbf{verb}-\textbf{direction},
\textbf{actor}-\textbf{object}-\textbf{direction}, and
\textbf{verb}-\textbf{object}-\textbf{direction}), and even the entire
sentence.

Data was gathered for seven subjects and a variety of different classification
analyses were performed using a linear support vector machine
(SVM; \citealp{cortes1995}).
For all of these, cross validation was employed to partition the dataset into
training and test sets, training classifiers on the training sets and measuring
their accuracy on the test sets.
For constituent pairs and triples, this was done with two kinds of classifiers,
ones trained jointly on the combination of the constituents and ones trained
independently on the constituents.
The accuracy obtained on within-subject analyses, training and testing the
classifiers on data from the same subject, was also compared to that obtained
on cross-subject analyses, training and testing the classifiers on data from
different subjects.
Two different methods were further employed to determine the brain regions used
by the classifiers and a variety of analyses were performed to measure the
degree of overlap.

\section{Results}

\begin{figure}
  \centering
  \begin{tabular}{@{}c@{\hspace*{2pt}}c@{}}
    \begin{tabular}[b]{@{}c@{}}
      \includegraphics[width=0.58\textwidth]
                      {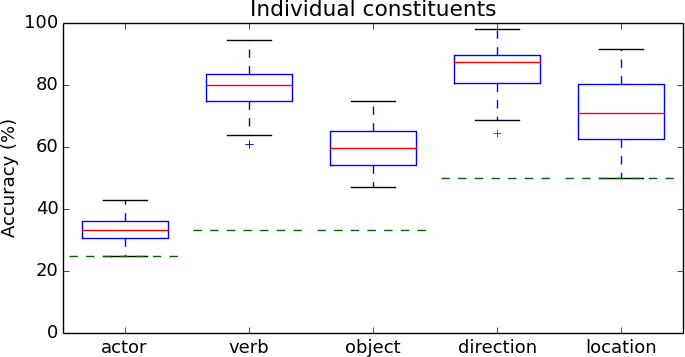}\\
      (a)\\*[3ex]
      \includegraphics[width=0.6\textwidth]
                      {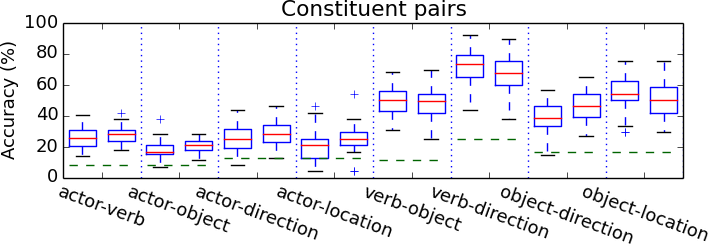}\\
      (b)\\*[3ex]
      \includegraphics[width=0.6\textwidth]
                      {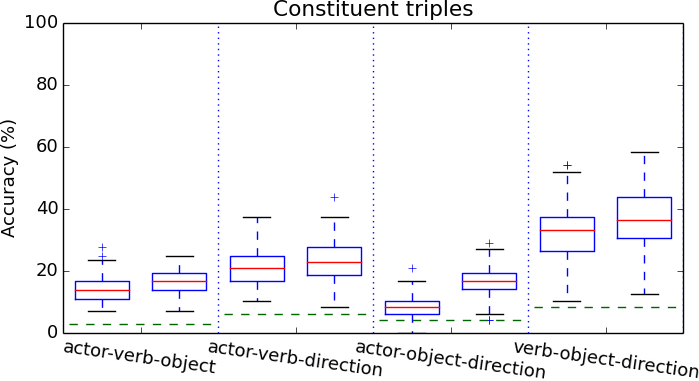}\\
      (c)
    \end{tabular}&
    \begin{tabular}[b]{@{}c@{}}
      \includegraphics[width=0.295\textwidth]
                      {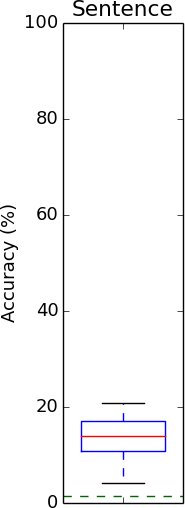}\\
      (d)
    \end{tabular}
  \end{tabular}
  \caption{%
    Classification accuracy, aggregated across subject and fold, for
    independent constituents~(a), constituent pairs~(b) and triples~(c), and
    entire sentences using independent per-constituent classifiers~(d).
    (b, c)~Comparison of joint~(left) \vs\ independent~(right) classification
    accuracy aggregated across subject and fold for constituent pairs and
    triples.
    Red lines indicate medians, box extents indicate upper and lower quartiles,
    error bars indicate maximal extents, and crosses indicate outliers.
    The dashed green lines indicates chance performance.}
  \label{fig:results}
\end{figure}

Table~\ref{tab:results}(top) presents the per-constituent classification
accuracies, both per-subject and aggregated across subject, for the
within-subject analyses.
Fig.~\ref{fig:results}(a) presents the per-constituent classification
accuracies aggregated across subject.
Performance well above chance was achieved on all five constituents, with only a
single fold for subject~1 and two folds for subject~2 at chance for the actor
analysis and two folds for subject~2 at chance for the location analysis.
Average performance across subject is also well above chance: \textbf{actor}
33.33\%$^{***}$ (chance 25.00\%), \textbf{verb} 78.92\%$^{***}$ (chance 33.33\%).
\textbf{object} 59.80\%$^{***}$ (chance 33.33\%), \textbf{direction}
84.60\%$^{***}$ (chance 50.00\%), and \textbf{location} 71.28\%$^{***}$ (chance
50.00\%).
(For all classification accuracies, `*' indicates $p<0.05$, `**' indicates
$p<0.005$, and `***' indicates $p<0.0005$.)

\begin{table}
  \centering
  \sisetup{
    table-format=2.2,
    table-space-text-post=\%$^{***}$
  }
  \begin{tabular}[b]{cSSSSS@{\hspace*{6ex}}S}
    \toprule
    &\multicolumn{6}{c}{\textbf{Classification Accuracy}}\\
    \cmidrule(l){2-7}
    &\multicolumn{6}{c}{\textbf{Within Subject}}\\
    \cmidrule(l){2-7}
    \textbf{Subject}
    &{\actor}&{\Verb}&{\object}&{\direction}&{\location}&{sentence}\\
    \cmidrule{1-1}\cmidrule(l){2-7}
    1& 30.4\%${}^{**}$& 77.6\%${}^{***}$& 55.4\%${}^{***}$& 84.6\%${}^{***}$& 69.8\%${}^{***}$& 11.3\%${}^{***}$\\
    2& 31.4\%${}^{***}$& 67.4\%${}^{***}$& 54.2\%${}^{***}$& 76.3\%${}^{***}$& 61.5\%${}^{**}$& 8.7\%${}^{***}$\\
    3& 35.6\%${}^{***}$& 83.0\%${}^{***}$& 62.3\%${}^{***}$& 93.0\%${}^{***}$& 67.7\%${}^{***}$& 14.9\%${}^{***}$\\
    4& 35.2\%${}^{***}$& 81.6\%${}^{***}$& 66.0\%${}^{***}$& 82.3\%${}^{***}$& 73.4\%${}^{***}$& 16.7\%${}^{***}$\\
    5& 33.2\%${}^{***}$& 87.0\%${}^{***}$& 66.3\%${}^{***}$& 88.0\%${}^{***}$& 75.5\%${}^{***}$& 16.8\%${}^{***}$\\
    6& 32.3\%${}^{***}$& 77.6\%${}^{***}$& 57.5\%${}^{***}$& 80.7\%${}^{***}$& 79.7\%${}^{***}$& 13.5\%${}^{***}$\\
    7& 35.2\%${}^{***}$& 78.3\%${}^{***}$& 56.9\%${}^{***}$& 87.2\%${}^{***}$& 71.4\%${}^{***}$& 14.9\%${}^{***}$\\[2ex]
    mean&33.33\%${}^{***}$&78.92\%${}^{***}$&59.80\%${}^{***}$&84.60\%${}^{***}$&71.28\%${}^{***}$&13.84\%${}^{***}$\\
    stddev& 4.46 & 7.88 & 6.66 & 7.57 & 10.97 & 4.3\\
    \cmidrule(l){2-7}
    &\multicolumn{6}{c}{\textbf{Cross Subject}}\\
    \cmidrule(l){2-7}
    1& 25.9\%& 51.4\%${}^{***}$& 37.7\%${}^{*}$& 65.4\%${}^{***}$& 66.1\%${}^{***}$& 3.5\%${}^{***}$\\
    2& 26.7\%& 39.9\%${}^{**}$& 39.9\%${}^{**}$& 59.1\%${}^{***}$& 53.6\%& 3.8\%${}^{***}$\\
    3& 30.0\%${}^{**}$& 43.1\%${}^{***}$& 40.1\%${}^{***}$& 69.5\%${}^{***}$& 67.2\%${}^{***}$& 2.8\%${}^{*}$\\
    4& 32.1\%${}^{***}$& 51.4\%${}^{***}$& 41.0\%${}^{***}$& 70.6\%${}^{***}$& 66.1\%${}^{***}$& 6.3\%${}^{***}$\\
    5& 31.3\%${}^{***}$& 51.7\%${}^{***}$& 49.0\%${}^{***}$& 63.3\%${}^{***}$& 62.0\%${}^{**}$& 6.1\%${}^{***}$\\
    6& 28.1\%${}^{*}$& 53.0\%${}^{***}$& 47.0\%${}^{***}$& 63.0\%${}^{***}$& 54.7\%& 5.2\%${}^{***}$\\
    7& 29.2\%${}^{*}$& 45.7\%${}^{***}$& 39.2\%${}^{**}$& 67.4\%${}^{***}$& 54.2\%& 3.3\%${}^{**}$\\[2ex]
    mean&29.04\%${}^{***}$&48.02\%${}^{***}$&41.99\%${}^{***}$&65.48\%${}^{***}$&60.57\%${}^{***}$&4.41\%${}^{***}$\\
    stddev & 5.69 & 7.06 & 6.97 & 8.42 & 11.57 & 2.50\\
    chance & 25.00\% & 33.33\% & 33.33\% & 50.00\% & 50.00\% & 1.39\%\\
    \bottomrule
  \end{tabular}
  \caption{%
    Results of per-constituent classification.
    Per-subject mean classification accuracy for each constituent, along with
    independent sentence classification, averaged across fold.
    Note that all six analyses perform above chance.
    A `*' indicates $p<0.05$, a `**' indicates $p<0.005$, and a `***' indicates
    $p<0.0005$.}
  \label{tab:results}
\end{table}

An additional analysis was conducted to measure the independence of the
representations for these constituents.
Classifiers were trained jointly for all constituent pairs, except for
\textbf{verb} and \textbf{location} (because \textbf{location} only applied to
a single \textbf{verb} \emph{fold}) and the classification accuracy was compared
against independent application of the classifiers trained on the constituents
in isolation (Fig.~\ref{fig:results}b).
Classifiers were similarly trained jointly for all constituent triples, except
for \textbf{actor}, \textbf{object}, and \textbf{location} due to lack of
sufficient training data, and a similar comparison was performed
(Fig.~\ref{fig:results}c).
An independent classification was deemed correct if it correctly classified all
of the constituents in the pair or triple.

A further analysis was conducted to measure the accuracy of decoding an entire
sentence from a single stimulus.
Training a joint classifier on entire sentences would require a sufficiently
large number of samples for each of the 72 possible sentences
($\text{4}\times\text{3}\times\text{3}\times\text{2}$), which would be
unfeasible to gather due to subject fatigue.
However, each sample was independently classified with the per-constituent
classifiers and the results were combined as described above
(Fig.~\ref{fig:results}d).
Average performance across subject is well above chance (13.84\%$^{***}$, chance
1.39\%).

The degree of independence of the classifiers was quantified by comparing the
individual classification results of the independent classifiers to those
produced by the joint classifiers, for all constituent pairs and triples,
by computing the accuracy and Matthews correlation coefficient (MCC), for
multi-class classification \citep{gorodkin2004}, over the samples where
the joint classifier was correct, yielding an average accuracy of 0.7056 and
an average correlation of 0.4139 across all analyses (Table~\ref{tab:mcc}).

\begin{table}
  \centering
  \begin{tabular}{@{}lllllllll@{}}
    &\actor&\actor&\actor&\actor&\Verb&\Verb&\object&\object\\
    &\Verb&\object&\direction&\location&\object&\direction&\direction&\location\\[1ex]
    accuracy&
    0.6607&
    0.7059&
    0.6336&
    0.6763&
    0.6709&
    0.7422&
    0.6544&
    0.6235\\
    MCC&
    0.3724&
    0.3430&
    0.3603&
    0.2807&
    0.5959&
    0.7048&
    0.5560&
    0.5067\\[2ex]
    &\actor&\actor&\actor&\Verb&&&&\\
    &\Verb&\Verb&\object&\object&&&&\\
    &\object&\direction&\direction&\direction&&&&\\[1ex]
    accuracy&
    0.8093&
    0.7403&
    0.8344&
    0.7154&&&&\\
    MCC&
    0.2521&
    0.3004&
    0.2352&
    0.4594&&&&
  \end{tabular}
  \caption{%
    Accuracy and MCC between independent and joint classification for
    constituent pairs (top) and triples (bottom), over the samples where the
    joint classifier was correct, aggregated across subject and fold.}
  \label{tab:mcc}
\end{table}

\begin{figure}
  \centering
  \begin{tabular}{c}
    \includegraphics[width=0.95\textwidth]{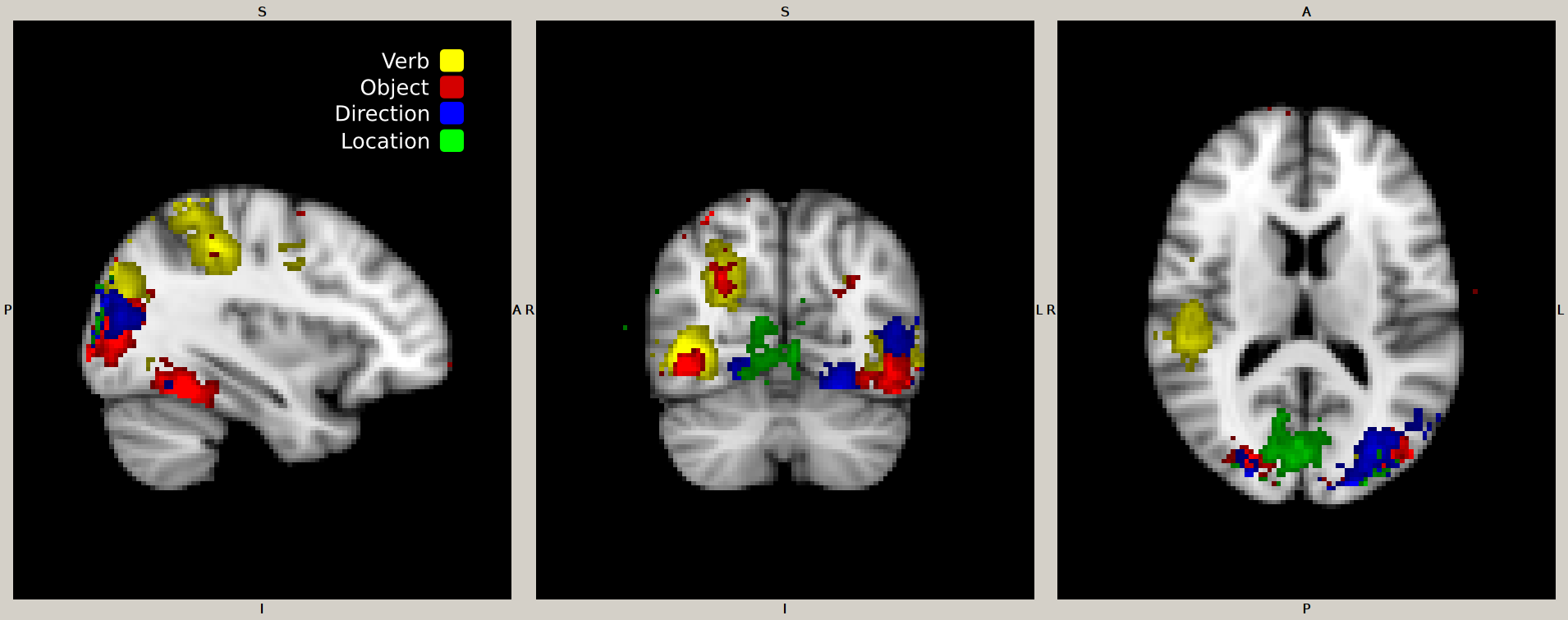}\\
    \includegraphics[width=0.95\textwidth]{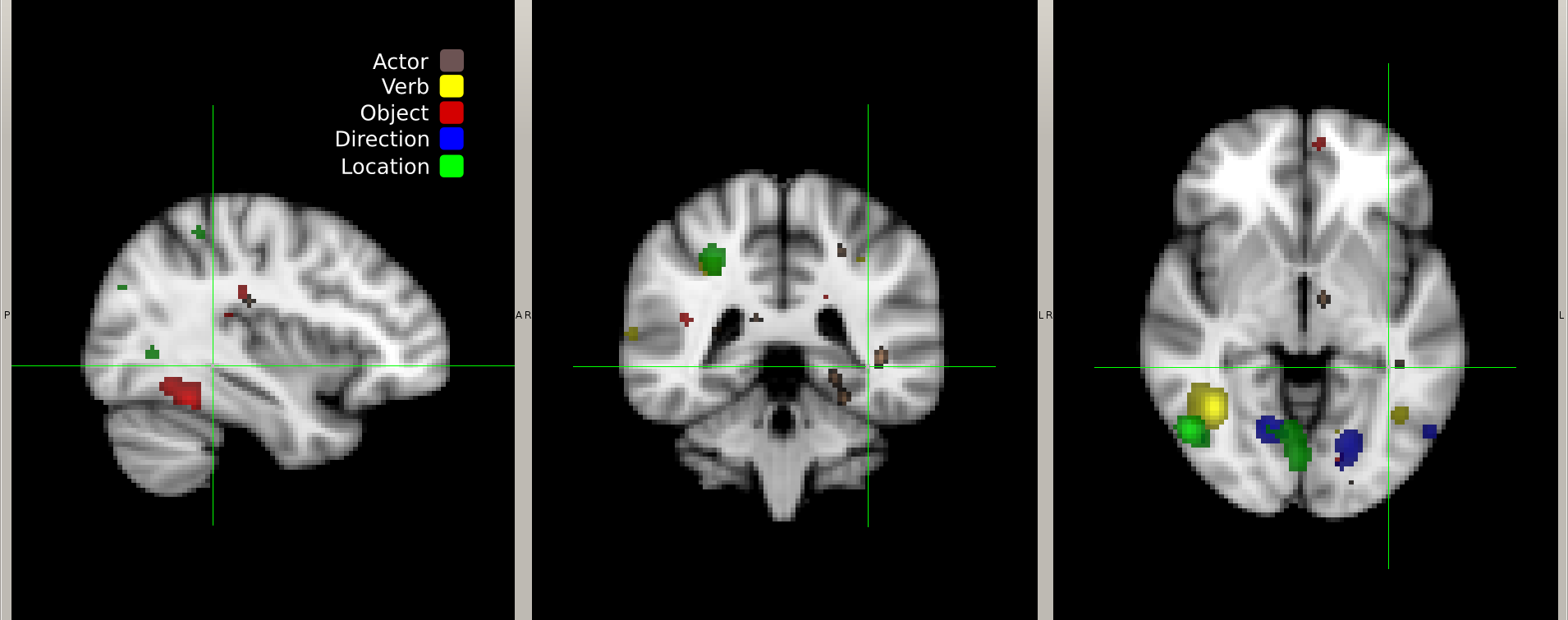}
  \end{tabular}
  \caption{%
    (top)~Searchlight analysis indicating the classification accuracy of
    different brain regions on the anatomical scans from subject~1 averaged
    across stimulus, class, and run.
    (bottom)~Thresholded SVM coefficients for subject~1, back-projected onto the
    anatomical scan, aggregated across run.}
  \label{fig:searchlight}
\end{figure}

Two distinct methods were used to locate brain regions used in the previous
analyses.
A spatial-searchlight \citep{kriegeskorte2006} linear-SVM method was
first employed on all subjects.
The accuracy was used to determine the sensitivity of each voxel and thresholded
upward to less than 10\% of the cross-validation measures.
These measures are overlaid and (2-stage) registered to MNI152 2mm
anatomicals.
This searchlight analysis was performed independently for all of the constituent
and joint classifiers.
The resulting constituent regions (omitting actor) are colour coded according to
the specific constituent being decoded.
The thresholded SVM coefficients were also back-projected for all constituents,
including actor, produced by the analysis in Table~\ref{tab:results}, for all
subjects, onto the anatomical scan, aggregated across run.
The resulting regions produced by both analyses for subject~1 are shown in
Fig.~\ref{fig:searchlight}.
(Figures for all subjects are included at the end.)

To further quantify the degree of spatial independence, the brain regions
indicated by searchlight and by the thresholded SVM coefficients of the
independent classifiers were compared to those of the joint classifiers, for
all constituent pairs and triples.
First, the percentage of voxels in the union of the constituents for
the independent classifier that were also in the intersection
was computed (Table~\ref{tab:quantitative-searchlight} top).
Next, the percentage of voxels in the joint classifier that are shared with the
independent classifier was also computed
(Table~\ref{tab:quantitative-searchlight} bottom).

\begin{table}
  \centering
  \begin{tabular}{@{}llllllll@{}}
    \multicolumn{8}{c}{$\displaystyle\frac{\displaystyle\left\lvert\bigcap_i\text{independent}_i\right\rvert}{\displaystyle\left\lvert\bigcup_i\text{independent}_i\right\rvert}$}\\
    \rule{0pt}{2ex}&&&&&&&\\
    \actor&\actor&\actor&\actor&\Verb&\Verb&\object&\object\\
    \Verb&\object&\direction&\location&\object&\direction&\direction&\location\\[1ex]
    3.30\%&
    6.74\%&
    1.74\%&
    1.32\%&
    14.98\%&
    1.20\%&
    8.43\%&
    4.48\%\\
    2.84\%&
    2.54\%&
    1.16\%&
    2.06\%&
    6.05\%&
    3.61\%&
    3.70\%&
    2.36\%\\[2ex]
    \actor&\actor&\actor&\Verb&&&&\\
    \Verb&\Verb&\object&\object&&&&\\
    \object&\direction&\direction&\direction&&&&\\[1ex]
    1.26\%&
    0.06\%&
    0.65\%&
    0.49\%&&&&\\
    0.42\%&
    0.01\%&
    0.00\%&
    0.20\%&&&&\\[3ex]
    \hline\\
    \multicolumn{8}{c}{$\displaystyle\frac{\displaystyle\left\lvert\left(\bigcup_i\text{independent}_i\right)\cap\text{joint}\right\rvert}{\displaystyle\left\lvert\text{joint}\right\rvert}$}\\
    \rule{0pt}{2ex}&&&&&&&\\
    \actor&\actor&\actor&\actor&\Verb&\Verb&\object&\object\\
    \Verb&\object&\direction&\location&\object&\direction&\direction&\location\\[1ex]
    67.42\%&
    48.64\%&
    68.53\%&
    69.37\%&
    79.53\%&
    74.53\%&
    65.97\%&
    79.15\%\\
    58.85\%&
    51.22\%&
    42.42\%&
    27.71\%&
    66.05\%&
    62.38\%&
    52.70\%&
    38.81\%\\[2ex]
    \actor&\actor&\actor&\Verb&&&&\\
    \Verb&\Verb&\object&\object&&&&\\
    \object&\direction&\direction&\direction&&&&\\[1ex]
    24.48\%&
    59.43\%&
    37.78\%&
    55.13\%&&&&\\
    60.68\%&
    56.51\%&
    38.35\%&
    58.25\%&&&&\\
  \end{tabular}
  \caption{%
    Quantitative comparison of the brain regions indicated by searchlight
    (upper rows) and thresholded SVM coefficients (lower rows) of the
    independent classifiers to the joint classifiers, for all constituent pairs
    and triples, averaged across subject.
    (top)~The percentage of voxels in the union of the constituents for the
    independent classifier that are also in the intersection.
    (bottom)~The percentage of voxels in the joint classifier that are shared
    with the independent classifier.}
  \label{tab:quantitative-searchlight}
\end{table}

\begin{figure}
  \centering
  \begin{tabular}{@{}c@{\hspace*{2pt}}c@{}}
    \begin{tabular}[b]{@{}c@{}}
      \includegraphics[width=0.58\textwidth]
                      {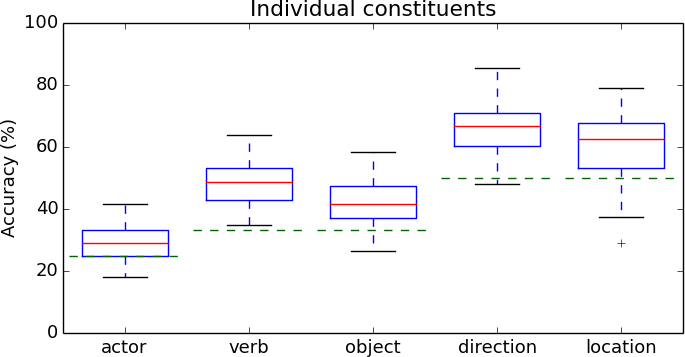}\\
      (a)\\*[3ex]
      \includegraphics[width=0.6\textwidth]
                      {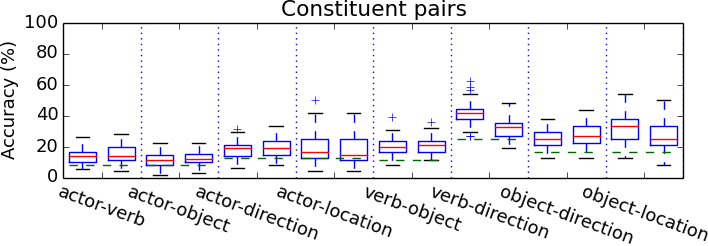}\\
      (b)\\*[3ex]
      \includegraphics[width=0.6\textwidth]
                      {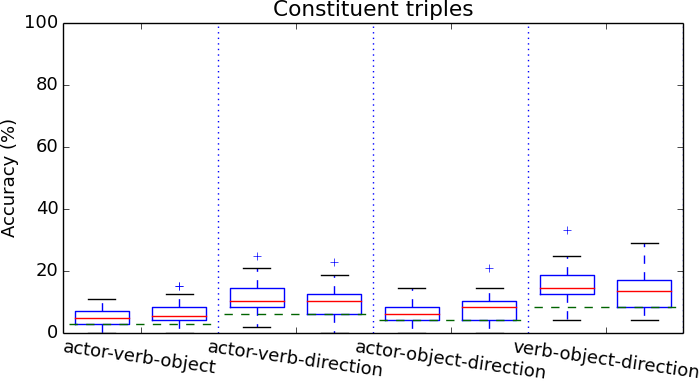}\\
      (c)
    \end{tabular}&
    \begin{tabular}[b]{@{}c@{}}
      \includegraphics[width=0.3\textwidth]
                      {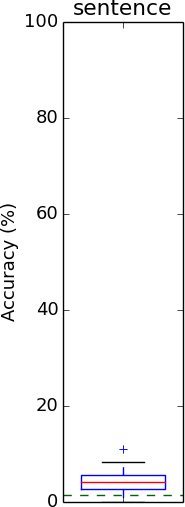}\\
      (d)
    \end{tabular}
  \end{tabular}
  \caption{%
    Cross-subject variant of Fig.~\ref{fig:results} with cross validation by
    run.}
  \label{fig:cross-folds-results}
\end{figure}

A further set of analyses was conducted to investigate the degree to which
different subjects employ different representations, in different brain
regions, of the constituents under study.
Cross-subject variants of the analyses in Table~\ref{tab:results}(top) and
Fig.~\ref{fig:results} were performed where the classifiers used to test on a
given run for a given subject were trained on data from all runs \emph{except}
the given run for all subjects \emph{except} the given subject.
These results are shown in Table~\ref{tab:results}(bottom) and
Fig.~\ref{fig:cross-folds-results}.
%
%
While classification accuracy is lower than the corresponding within-subject
analyses, all analyses aggregated across subject, all per-subject
single-constituent analyses, all per-subject independent-sentence analyses, and
all but one of the remaining per-subject analyses are above chance; the vast
majority significantly so.

\section{Discussion}

The stimuli were purely visual.
There were no words, phrases, or sentences presented, either auditorily or
visually (orthographic, signed).
Subjects were not asked to provide a specific behavioural response other than
to watch the video and think about its content.
No behavioural or motor response of any kind was elicited.
Specifically, subjects were not asked to produce words, phrases, or sentences,
either oral or visual (written, signed).
Thus neither the stimuli nor the (nonexistent) behavioural response were
overtly linguistic.
Nonetheless, the experimental setup was implicitly linguistic in a number of
ways.
Subjects were shown sample video prior to imaging and were informed of the
structure of the stimuli and the intended collection of actors, verbs, objects,
directions, and locations.
All subjects were aware of the experiment design, were informed of the general
intended depiction of the stimuli prior to the scan, and were instructed to
think of the intended depiction after each presentation.
While no specific behavioural response was elicited, they were asked to think
about the sentence depicted by each video.
It is conceivable that such subject instruction introduced a linguistic aspect
to the task and is what induced a decomposable representation.

This would be interesting in its own right, as it would indicate generation of
internal linguistic representations even given a lack of overt linguistic
behavioural and motor response.
Nonetheless, it would be interesting to see if such representations arose
even when subjects were not given such explicit instruction and perhaps were
not even primed as to the experiment design, the set of target constituents, and
the set of classes within each constituent.
Moreover, it would be interesting to see if such representations also arise
for stimuli that are less conducive to sentential description, such as more
abstract, perhaps synthetic, video of moving shapes that nonetheless could be
conceptually decomposed into shape \vs\ motion patterns \vs\ direction and
location that would not be described as nouns, verbs, and prepositions.

The results indicate that brain activity corresponding to each of the
constituents, \textbf{actor}, \textbf{verb}, \textbf{object},
\textbf{direction}, and \textbf{location}, can be reliably decoded from fMRI
scans, both individually, and in combination.
Given neural activation, one can decode what the subjects are thinking about.

Furthermore, the analysis indicates that a decomposable neural representation
for each of these five constituents exists in the brain.
This is surprising; intermediate neural representation could have been all
interdependent, just like the inputs and outputs.
People engage in distinct motions when \emph{fold}ing \emph{chair}s,
\emph{shirt}s, and \emph{tortilla}s.
If the representation of a verb, like \emph{fold}, was neurally encoded for a
particular object, for example, to reflect the particular motion involved when
performing the action denoted by that verb, it would not be possible to decode
this verb with performance above chance in the experiment design, because it is
counterbalanced with respect to the objects with which the action is being
performed.
Moreover, if there were some level of object specificity in verbs, one would
expect this to be reflected in marked decrease in classification accuracy of
independent classifiers for verb and object over a joint classifier for the
pair.
This, however, does not appear to be the case: averaged across subject, the
joint verb-object classifier has 49.58\%$^{***}$ accuracy while the independent
one has 47.94\%$^{***}$.
The relative performance of joint \vs\ independent classification appears
similar across all combinations of constituents, not just verbs and objects
(Figs.~\ref{fig:results}b, c), so much so that one can decode an
\textbf{\emph{entire sentence}} from a single stimulus, with accuracy far above
chance, using per-constituent classifiers trained independently on those
constituents.
Moreover, joint and independent classification are highly correlated
(Table~\ref{tab:mcc}), indicating that the joint classifiers are not making
significant use of information beyond that available to the independent
classifiers.

In general, the searchlight analysis and the back-projected SVM coefficients
(Fig.~\ref{fig:searchlight}) indicate that such decoding relies on different
brain regions for different constituents.
\textbf{Actor} activity is present in the fusiform face
area \citep{kanwisher1997}.
\textbf{Verb} activity is present in visual-pathway areas (lateral
occipital-LO, lingual gyrus-LG, and fusiform gyrus) as well as prefrontal areas
(inferior frontal gyrus, middle frontal gyrus, and cingulate) and areas
consistent with the hypothetical `mirror system' \citep{arbib2006} and the
hypothetical `theory of mind' (pre-central gyrus, angular gyrus-AG, and
superior parietal lobule-SPL) areas \citep{dronkers2004, turken2011}.
\textbf{Object} activity is present in the temporal cortex, and agrees with
previous work on object-category encoding \citep{gazzaniga2008}.
\textbf{Direction} and \textbf{location} activity is present in the visual
cortex with significant \textbf{location} activity occurring in the early
visual cortex.
More specifically, quantitative analysis of the brain regions indicated by both
searchlight and the thresholded SVM coefficients indicates that the brain
regions used for independent constituent classification are largely disjoint
(3.72\% for searchlight and 2.08\% for thresholded SVM weights, averaged across
both subject and analysis) and largely cover (60.83\% for searchlight and
51.16\% for thresholded SVM weights, averaged across both subject and analysis)
those used for joint classification (Table~\ref{tab:quantitative-searchlight}).

Note that we are claiming that the brain independently processes constituents,
\eg\ \textbf{verb} and \textbf{object}, \emph{not} that the output of such
processing is independent.
In particular, we are \emph{not} claiming that the outputs of the classifiers
are independent across constituent.
Classification results are produced by a long pipeline: the stimulus, the
evoked brain activity, its indirect measurement via fMRI, and its analysis via
classification.
Cross-constituent dependence can be introduced at any stage in this pipeline
and could also be masked by any subsequent stage.
Moreover, the classifiers are imperfect.
The confusion matrices are not diagonal.
Since the design is counterbalanced, in order for a~$\chi^2$ test not to
reject the null hypothesis, the contingency table must be uniform.
However, if the verb classifier exhibits a misclassification bias where, for
example, \emph{carry} is misclassified as \emph{fold} more frequently than as
\emph{leave}, and the object classifier exhibits a similar misclassification
bias, where, for example, \emph{chair} is misclassified as \emph{shirt} more
frequently than as \emph{tortilla}, this would manifest as dependence between
verb and object in the classifier output that would have no bearing on
classification accuracy.
Nor would it indicate joint usage of \textbf{verb} and \textbf{object}
information during classification.
Thus it makes no sense to perform a standard~$\chi^2$ independence test
between pairs of constituent classifier outputs.

What we are claiming is that the brain largely makes classification decisions
for one constituent independent of those for other constituents.
We take as evidence for this:
%
%
\begin{compactitem}
\item Classification accuracy using independent classifiers is largely the same
  as that for corresponding joint classifiers.
\item The brain regions employed by the per-constituent classifiers are largely
  pairwise disjoint.
\item The brain regions employed by the joint classifiers largely consist of
  the unions of the brain regions employed by the component constituent
  classifiers.
\end{compactitem}
Moreover, one can train a classifier on the \emph{words} that one set of
subjects think of when watching a video to recognise \emph{sentences} that a
\emph{different} subject thinks of when watching a \emph{different} video, with
accuracy \emph{far} better than chance.
This suggests that there must be sufficient commonality, across subjects,
between representations and brain regions used for the constituents under
study, to allow such.

Compositionality is a rich notion.
Not only must it be possible to determine~2 and~3 from `$\text{2}+\text{3}$',
it must be possible to determine that 2 is an argument of this addition
but not the addition in `$\text{4}+(\text{3}\times\text{2})$', even though it
appears elsewhere in the formula.
For language and vision, it must be possible to determine that a person is
folding the chair and not the shirt, when a shirt is present in the field of
view but is not being folded.
The present analysis can be construed as computational identification of
associative-linguistic representations, a form of syntax-less language
learning, without prefrontal cortex (PFC) engagement \citep{friederici2013}.
Further, not all operations are commutative or symmetric: it must be possible
to distinguish `$\text{2}\div\text{3}$' from `$\text{3}\div\text{2}$.'
For language and vision, some predicates are also asymmetric; it must be
possible to distinguish between a person approaching a dog from a dog
approaching a person.
Making such distinctions will require analysing fine-grained PFC activity,
likely using a region-of-interest approach \citep{jeon2013}.
Finally, the individual constituents may themselves be decomposable.
Verbs like \emph{raise} and \emph{lower} may decompose into lower-level
constituents indicating causation of upward \vs\ downward motion where the
lower-level constituents denoting causation and motion are shared between the
two verbs but those denoting direction are
not \citep{miller1976, jackendoff1983, pinker1989}.
For now, the findings are agnostic to these issues.

\section{Conclusion}

It has been demonstrated that it is possible to decode a subject's brain
activity into constituents, which, when combined, yield a sentential
description of a video stimulus.
To do so, the first study was conducted which decodes brain activity associated
with \textbf{actor}s, \textbf{verb}s, \textbf{object}s, \textbf{direction}s,
and \textbf{location}s from video stimuli, both independently and jointly.
These results are the first to indicate that the neural representations for
these constituents compose together to form the meaning of a sentence,
apparently without modifying one another, even when evoked by purely visual,
nonlinguistic stimuli, using what appear to be common representations and brain
regions that vary little across subject, at least at the granularity
investigated.
These results are in concord with Jackendoff's Cognitive Constraint and
Conceptual Structure Hypothesis and indicate that representations which attempt
to decompose meaning into constituents may have a neural basis.

\section{Methods Summary}

Subjects were shown video depicting events described by entire sentences
composed of an \textbf{actor}, a \textbf{verb}, an \textbf{object}, and a
\textbf{direction} of motion or a \textbf{location} of the event in the field of
view.
Subjects were shown sample video prior to imaging and were informed of the
structure of the stimuli and the intended collection of four \textbf{actor}s,
three \textbf{verb}s (\emph{carry}, \emph{fold}, and \emph{leave}), three
\textbf{object}s (\emph{chair}, \emph{shirt}, and \emph{tortilla}), two
\textbf{direction}s (\emph{leftward} and \emph{rightward}), and two
\textbf{location}s (\emph{on the left} and \emph{on the right})).
They were asked to think about the sentence depicted by each video, but no overt
behavioural response was elicited.

Subjects were scanned (fMRI) while watching the stimuli.
Each subject underwent eight runs, each run comprising 72 stimulus
presentations in a rapid event-related design \citep{just2010}.
The presentations were counterbalanced within and across runs, for all
constituent categories.
Scan data was preprocessed using AFNI \citep{cox1996} to skull-strip each
volume, motion correct and detrend each run, and perform alignment.
Within-subject experiments were carried out in the native coordinate space
while cross-subject experiments were aligned to MNI152 using an affine
transform.
A subset of significant voxels, determined by Fisher scores and Linear
Discriminant Dimensionality Reduction \citep{gu2011}, was selected to
perform classification using a linear support vector machine (SVM)
classifier \citep{cortes1995}.
Classification was performed on individual constituents and constituent
aggregates: pairs, triples, and entire sentences.
Two kinds of classifiers were used when classifying constituent pairs and
triples: ones trained independently on the component constituents and ones
trained jointly.
Two kinds of analyses were performed: within subject and cross subject.
For within-subject analyses, leave-one-out cross validation was performed by
run, training and testing on the same subject.
When testing on run~$r$, the classifiers were trained on all runs \emph{except}
run~$r$.
For cross-subject analyses, leave-one-out cross validation was performed by
subject and run.
When testing on run~$r$ for subject~$s$, the classifiers were trained on all
runs \emph{except} run~$r$ for all subjects \emph{except} subject~$s$.

\section{Methods}

\paragraph{Stimuli}

Videos depicting one of four human \textbf{actor}s performing one of three
\textbf{verb}s (\emph{carry}, \emph{fold}, and \emph{leave}), each with one of
three \textbf{object}s (\emph{chair}, \emph{shirt}, and \emph{tortilla}), were
filmed for this task.
The verbs were chosen to be discriminable based on the following
features \citep{kemmerer2008}:
\begin{quote}
  \begin{tabular}{lrr}
    \emph{carry} & $-$state-change & $+$contact\\
    \emph{fold}  & $+$state-change & $+$contact\\
    \emph{leave} & $-$state-change & $-$contact\\
  \end{tabular}
\end{quote}
Objects were chosen based on categories previously found to be discriminable:
\emph{chair} (furniture), \emph{shirt} (clothing), and \emph{tortilla} (food)
and also selected to allow each verb to be performed with each
object \citep{just2010}.
All stimuli enactments were filmed against the same uncluttered uniform
nonvarying background, which contained no other objects except for a table.
The action depiction was intentionally varied to be unconventional (humorous) to
keep subjects awake, attentive, and unhabituated.

In addition to depicting an actor, a verb, and an object, each stimulus
also depicted a \textbf{direction} or a \textbf{location}.
Direction was only depicted for the two verbs, \emph{carry} and \emph{leave},
while location was only depicted for the verb \emph{fold}.
The variation in direction and location was accomplished by mirroring videos
about the vertical axis.
Such mirroring induces variation in direction of motion (\emph{leftward}
\vs\ \emph{rightward}) for the verbs \emph{carry} and \emph{leave} and induces
variation in the location in the field of view where the verb \emph{fold} occurs
(\emph{on the left} \vs\ \emph{on the right}).
All other variation was accomplished by filming a combination of actor, verb,
and object.
There were four actors, three verbs, three objects, two directions, and two
locations, leading to
$\text{4}\times\text{3}\times\text{3}\times\text{2}=\text{72}$ possible
distinct depictions, for which between~3 and~7 (mean~5.5) videos were employed
for each such depiction.

Each subject viewed a total of 576 stimulus presentations, divided into eight
runs of equal length.
The runs were individually counterbalanced.
Each run comprised 72 stimulus presentations, exactly one for each possible
depiction.
The particular video chosen for the depiction was randomly drawn from a uniform
distribution.
Some stimuli may have been chosen for multiple runs.
All subjects were presented with the same stimuli and presentation order
within and across runs.

\paragraph{Study Subjects}

Informed consent was obtained from all subjects.
All protocols, experiments, and analyses were carried out with approval of the
Institutional Review Board at Purdue University.
Data was gathered for eight subjects, two women and six men.
Six subjects were between 20 and 30 years old, two were between 50 and 60 years
old.
Seven subjects were students and faculty.
One subject was recruited from the general population of West Lafayette, IN.

\paragraph{Data Collection}

A rapid event-related design \citep{just2010} was employed.
Two-second video clips were presented at 10fps followed by an average of 4s
(minimum 2s) fixation.
Each run comprised 72 stimulus presentations spanning 244 captured brain
volumes and ended with 24s of fixation.
Runs were separated by several minutes, during which no stimuli were presented,
no data was gathered, and subjects engaged in unrelated conversation with the
experimenters.
This separation between runs allowed runs to constitute folds for cross
validation without introducing spurious correlation in brain activity between
runs.

Imaging was performed at Purdue University using a 3T GE Signa HDx scanner
(Waukesha, Wisconsin) with a Nova Medical (Wilmington, Massachusetts)
16 channel brain array to collect whole-brain volumes via a
gradient-echo EPI sequence with 2000ms TR, 22ms TE, 200mm$\times$200mm
FOV, and 77$^{\circ}$ flip angle.
Thirty-five axial slices were acquired with a 3.0mm slice thickness using a
64$\times$64 acquisition matrix resulting in
3.125mm$\times$3.125mm$\times$3.0mm voxels.

Data was collected for eight subjects but the data for one subject was
discarded due to excessive motion.
One subject did eight runs without exiting the scanner.
All other subjects exited the scanner at various points during the set of
eight runs, which required cross-session registration.
All subjects were aware of the experiment design, shown sample stimuli,
informed of the structure of the stimuli and the intended collection of actors,
verbs, objects, directions, and locations, prior to imaging, and instructed to
think of the intended depiction after each presentation, but no overt
behavioural response was elicited.

\paragraph{Preprocessing and Dimensionality Reduction}

Whole-brain scans were processed using AFNI \citep{cox1996} to skull-strip
each volume, motion correct and detrend each run, and align all scans for a
given subject to a subject-specific reference volume.
Voxels within a run were z-scored, subtracting the mean value of that voxel for
the run and dividing by its variance.
Since each brain volume has very high dimension, 143,360 voxels, voxels were
eliminated by computing a per-voxel Fisher score on the training set and
keeping the 4,000 highest-scoring voxels (12,000 for the cross-subject
analyses).
The Fisher score of a voxel~$v$ for a classification task with $C$~classes
where each class~$c$ has~$n_c$ examples was computed as
\begin{equation*}
  \frac{\displaystyle\sum_{c=1}^C n_c(\mu_{c,v}-\mu)^2}
       {\displaystyle\sum_{c=1}^C n_c \sigma_{c,v}^2}
\end{equation*}
where~$\mu_{c,v}$ were~$\sigma_{c,v}$ are the per-class per-voxel means and
variances and~$\mu$ was the mean for the entire brain volume.
The resulting voxels were then analysed with Linear Discriminant Dimensionality
Reduction \citep{gu2011} to select a smaller number of potentially-relevant
voxels, selecting on average 1,084 voxels per-subject per-fold (12,000 for the
cross-subject analyses).
Both stages of voxel selection were performed independently for each fold of
each subject.
The set of voxels to consider was determined solely from the training set.
That same subset of voxels was used in the test set for classification.

\paragraph{Classifier}

A linear support vector machine (SVM) was employed to classify the selected
voxels \citep{cortes1995}.
Because fMRI acquisition times are slow, equal to the length of the video
stimuli, a single brain volume that corresponds to the peak brain activation
induced by that video stimulus was classified to recover the features that the
subjects were asked to think about.
The third brain volume after the onset of each stimulus was used, because fMRI
does not measure neural activation but instead measures the flow of oxygenated
blood, the blood-oxygen-level-dependent (BOLD) signal, which correlates with
increased neural activation.
It takes roughly five to six seconds for this signal to peak, which puts the
peak in the third volume after the stimulus presentation.

\paragraph{Cross Validation}

Two kinds of analyses were performed: within subject and cross subject.
The within-subject analyses trained and tested each classifier on the same
subject.
In other words, classifiers were trained on the data for subject~$s$ and also
tested on the data for subject~$s$.
This was repeated for all seven subjects.
For these, leave-one-out cross validation was performed by run: when testing on
run~$r$, the classifiers were trained on all runs \emph{except} run~$r$.
Such cross validation precludes training on the test data.
Partitioning by run ensures that information could not flow from the training
set to the test set through the hemodynamic response function (HRF).\@
This was repeated for all eight runs, thus performing eight-fold cross
validation.

The cross-subject analyses trained and tested each classifier on different
subjects.
In particular, a classifier was trained on the data for all subjects except
subject~$s$ and then tested on the data for subject~$s$.
This was repeated for all seven subjects.
For these, leave-one-out cross validation was performed by both subject and
run: when testing on run~$r$ for subject~$s$, the classifiers were trained on
all runs \emph{except} run~$r$ for all subjects \emph{except} subject~$s$.
While there is no potential for training on the test data, even without cross
validation by run, there is potential for a different HRF-based confound.
Due to the HRF, each scan potentially contains information from prior stimuli in
the same run.
Since the presentation order did not vary by subject, it is conceivable that
classifier performance is due, in part, to the current stimulus in the context
of previous stimuli in the same run, not just the current stimulus.
One could control for this confound by randomising presentation order across
subject, but this was not part of the experiment design.
Cross validation by run is an alternative control for this confound.

\paragraph{Analysis}

Classification was performed on individual constituents and constituent
aggregates: pairs, triples, and entire sentences.
Two kinds of classifiers were used when classifying constituent pairs and
triples: ones trained independently on the component constituents and ones
trained jointly.

Thirty classification analyses were conducted in total: five single-constituent
analyses, eight constituent-pair analyses, both independent and joint, four
constituent-triple analyses, both independent and joint, and an independent
sentence analysis.
These analyses varied in training- and test-set sizes because of several
properties of the design.
First, \textbf{verb} does not combine with \textbf{location} since location
only applies to a single verb, \emph{fold}.
Second, a joint classifier was not trained for \textbf{actor}, \textbf{object},
and \textbf{location} because there would be only seven training samples per
subject, fold, and class.
Similarly, only an independent classifier was employed for sentence and a joint
classifier was not trained because there would be only seven training samples
per subject, fold, and class.
Thus per-subject classification results are over 192 trials for analyses that
involve \textbf{location}, 384 trials for analyses that involve
\textbf{direction}, and 576 trials for all other analyses.
Classification results aggregated across subjects are over 1,344 trials for
analyses that involve \textbf{location}, 2,688 trials for analyses that involve
\textbf{direction}, and 4,032 trials for all other analyses.
For within-subject analyses, the training set was seven times as large as the
test set and contained exactly seven times as many depictions for any
combination of particular constituents as the test set.
However, the particular stimulus video for a given depiction may have appeared
more than once in the training set and may have been shared between the
training and test sets.
Cross-subject analyses were similar except that the training set was 42 times
as large as the test set.

\paragraph{Statistical Significance}

For all classification accuracies, `*' indicates $p<0.05$, `**' indicates
$p<0.005$, and `***' indicates $p<0.0005$.
Such~$p$ values were computed for all classification results, taking a
one-sided binomial distribution (repeated independent Bernoulli trials with a
uniform distribution over possible outcomes) to be the null hypothesis.
In most cases, this leads to extremely small~$p$ values.
Assuming independence between trials, where each trial is uniformly
distributed, is warranted because all runs were counterbalanced.
All within-subject and cross-subject analyses that aggregate across subject are
highly significant; the largest~$p$ value was less than $\text{10}^{-\text{8}}$.
The `***' annotations are omitted in plots for such results.
%
%
%
Of the 210 per-subject analyses, only three instances have~$p$ values that
exceed 0.05 for within subject and only eighteen instances have~$p$ values that
exceed 0.05 for cross subject.
We know of no way to determine statistical significance of the
non-classification-accuracy results.

\paragraph{Determining Brain Regions Used by Classifiers}

Two different techniques were employed to determine the brain regions used by
the classifiers.
The first was a spatial searchlight which slides a small sphere across the
entire brain volume and performs training and test using only the voxels inside
that sphere.
A sphere of radius three voxels, densely placed at the centre of every voxel,
was used and no dimensionality reduction was performed on the remaining
voxels.
An eight-fold cross validation was then performed, as described above, for each
position of the sphere and those spheres whose classification accuracies
exceeded a specified threshold were back-projected onto the anatomical scans.
The second method back-projected the SVM coefficients for the trained
classifiers onto the anatomical scans using a classifier \citep{hanson2009} with
a different metric, $w(i)^2$.
The higher the absolute value of the coefficient the more that voxel
contributes to the classification performance of the SVM.

\section*{Addendum}

\begin{compactdesc}
\item[Acknowledgements] AB, NS, and JMS were supported, in part, by Army
  Research Laboratory (ARL) Cooperative Agreement W911NF-10-2-0060.
  AB was supported, in part, by the Center for Brains, Minds and Machines
  (CBMM), funded by NSF STC award CCF-1231216.
  CX and JJC were supported, in part, by ARL Cooperative Agreement
  W911NF-10-2-0062 and NSF CAREER grant IIS-0845282.
  CDF was supported, in part, by NSF grant CNS-0855157.
  CH and SJH were supported, in part, by the McDonnell Foundation.
  BAP was supported, in part, by Science Foundation Ireland grant
  09/IN.1/I2637.
  The views and conclusions contained in this document are those of the authors
  and should not be interpreted as representing the official policies, either
  express or implied, of the supporting institutions.
  The U.S. Government is authorised to reproduce and distribute reprints for
  Government purposes, notwithstanding any copyright notation herein.
  Dr.\ Gregory G. Tamer, Jr.\ provided assistance with imaging and analysis.
\item[Author Contributions] AB, NS, and JMS designed the experiments, prepared
  stimuli, ran subjects, conducted analyses, and wrote the manuscript.
  CX and JJC implemented dimensionality reduction and the classifier.
  CDF, SH, EM, and RBW contributed to experiment design and conducting analyses.
  CH, SJH, and BAP contributed to experiment design, running pilot
  subjects, and conducting analyses.
  TMT contributed to experiment design, running subjects, and conducting
  analyses.
\item[Data Deposition] Stimuli, presentation files and software, and scan data
  are available at\\%
  \url{http://upplysingaoflun.ecn.purdue.edu/~qobi/fmri2014.tgz}
\item[Competing Interests] The authors declare that they have no competing
  financial interests.
\item[Correspondence] Correspondence and requests for materials should be
  addressed to \href{mailto:qobi@purdue.edu}{Jeffrey Mark Siskind}.
\end{compactdesc}

\bibliographystyle{abbrvnat}
\setlength{\bibsep}{0.25ex}
\bibliography{arxiv2015}

\begin{figure}
  \centering
  {\setlength{\tabcolsep}{1pt}
  \begin{tabular}{@{}ccc@{\hspace{10pt}}ccc@{\hspace{10pt}}ccc@{}}
    \includegraphics[width=0.1\textwidth]{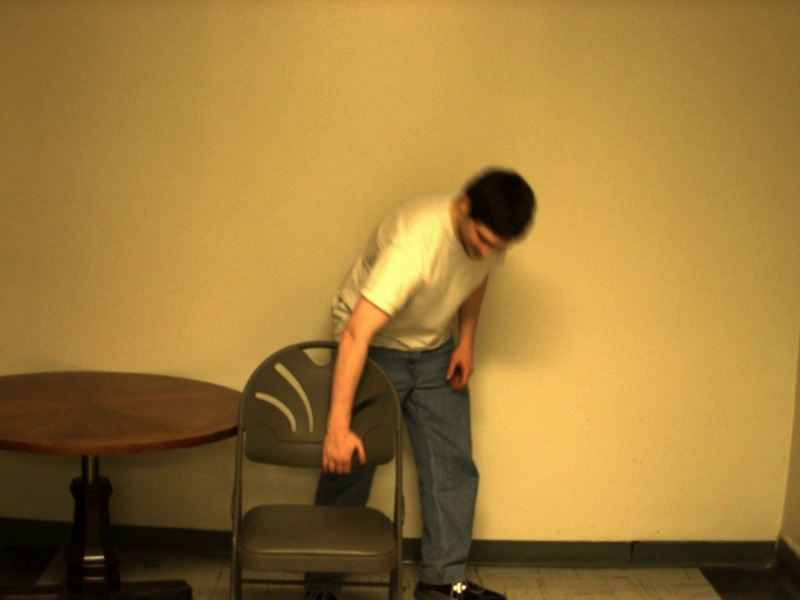}&
    \includegraphics[width=0.1\textwidth]{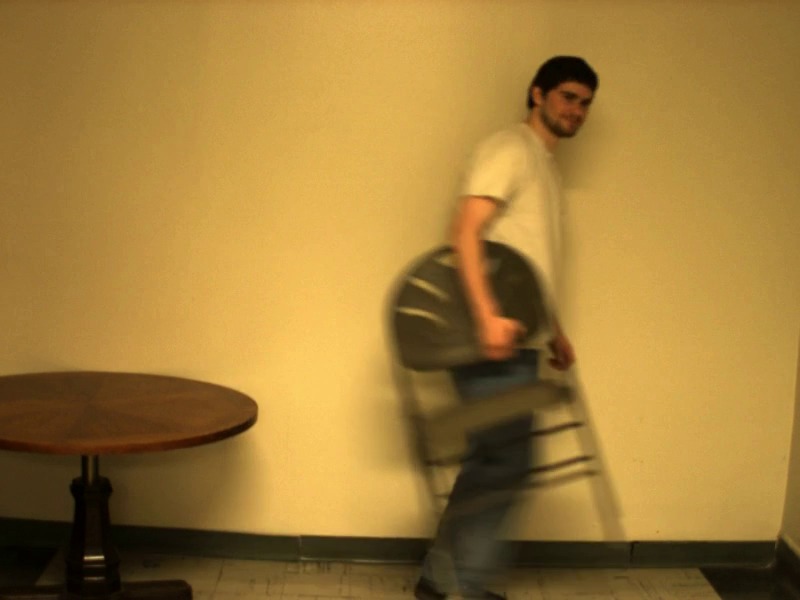}&
    \includegraphics[width=0.1\textwidth]{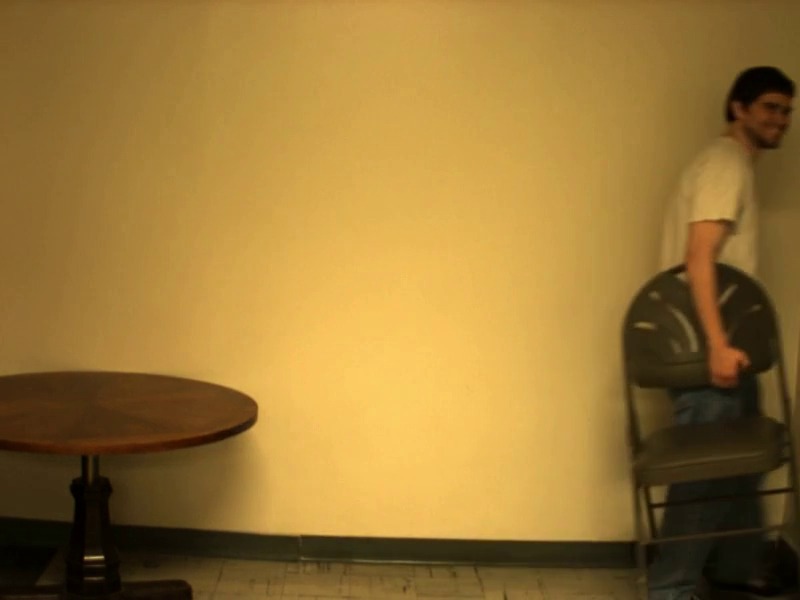}&
    \includegraphics[width=0.1\textwidth]{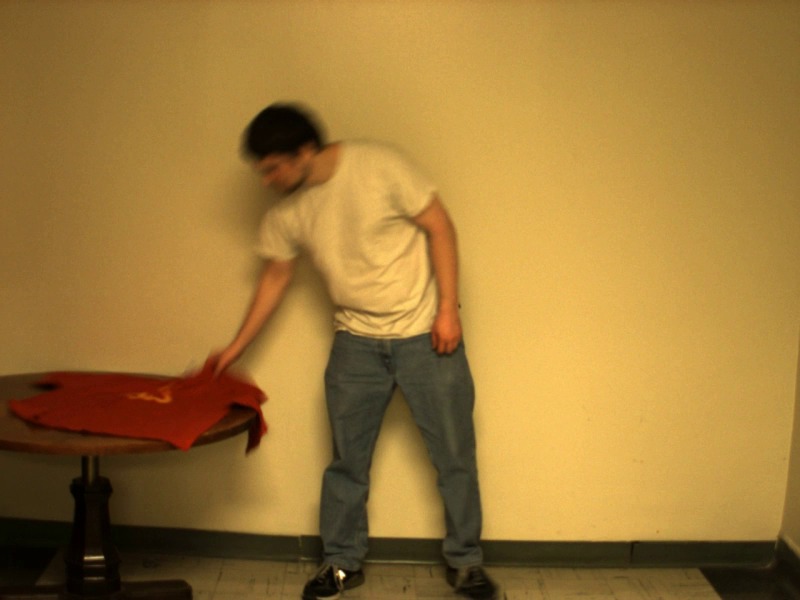}&
    \includegraphics[width=0.1\textwidth]{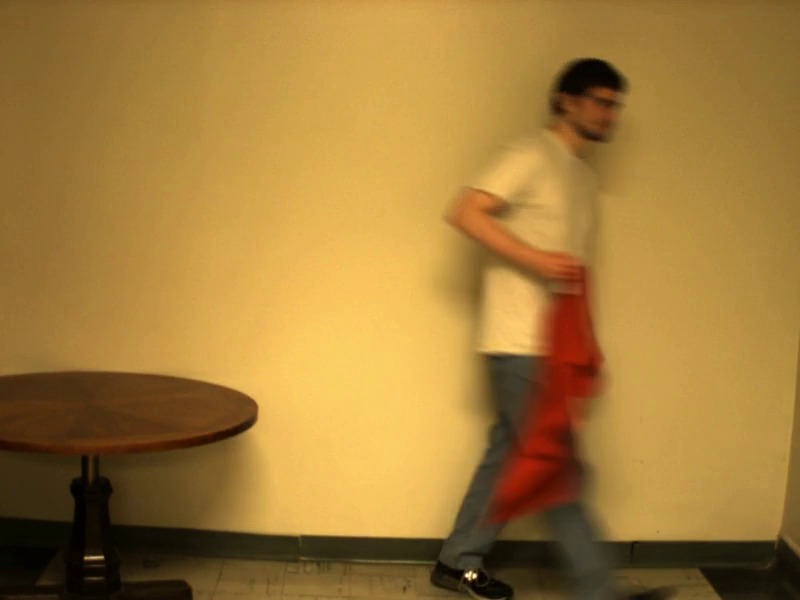}&
    \includegraphics[width=0.1\textwidth]{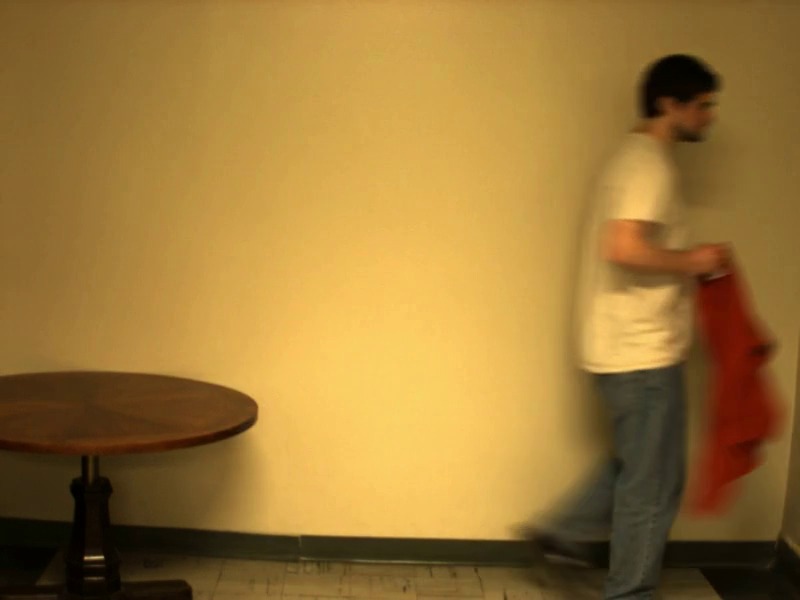}&
    \includegraphics[width=0.1\textwidth]{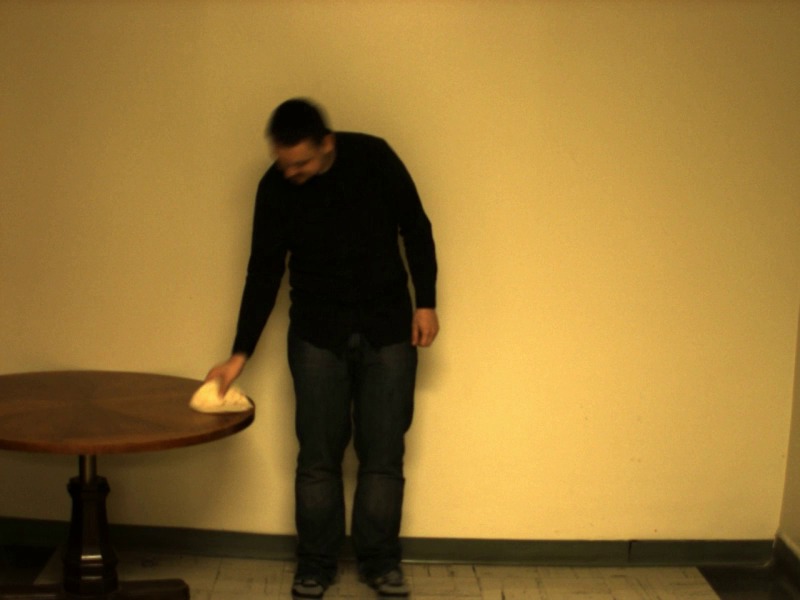}&
    \includegraphics[width=0.1\textwidth]{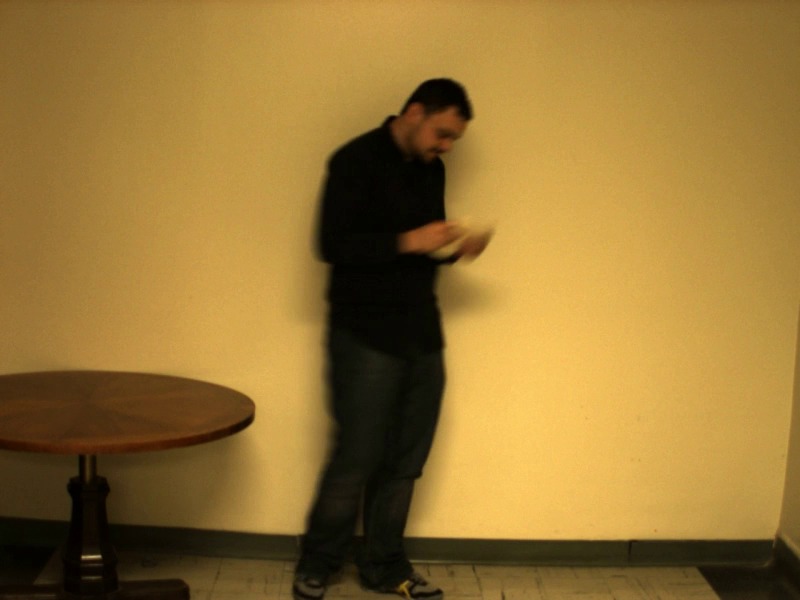}&
    \includegraphics[width=0.1\textwidth]{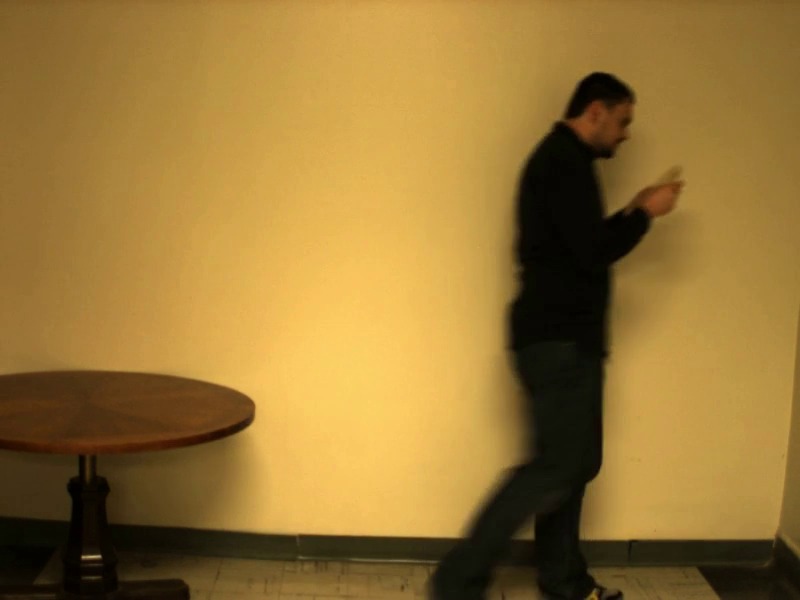}\\[-0.8ex]
    \multicolumn{3}{c@{\hspace{10pt}}}{\emph{carry chair}}&
    \multicolumn{3}{c@{\hspace{10pt}}}{\emph{carry shirt}}&
    \multicolumn{3}{c}{\emph{carry tortilla}}\\[0.5ex]
    \includegraphics[width=0.1\textwidth]{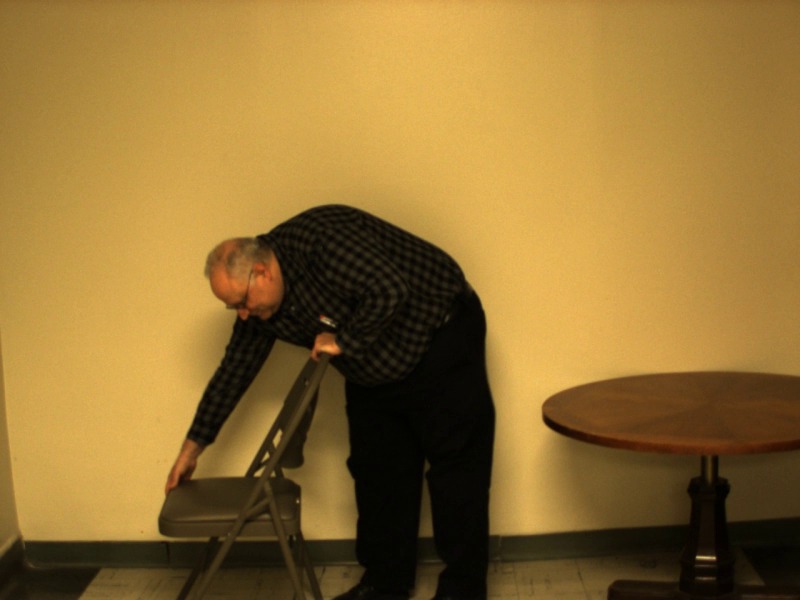}&
    \includegraphics[width=0.1\textwidth]{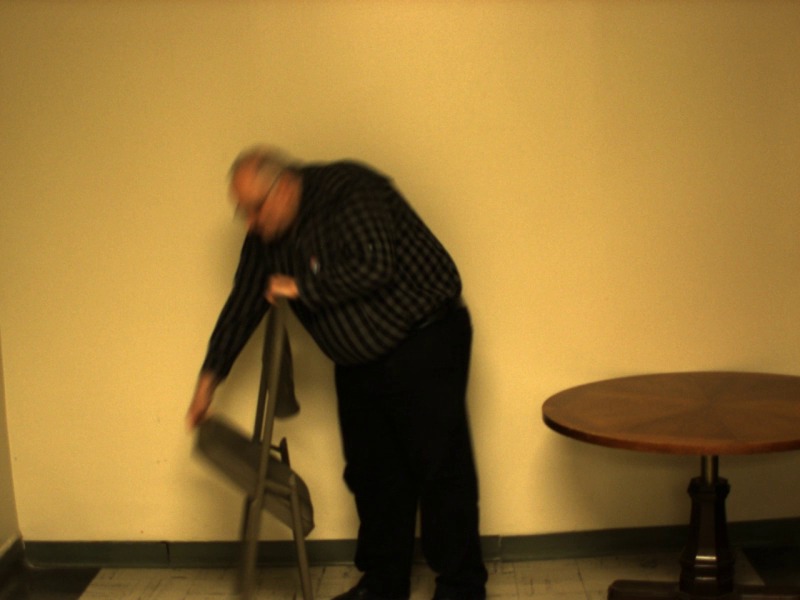}&
    \includegraphics[width=0.1\textwidth]{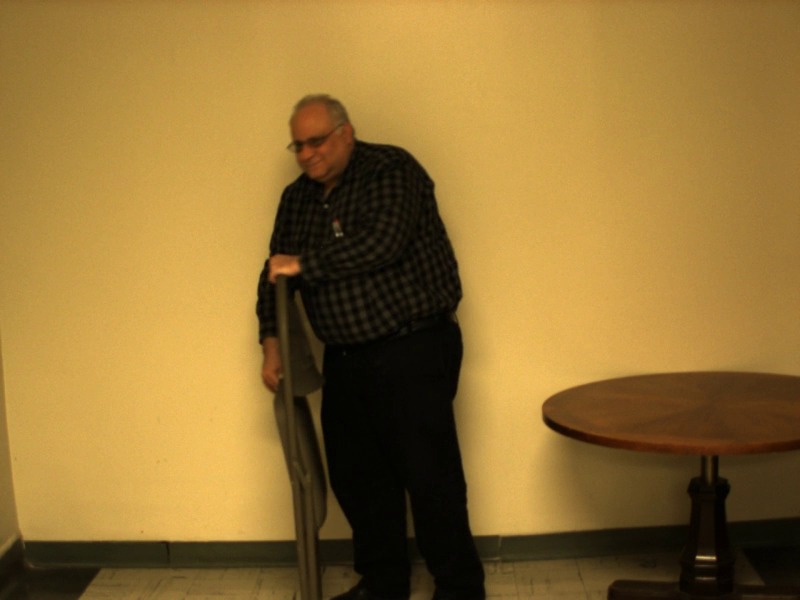}&
    \includegraphics[width=0.1\textwidth]{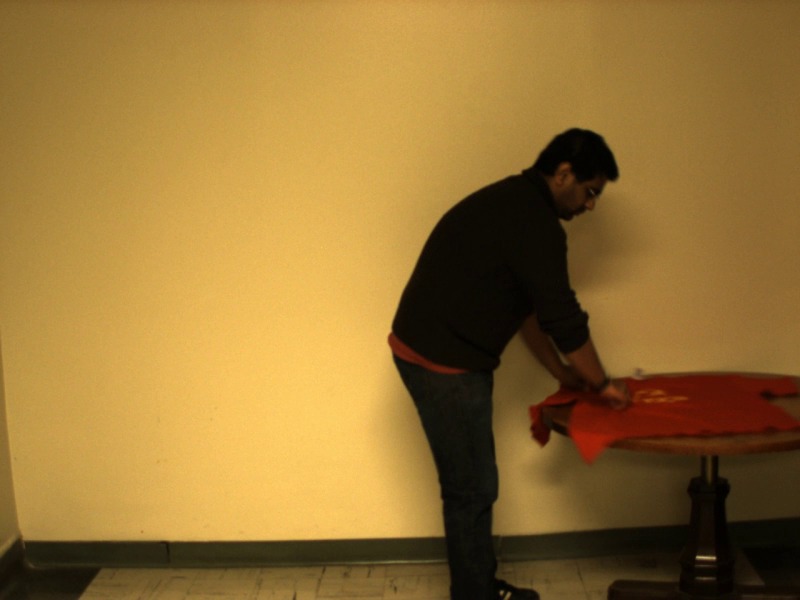}&
    \includegraphics[width=0.1\textwidth]{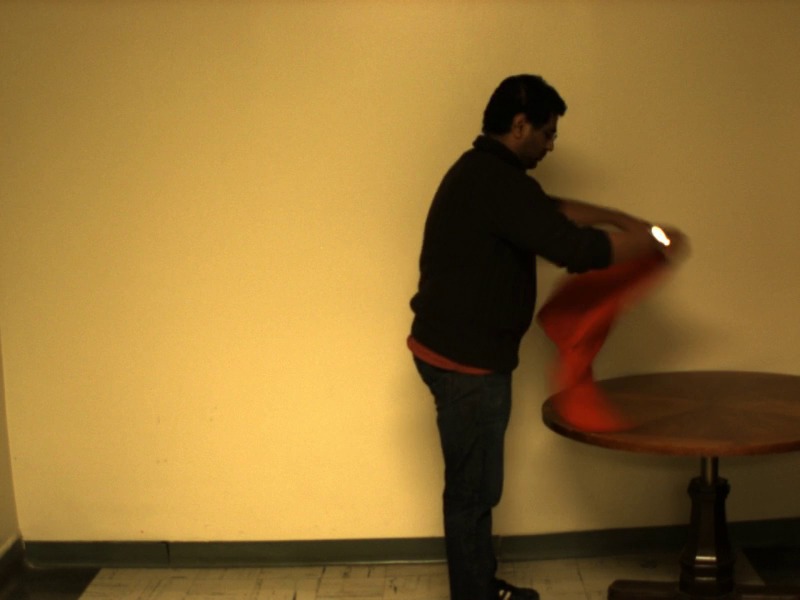}&
    \includegraphics[width=0.1\textwidth]{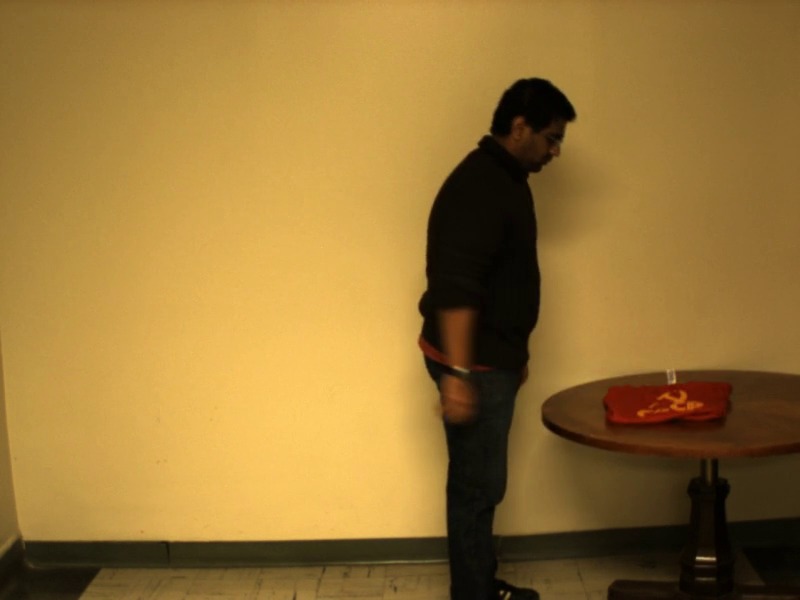}&
    \includegraphics[width=0.1\textwidth]{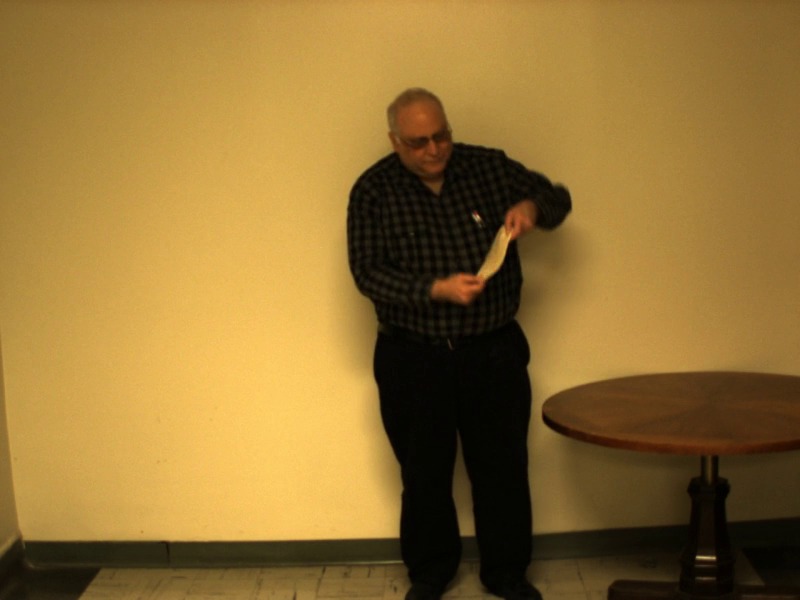}&
    \includegraphics[width=0.1\textwidth]{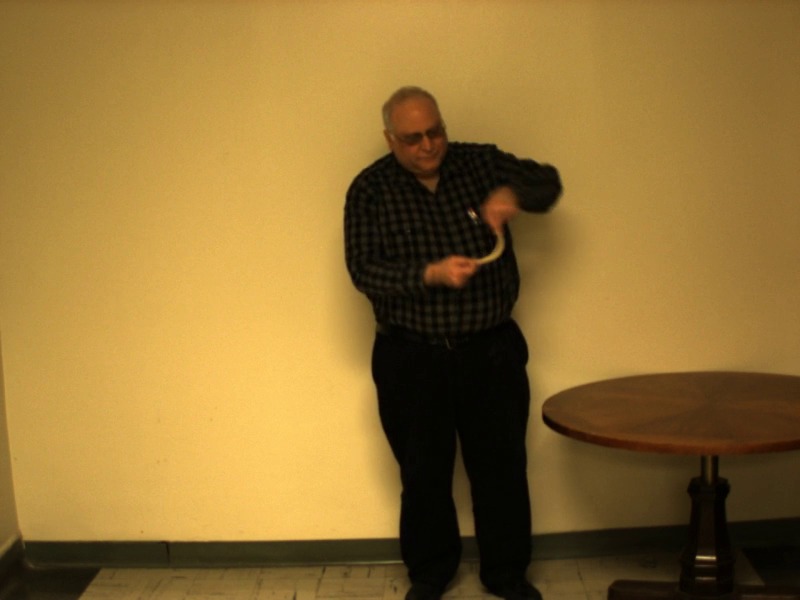}&
    \includegraphics[width=0.1\textwidth]{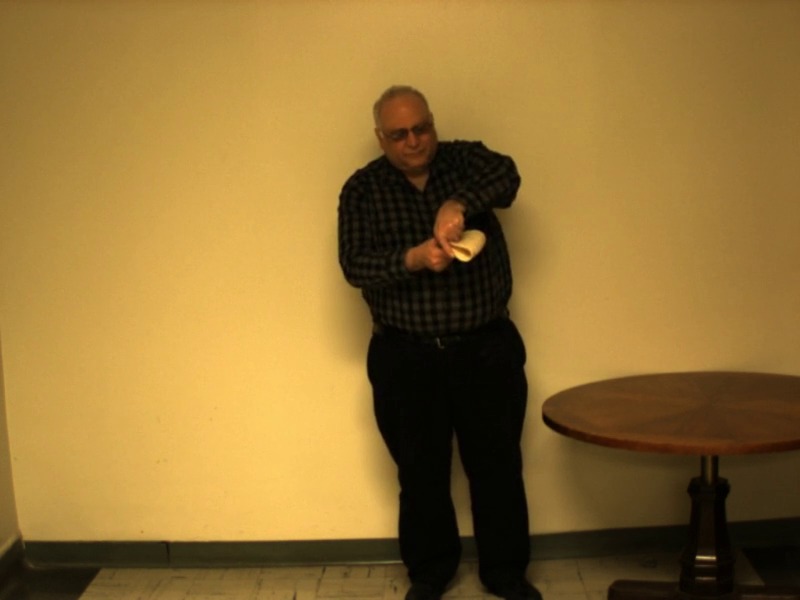}\\[-0.8ex]
    \multicolumn{3}{c@{\hspace{10pt}}}{\emph{fold chair}}&
    \multicolumn{3}{c@{\hspace{10pt}}}{\emph{fold shirt}}&
    \multicolumn{3}{c}{\emph{fold tortilla}}\\[0.5ex]
    \includegraphics[width=0.1\textwidth]{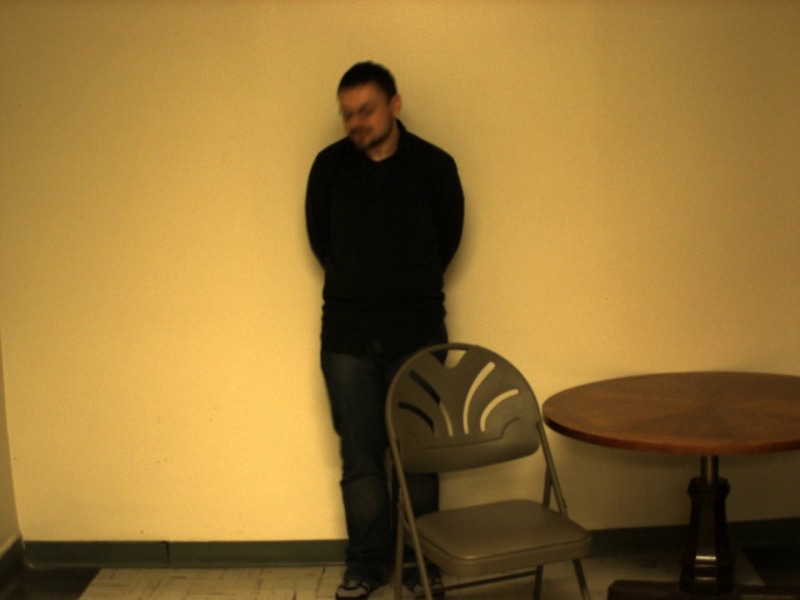}&
    \includegraphics[width=0.1\textwidth]{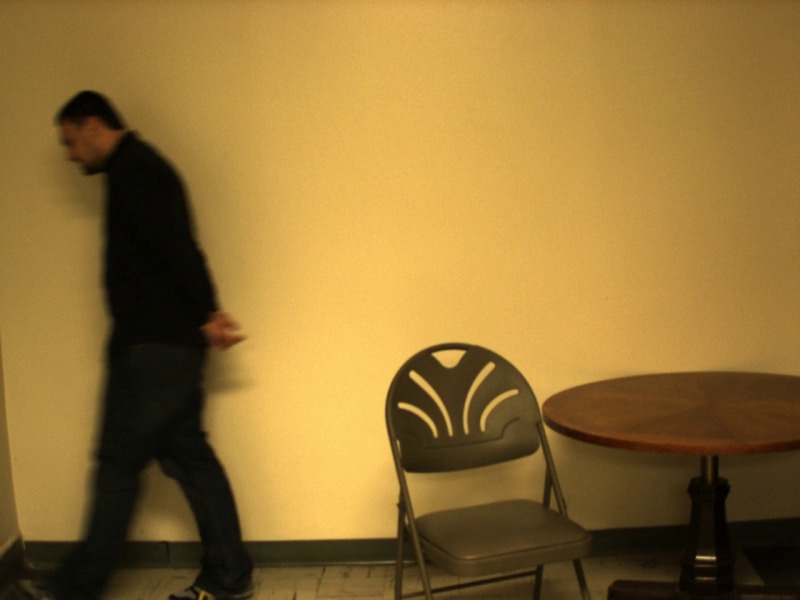}&
    \includegraphics[width=0.1\textwidth]{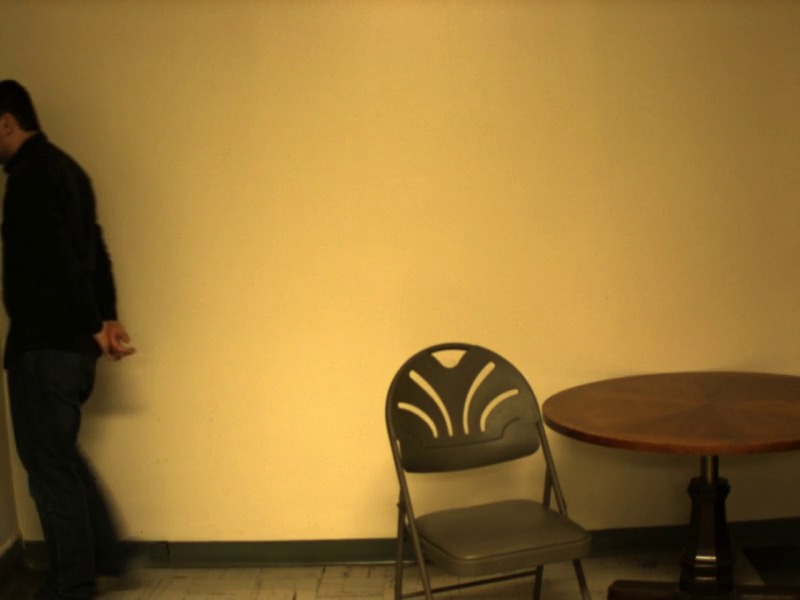}&
    \includegraphics[width=0.1\textwidth]{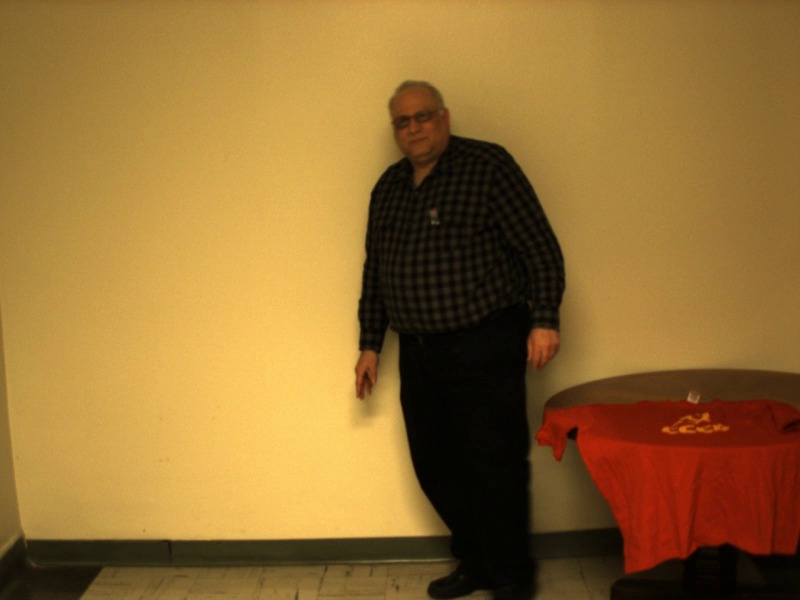}&
    \includegraphics[width=0.1\textwidth]{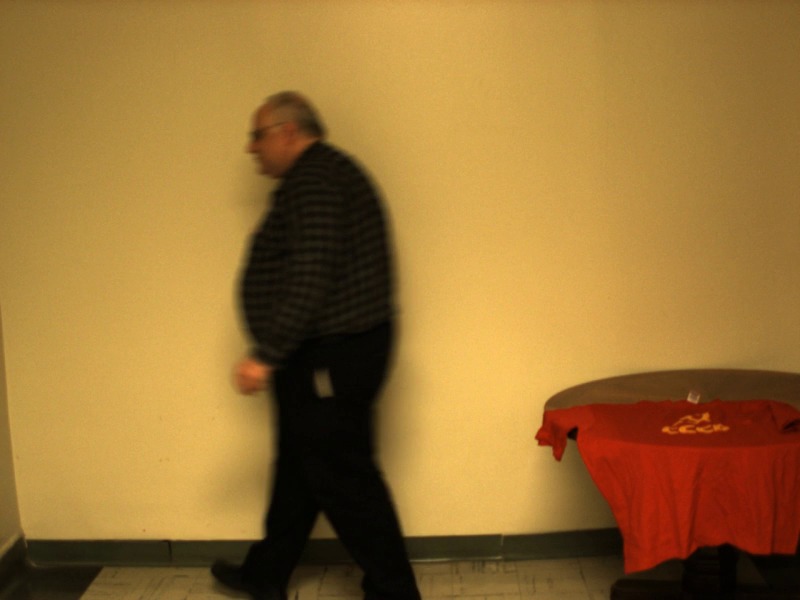}&
    \includegraphics[width=0.1\textwidth]{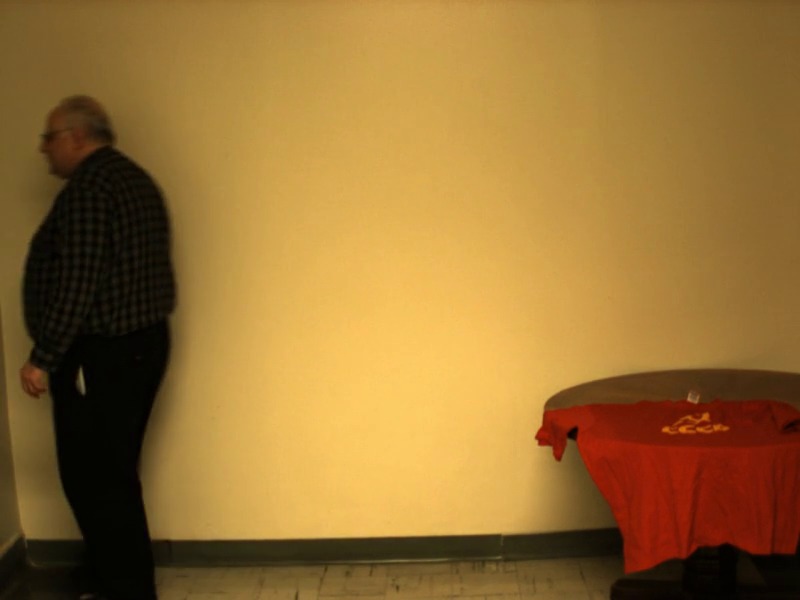}&
    \includegraphics[width=0.1\textwidth]{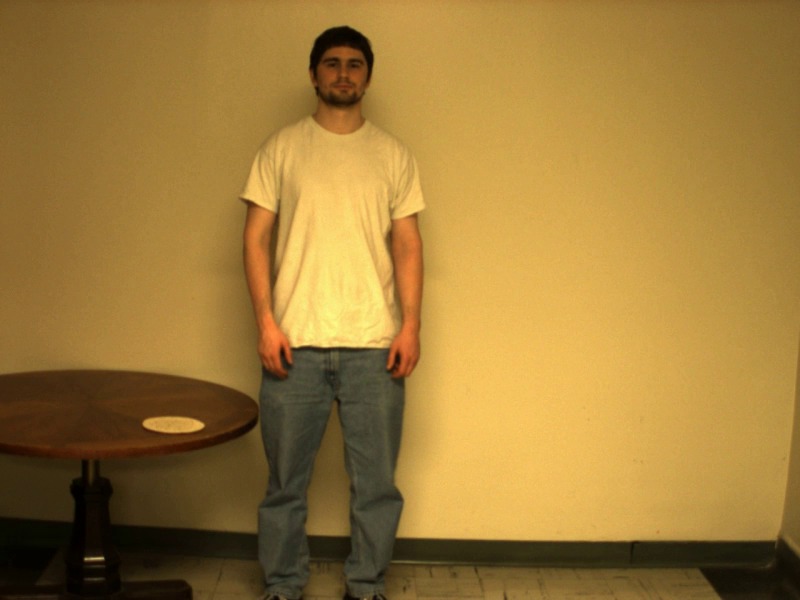}&
    \includegraphics[width=0.1\textwidth]{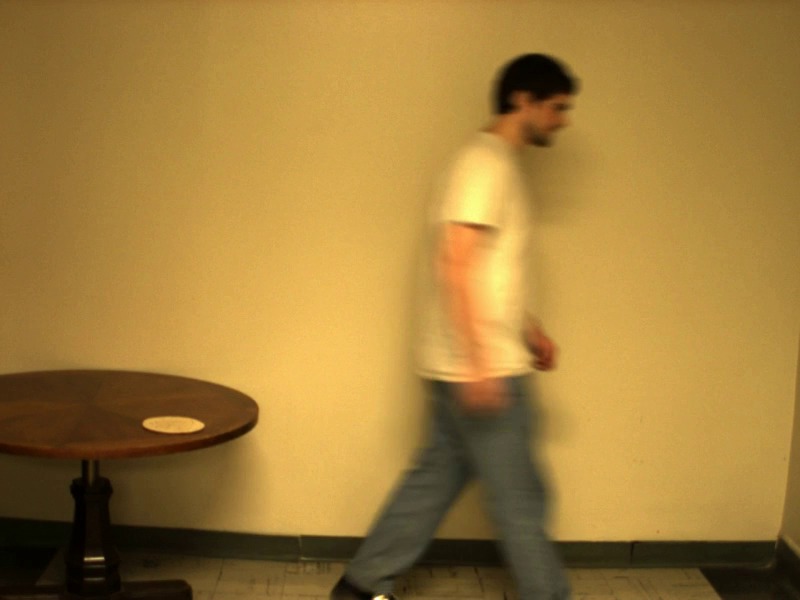}&
    \includegraphics[width=0.1\textwidth]{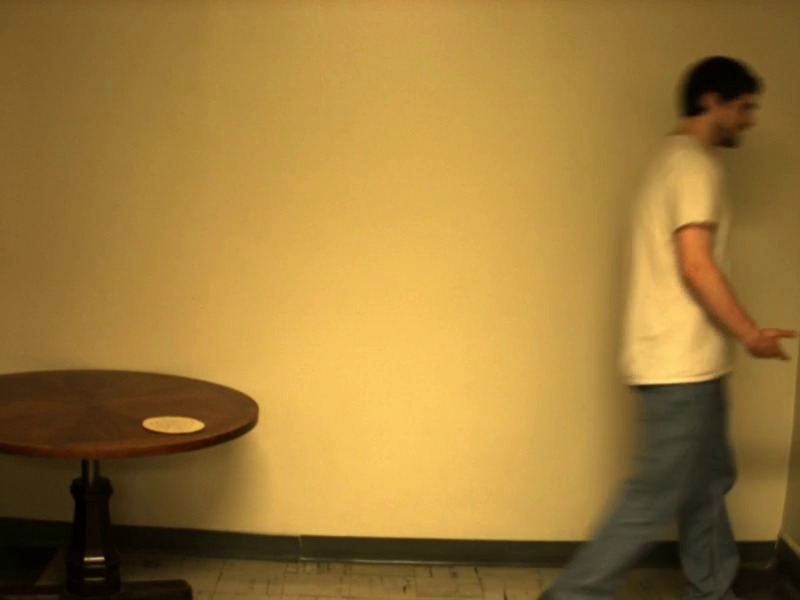}\\[-0.8ex]
    \multicolumn{3}{c@{\hspace{10pt}}}{\emph{leave chair}}&
    \multicolumn{3}{c@{\hspace{10pt}}}{\emph{leave shirt}}&
    \multicolumn{3}{c}{\emph{leave tortilla}}
  \end{tabular}}\\*[1ex]
  \begin{tabular}{@{}l@{\hspace*{6ex}}rrrrrrr@{}}
      \toprule
      analysis                      &     a&     b&     c&     d&     e&     f&     g\\
      \midrule
      \actor                        &     4&   126&   504&    18&    72&   576&  4032\\
      \Verb                         &     3&   168&   504&    24&    72&   576&  4032\\
      \object                       &     3&   168&   504&    24&    72&   576&  4032\\
      \direction                    &     2&   168&   336&    24&    48&   384&  2688\\
      \location                     &     2&    84&   168&    12&    24&   192&  1344\\
      \midrule
      \actor-\Verb                  &    12&    42&   504&     6&    72&   576&  4032\\
      \actor-\object                &    12&    42&   504&     6&    72&   576&  4032\\
      \actor-\direction             &     8&    42&   336&     6&    48&   384&  2688\\
      \actor-\location              &     8&    21&   168&     3&    24&   192&  1344\\
      \Verb-\object                 &     9&    56&   504&     8&    72&   576&  4032\\
      \Verb-\direction              &     4&    84&   336&    12&    48&   384&  2688\\
      \object-\direction            &     6&    56&   336&     8&    48&   384&  2688\\
      \object-\location             &     6&    28&   168&     4&    24&   192&  1344\\
      \midrule
      \actor-\Verb-\object          &    36&    14&   504&     2&    72&   576&  4032\\
      \actor-\Verb-\direction       &    16&    21&   336&     3&    48&   384&  2688\\
      \actor-\object-\direction     &    24&    14&   336&     2&    48&   384&  2688\\
      \Verb-\object-\direction      &    12&    28&   336&     4&    48&   384&  2688\\
      \midrule
      sentence                      &    72&     7&   504&     1&    72&   576&  4032\\
      \bottomrule
  \end{tabular}
  \caption{%
    (top)~Key frames from sample stimuli.
    (bottom)~Dataset statistics for single constituent, joint constituent pair,
    joint constituent triple, and independent sentence analyses.
    (a)~Number of classes.
    (b)~Number of training samples per subject, fold, and class.
    (c)~Number of training samples per subject and fold ($\text{a}\times
    \text{b}=\text{e}\times 7$).
    (d)~Number of test samples per subject, fold, and class.
    (e)~Number of test samples per subject and fold ($\text{a}\times \text{d}$).
    (f)~Number of test samples per subject ($\text{e}\times 8$).
    (g)~Number of test samples ($\text{f}\times 7$).
    The number of classes and number of test samples for independent and joint
    analyses for corresponding constituent pairs and triples are the same.
    No classifiers were trained for the independent constituent pair and triple
    analyses as these used the single-constituent classifiers.
    The number of training samples for the sentence analysis is the
    hypothetical number for a joint classifier that was not trained; only
    independent classification was attempted due to insufficient training-set
    size.}
  \vspace*{-3ex}
\end{figure}

\begin{table}
  \centering
  \resizebox{\textwidth}{!}{\begin{tabular}{@{}llll@{\hspace*{4ex}}lllllll@{}}
      \toprule
analysis                &  chance&    mean&  stddev&       1&       2&       3&       4&       5&       6&       7\\
\midrule
\actor                        &  0.2500&  0.3333$^{***}$&   0.045&  0.304$^{**}$&  0.314$^{***}$&  0.356$^{***}$&  0.352$^{***}$&  0.332$^{***}$&  0.323$^{***}$&  0.352$^{***}$\\
\Verb                         &  0.3333&  0.7892$^{***}$&   0.079&  0.776$^{***}$&  0.674$^{***}$&  0.830$^{***}$&  0.816$^{***}$&  0.870$^{***}$&  0.776$^{***}$&  0.783$^{***}$\\
\object                       &  0.3333&  0.5980$^{***}$&   0.067&  0.554$^{***}$&  0.542$^{***}$&  0.623$^{***}$&  0.660$^{***}$&  0.663$^{***}$&  0.575$^{***}$&  0.569$^{***}$\\
\direction                    &  0.5000&  0.8460$^{***}$&   0.076&  0.846$^{***}$&  0.763$^{***}$&  0.930$^{***}$&  0.823$^{***}$&  0.880$^{***}$&  0.807$^{***}$&  0.872$^{***}$\\
\location                     &  0.5000&  0.7128$^{***}$&   0.110&  0.698$^{***}$&  0.615$^{**}$&  0.677$^{***}$&  0.734$^{***}$&  0.755$^{***}$&  0.797$^{***}$&  0.714$^{***}$\\
\midrule
\actor-\Verb                  &  0.0833&  0.2579$^{***}$&   0.063&  0.210$^{***}$&  0.214$^{***}$&  0.302$^{***}$&  0.264$^{***}$&  0.292$^{***}$&  0.259$^{***}$&  0.266$^{***}$\\
\actor\&\Verb                 &  0.0833&  0.2686$^{***}$&   0.054&  0.236$^{***}$&  0.220$^{***}$&  0.295$^{***}$&  0.293$^{***}$&  0.295$^{***}$&  0.252$^{***}$&  0.288$^{***}$\\[2ex]
\actor-\object                &  0.0833&  0.1756$^{***}$&   0.055&  0.123$^{**}$&  0.158$^{***}$&  0.193$^{***}$&  0.220$^{***}$&  0.184$^{***}$&  0.175$^{***}$&  0.175$^{***}$\\
\actor\&\object               &  0.0833&  0.2061$^{***}$&   0.041&  0.167$^{***}$&  0.170$^{***}$&  0.224$^{***}$&  0.229$^{***}$&  0.233$^{***}$&  0.208$^{***}$&  0.212$^{***}$\\[2ex]
\actor-\direction             &  0.1250&  0.2504$^{***}$&   0.084&  0.198$^{***}$&  0.195$^{***}$&  0.302$^{***}$&  0.263$^{***}$&  0.227$^{***}$&  0.273$^{***}$&  0.294$^{***}$\\
\actor\&\direction            &  0.1250&  0.2846$^{***}$&   0.071&  0.260$^{***}$&  0.258$^{***}$&  0.313$^{***}$&  0.323$^{***}$&  0.289$^{***}$&  0.234$^{***}$&  0.315$^{***}$\\[2ex]
\actor-\location              &  0.1250&  0.2031$^{***}$&   0.095&  0.161&  0.177$^{*}$&  0.245$^{***}$&  0.141&  0.203$^{**}$&  0.297$^{***}$&  0.198$^{**}$\\
\actor\&\location             &  0.1250&  0.2403$^{***}$&   0.079&  0.208$^{**}$&  0.182$^{*}$&  0.240$^{***}$&  0.224$^{***}$&  0.245$^{***}$&  0.302$^{***}$&  0.281$^{***}$\\[2ex]
\Verb-\object                 &  0.1111&  0.4958$^{***}$&   0.092&  0.523$^{***}$&  0.389$^{***}$&  0.540$^{***}$&  0.545$^{***}$&  0.595$^{***}$&  0.462$^{***}$&  0.417$^{***}$\\
\Verb\&\object                &  0.1111&  0.4794$^{***}$&   0.089&  0.439$^{***}$&  0.366$^{***}$&  0.514$^{***}$&  0.547$^{***}$&  0.589$^{***}$&  0.437$^{***}$&  0.464$^{***}$\\[2ex]
\Verb-\direction              &  0.2500&  0.7143$^{***}$&   0.111&  0.737$^{***}$&  0.581$^{***}$&  0.766$^{***}$&  0.727$^{***}$&  0.828$^{***}$&  0.596$^{***}$&  0.766$^{***}$\\
\Verb\&\direction             &  0.2500&  0.6711$^{***}$&   0.115&  0.661$^{***}$&  0.505$^{***}$&  0.784$^{***}$&  0.682$^{***}$&  0.766$^{***}$&  0.625$^{***}$&  0.674$^{***}$\\[2ex]
\object-\direction            &  0.1667&  0.3906$^{***}$&   0.094&  0.354$^{***}$&  0.276$^{***}$&  0.456$^{***}$&  0.453$^{***}$&  0.471$^{***}$&  0.365$^{***}$&  0.359$^{***}$\\
\object\&\direction           &  0.1667&  0.4621$^{***}$&   0.100&  0.427$^{***}$&  0.346$^{***}$&  0.544$^{***}$&  0.513$^{***}$&  0.542$^{***}$&  0.414$^{***}$&  0.448$^{***}$\\[2ex]
\object-\location             &  0.1667&  0.5513$^{***}$&   0.107&  0.604$^{***}$&  0.557$^{***}$&  0.599$^{***}$&  0.505$^{***}$&  0.536$^{***}$&  0.599$^{***}$&  0.458$^{***}$\\
\object\&\location            &  0.1667&  0.5000$^{***}$&   0.110&  0.453$^{***}$&  0.437$^{***}$&  0.469$^{***}$&  0.557$^{***}$&  0.563$^{***}$&  0.552$^{***}$&  0.469$^{***}$\\
\midrule
\actor-\Verb-\object          &  0.0278&  0.1434$^{***}$&   0.045&  0.125$^{***}$&  0.099$^{***}$&  0.179$^{***}$&  0.177$^{***}$&  0.161$^{***}$&  0.149$^{***}$&  0.113$^{***}$\\
\actor\&\Verb\&\object        &  0.0278&  0.1687$^{***}$&   0.042&  0.135$^{***}$&  0.123$^{***}$&  0.184$^{***}$&  0.193$^{***}$&  0.210$^{***}$&  0.161$^{***}$&  0.174$^{***}$\\[2ex]
\actor-\Verb-\direction       &  0.0625&  0.2139$^{***}$&   0.061&  0.188$^{***}$&  0.172$^{***}$&  0.279$^{***}$&  0.203$^{***}$&  0.253$^{***}$&  0.180$^{***}$&  0.224$^{***}$\\
\actor\&\Verb\&\direction     &  0.0625&  0.2333$^{***}$&   0.075&  0.201$^{***}$&  0.182$^{***}$&  0.271$^{***}$&  0.284$^{***}$&  0.260$^{***}$&  0.185$^{***}$&  0.250$^{***}$\\[2ex]
\actor-\object-\direction     &  0.0417&  0.0867$^{***}$&   0.040&  0.083$^{***}$&  0.068$^{*}$&  0.117$^{***}$&  0.060&  0.096$^{***}$&  0.099$^{***}$&  0.083$^{***}$\\
\actor\&\object\&\direction   &  0.0417&  0.1633$^{***}$&   0.055&  0.138$^{***}$&  0.107$^{***}$&  0.185$^{***}$&  0.206$^{***}$&  0.198$^{***}$&  0.138$^{***}$&  0.172$^{***}$\\[2ex]
\Verb-\object-\direction      &  0.0833&  0.3255$^{***}$&   0.102&  0.375$^{***}$&  0.182$^{***}$&  0.318$^{***}$&  0.401$^{***}$&  0.417$^{***}$&  0.289$^{***}$&  0.297$^{***}$\\
\Verb\&\object\&\direction    &  0.0833&  0.3679$^{***}$&   0.110&  0.339$^{***}$&  0.227$^{***}$&  0.445$^{***}$&  0.430$^{***}$&  0.474$^{***}$&  0.305$^{***}$&  0.357$^{***}$\\
\midrule
sentence\&                    &  0.0139&  0.1384$^{***}$&   0.043&  0.113$^{***}$&  0.087$^{***}$&  0.149$^{***}$&  0.167$^{***}$&  0.168$^{***}$&  0.135$^{***}$&  0.149$^{***}$\\
\bottomrule
  \end{tabular}}
  \caption{%
    Per-subject classification accuracy, including means and standard deviations
    across subjects, for different classifiers, averaged across fold.
    Joint classifiers are indicated with `-'.
    Independent classifiers are indicated with `\&'.}
  \label{xtab:results}
\end{table}

\begin{figure}
  \centering
  \begin{tabular}{@{}cccc@{}}
    \includegraphics[height=0.124\textheight]
                    {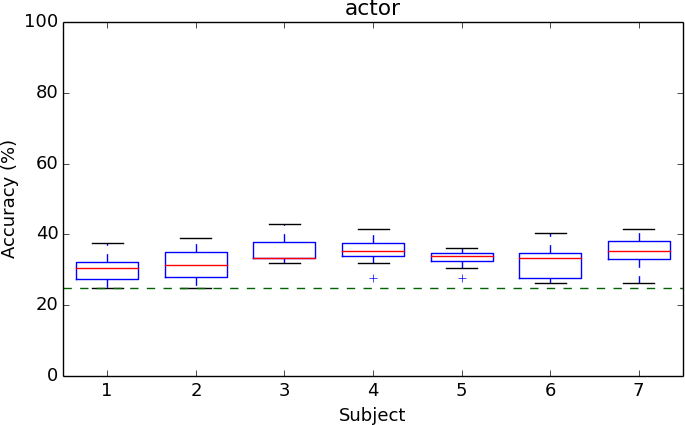}&
    \includegraphics[height=0.124\textheight]
                    {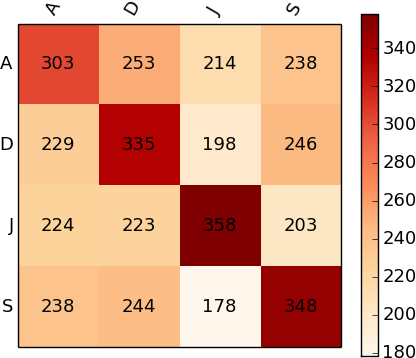}&
    \includegraphics[height=0.124\textheight]
                    {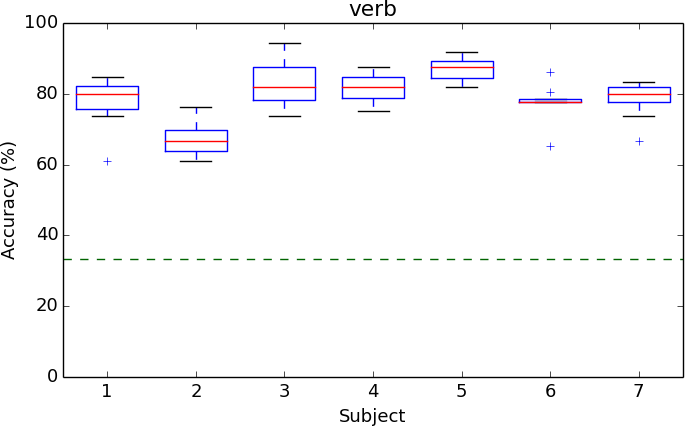}&
    \includegraphics[height=0.124\textheight]
                    {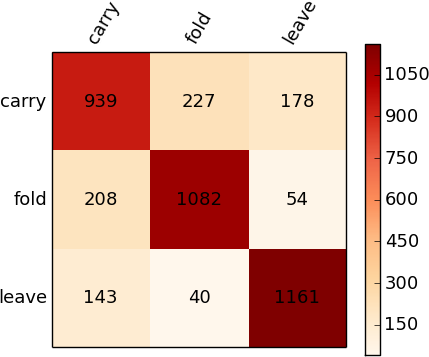}\\
    \includegraphics[height=0.124\textheight]
                    {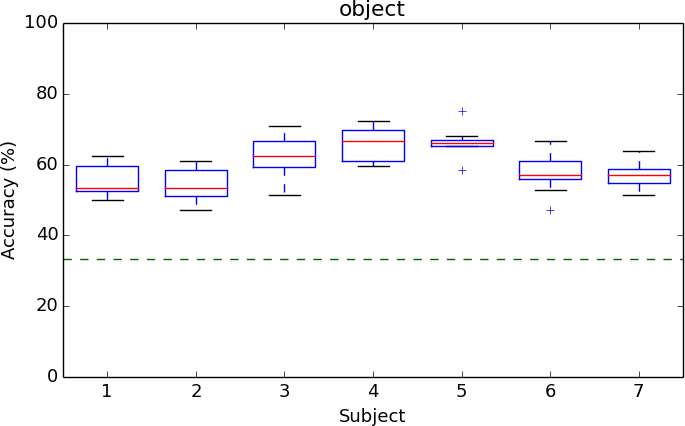}&
    \includegraphics[height=0.124\textheight]
                    {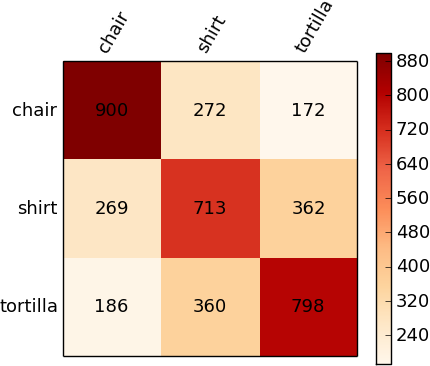}&
    \includegraphics[height=0.124\textheight]
                    {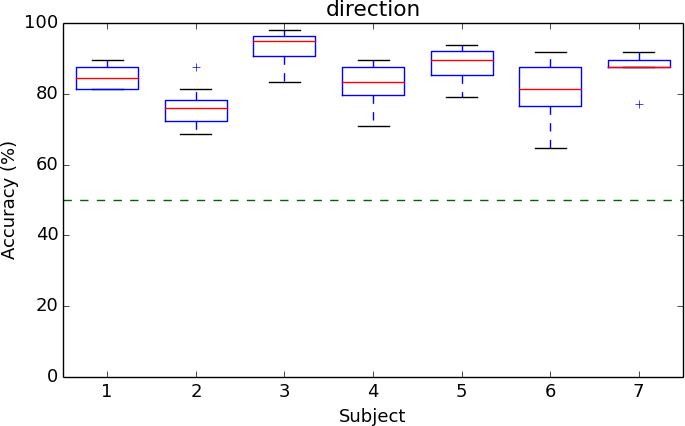}&
    \includegraphics[height=0.124\textheight]
                    {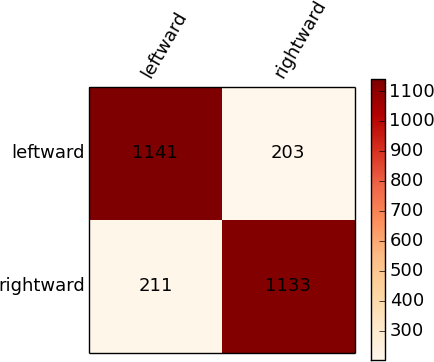}\\
    \includegraphics[height=0.124\textheight]
                    {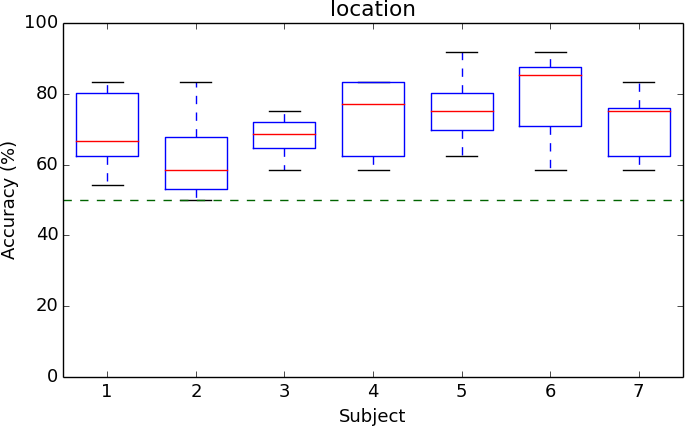}&
    \includegraphics[height=0.124\textheight]
                    {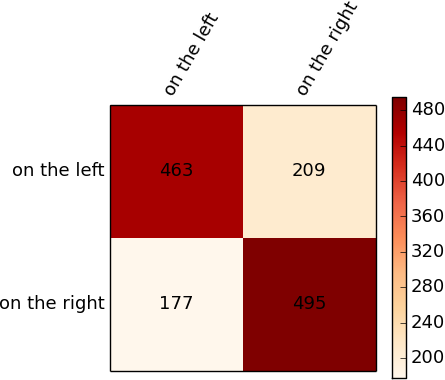}&
    \includegraphics[height=0.124\textheight]
                    {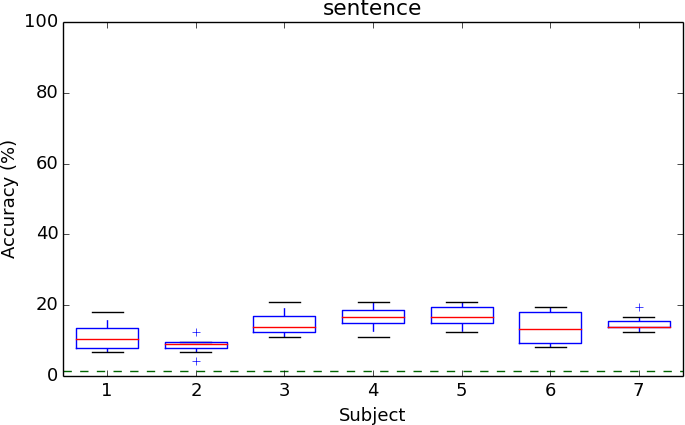}&\\*[6ex]
    \includegraphics[height=0.124\textheight]
                    {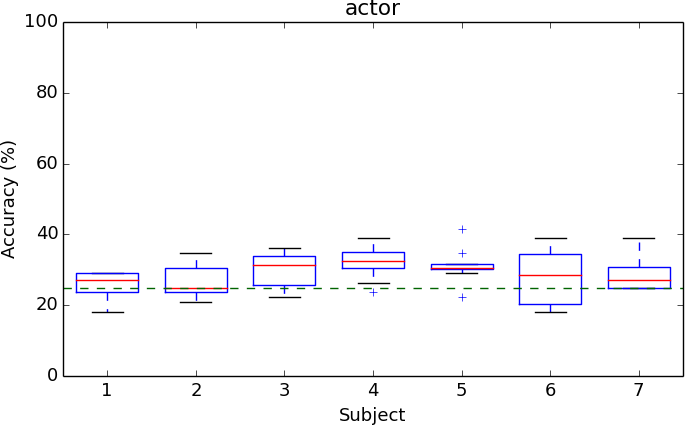}&
    \includegraphics[height=0.124\textheight]
                    {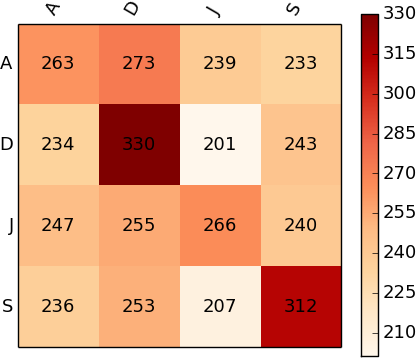}&
    \includegraphics[height=0.124\textheight]
                    {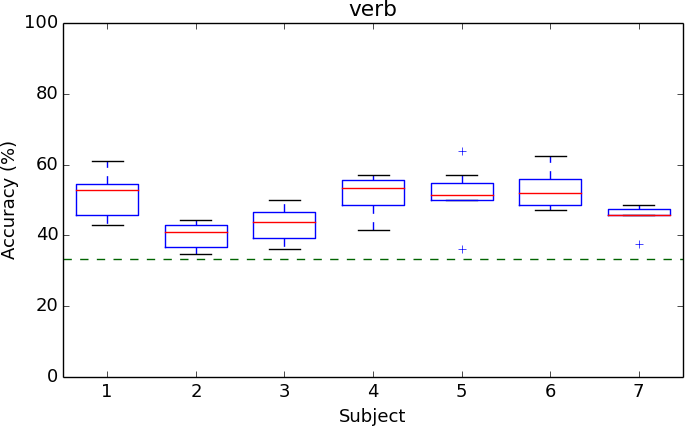}&
    \includegraphics[height=0.124\textheight]
                    {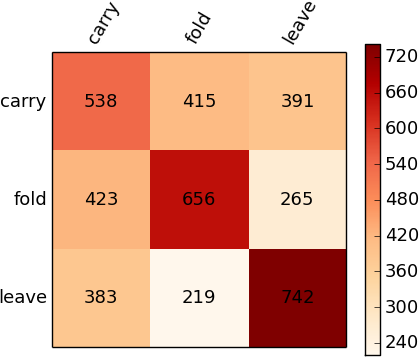}\\
    \includegraphics[height=0.124\textheight]
                    {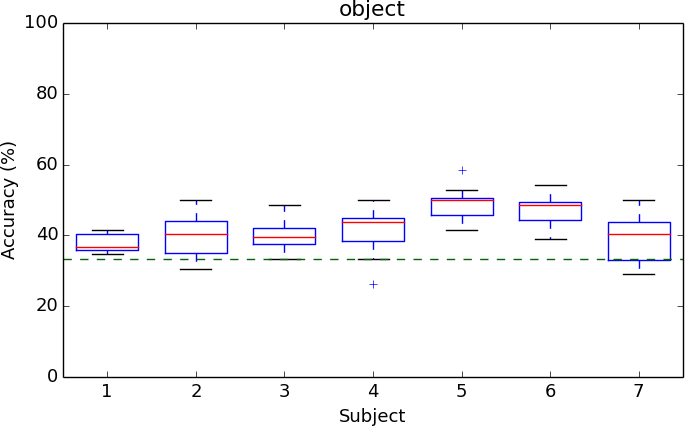}&
    \includegraphics[height=0.124\textheight]
                    {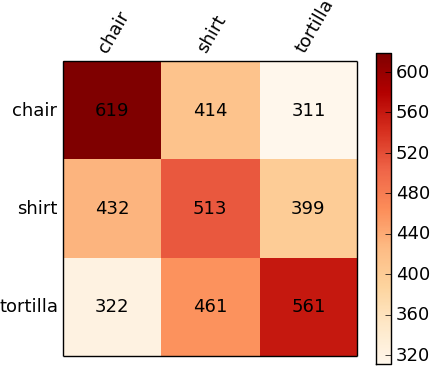}&
    \includegraphics[height=0.124\textheight]
                    {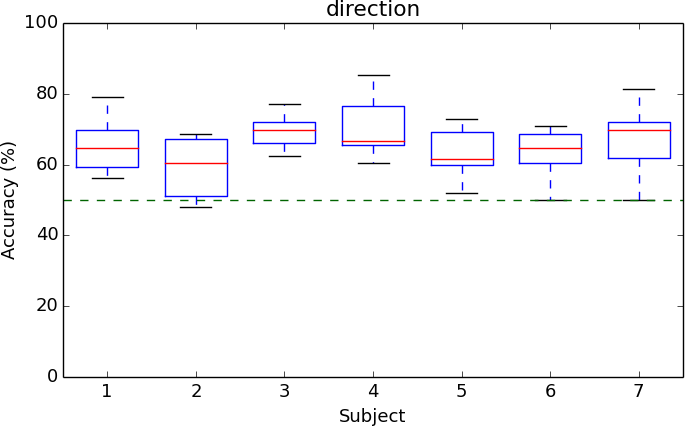}&
    \includegraphics[height=0.124\textheight]
                    {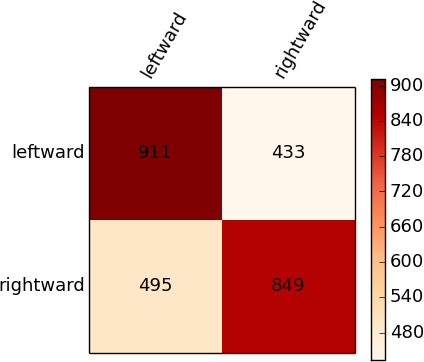}\\
    \includegraphics[height=0.124\textheight]
                    {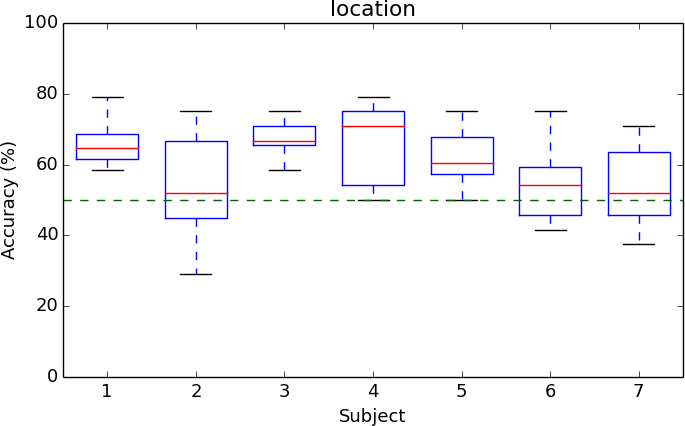}&
    \includegraphics[height=0.124\textheight]
                    {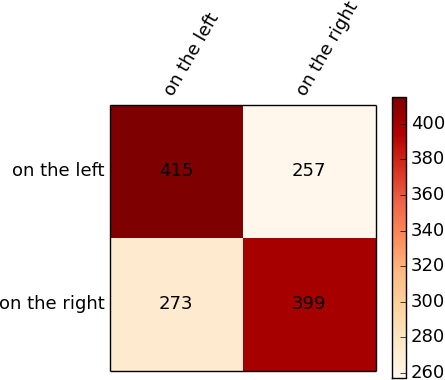}&
    \includegraphics[height=0.124\textheight]
                    {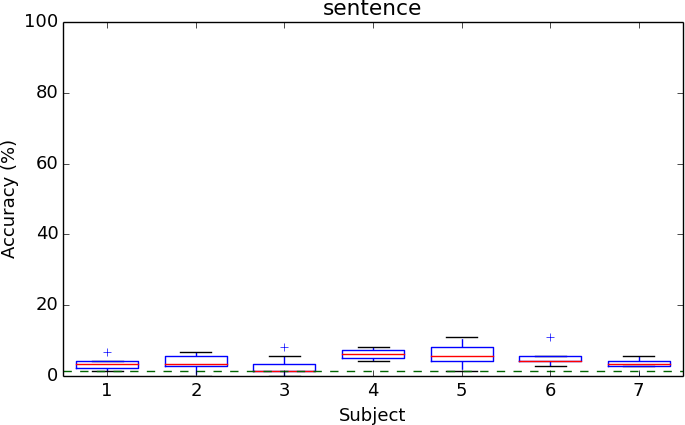}&

  \end{tabular}
  \caption{%
    (top)~Per-subject classification accuracy for \textbf{actor},
    \textbf{verb}, \textbf{object}, \textbf{direction}, \textbf{location}, and
    sentence across the different folds and corresponding confusion matrices
    aggregated across subject and fold.
    Note that they are largely diagonal.
    (bottom)~Cross-subject variants of top.}
\end{figure}

\begin{figure}
  \centering
  \begin{tabular}{@{}cc@{}}
  \includegraphics[height=0.134\textheight]
                  {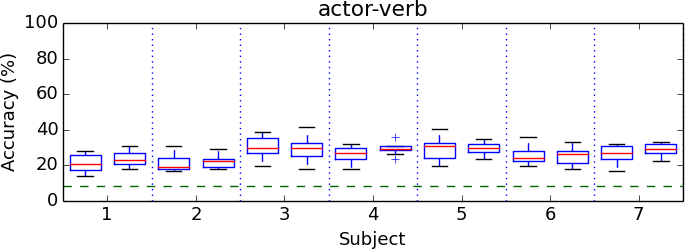}&
  \includegraphics[height=0.134\textheight]
                  {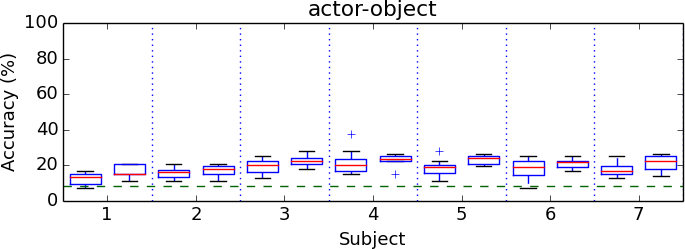}\\
  \includegraphics[height=0.134\textheight]
                  {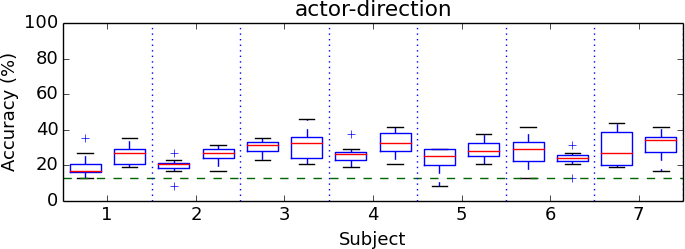}&
  \includegraphics[height=0.134\textheight]
                  {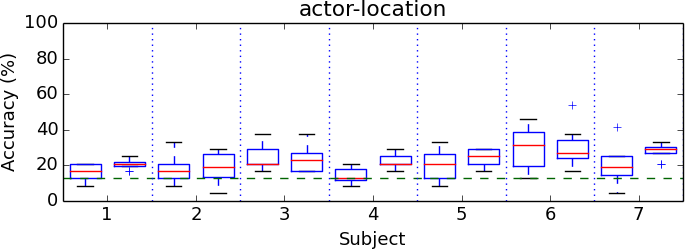}\\
  \includegraphics[height=0.134\textheight]
                  {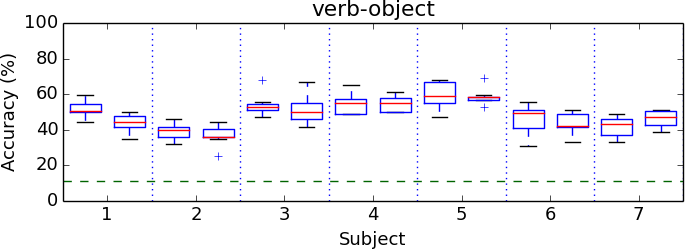}&
  \includegraphics[height=0.134\textheight]
                  {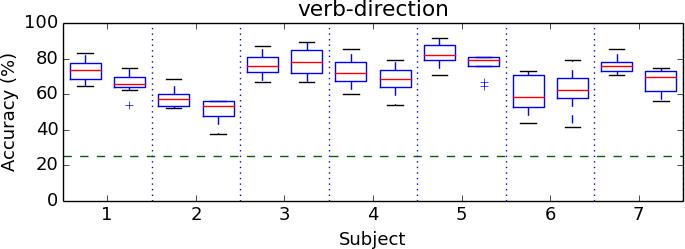}\\
  \includegraphics[height=0.134\textheight]
                  {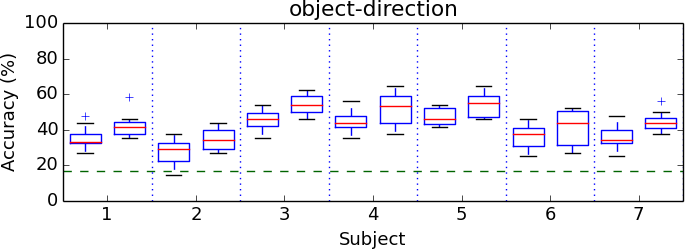}&
  \includegraphics[height=0.134\textheight]
                  {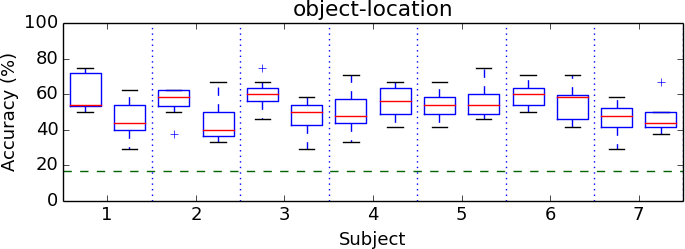}\\*[2ex]
  \includegraphics[height=0.134\textheight]
                  {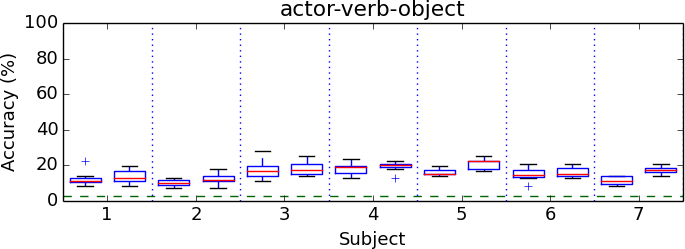}&
  \includegraphics[height=0.134\textheight]
                  {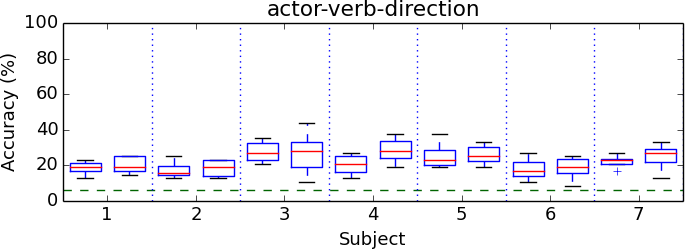}\\
  \includegraphics[height=0.134\textheight]
                  {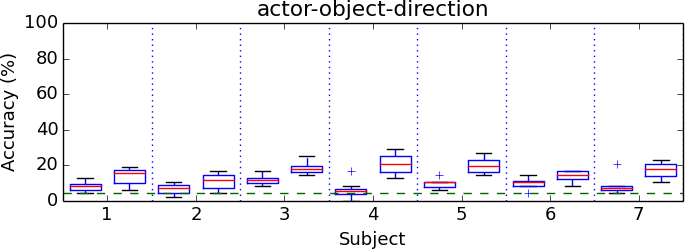}&
  \includegraphics[height=0.134\textheight]
                  {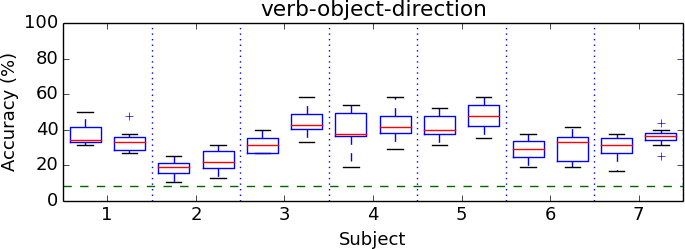}
  \end{tabular}
  \caption{%
    Per-subject comparison of joint~(left) \vs\ independent~(right)
    classification accuracy for constituent pairs and triples across the
    different folds.}
  \label{xfig:pair-triple}
\end{figure}

\begin{table}
  \centering
  \def\acc{\multicolumn{1}{c}{acc}}
  \def\mcc{\multicolumn{1}{c}{mcc}}
  \begin{tabular}{@{}l@{\hspace*{6ex}}rrrrrr@{}}
    \toprule
    & \multicolumn{2}{c}{bit}  & \multicolumn{2}{c}{all} & \multicolumn{2}{c}{good}\\
    \cmidrule(r){2-3} \cmidrule(lr){4-5} \cmidrule(l){6-7}
    analysis                      & \acc& \mcc& \acc& \mcc& \acc& \mcc\\
    \midrule
    \actor-\Verb                  &  0.6607&  0.1600&  0.3006&  0.2371&   0.4250&   0.3724\\
    \actor-\object                &  0.7059&  0.1475&  0.2584&  0.1910&   0.3983&   0.3430\\
    \actor-\direction             &  0.6336&  0.1266&  0.3363&  0.2412&   0.4398&   0.3603\\
    \actor-\location              &  0.6763&  0.1149&  0.2723&  0.1665&   0.3736&   0.2807\\
    \Verb-\object                 &  0.6709&  0.3419&  0.4712&  0.4051&   0.6433&   0.5959\\
    \Verb-\direction              &  0.7422&  0.4172&  0.6440&  0.5433&   0.7703&   0.7048\\
    \object-\direction            &  0.6544&  0.2946&  0.4475&  0.3370&   0.6305&   0.5560\\
    \object-\location             &  0.6235&  0.2537&  0.4754&  0.3702&   0.5897&   0.5067\\
    \midrule
    \actor-\Verb-\object          &  0.8093&  0.1833&  0.1262&  0.1010&   0.2751&   0.2521\\
    \actor-\Verb-\direction       &  0.7403&  0.1979&  0.1916&  0.1432&   0.3409&   0.3004\\
    \actor-\object-\direction     &  0.8344&  0.1314&  0.1250&  0.0867&   0.2661&   0.2352\\
    \Verb-\object-\direction      &  0.7154&  0.3318&  0.2850&  0.2267&   0.5006&   0.4594\\
    \bottomrule
  \end{tabular}
  \caption{%
    Comparison of independent classifiers with joint classifiers, aggregated
    across subject and fold.
    `Acc' denotes accuracy and `mcc' denotes Matthews correlation coefficient
    (MCC).
    The `bit' values involve computing a binary correct/incorrect label for each
    sample with both the independent and joint classifiers and computing the
    accuracy and MCC over the samples between the independent and joint
    classifiers.
    The `all' values involve computing a (nonbinary) class label for each
    sample with both the independent and joint classifiers and computing the
    accuracy and MCC over the samples between the independent and joint
    classifiers.
    The `good' values involved computing accuracy and MCC over the samples
    between the independent and joint classifiers for only those `all' samples
    where the joint classifier is correct.}
\end{table}

\begin{figure}
  \centering
  \begin{tabular}{@{}cc@{}}
    \includegraphics[height=0.125\textheight]{subject1-9events-searchlight}&
    \includegraphics[height=0.125\textheight]{subject1-9events-svm}\\
    \includegraphics[height=0.125\textheight]{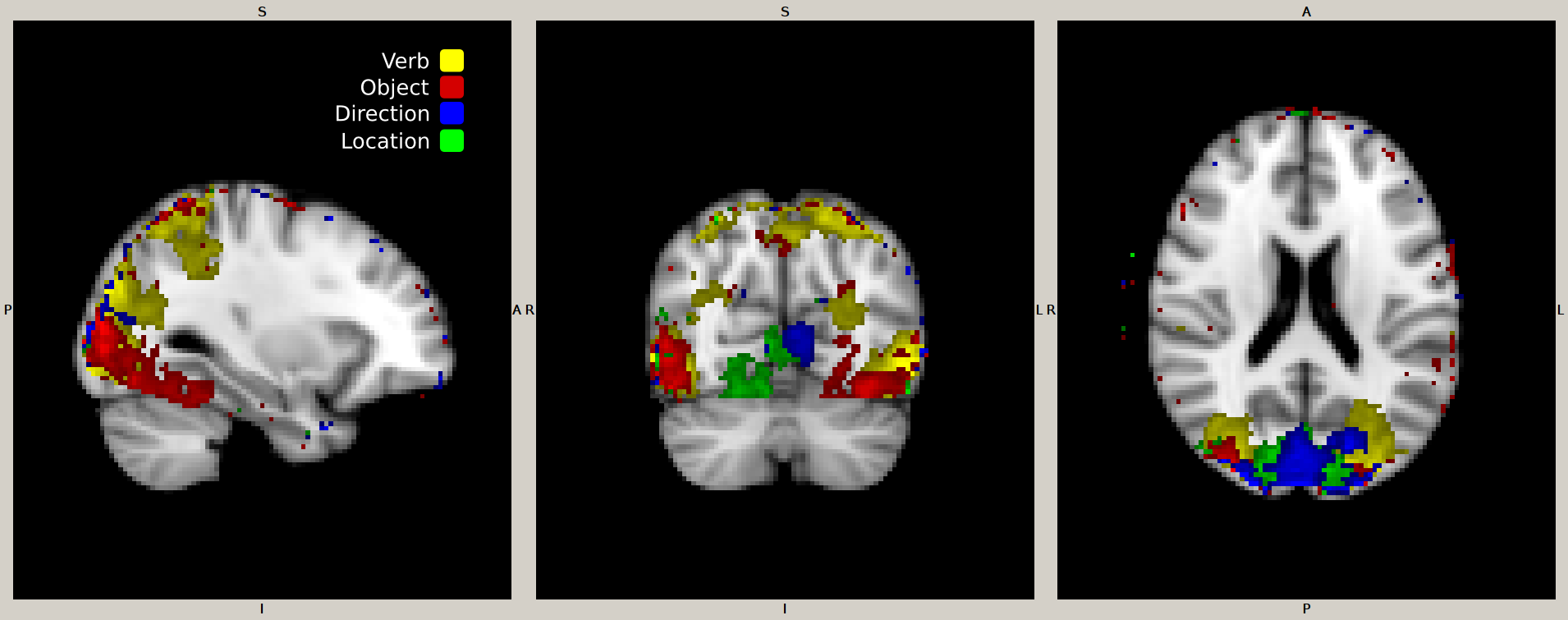}&
    \includegraphics[height=0.125\textheight]{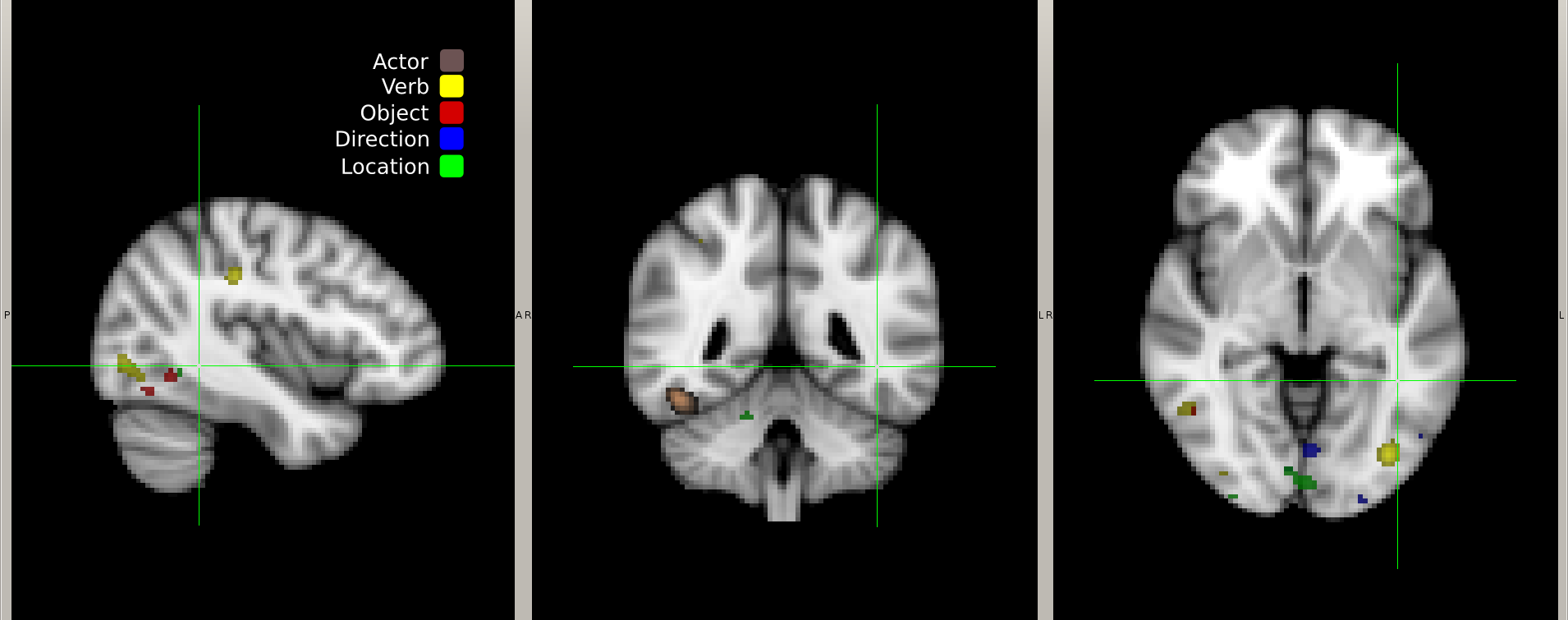}\\
    \includegraphics[height=0.125\textheight]{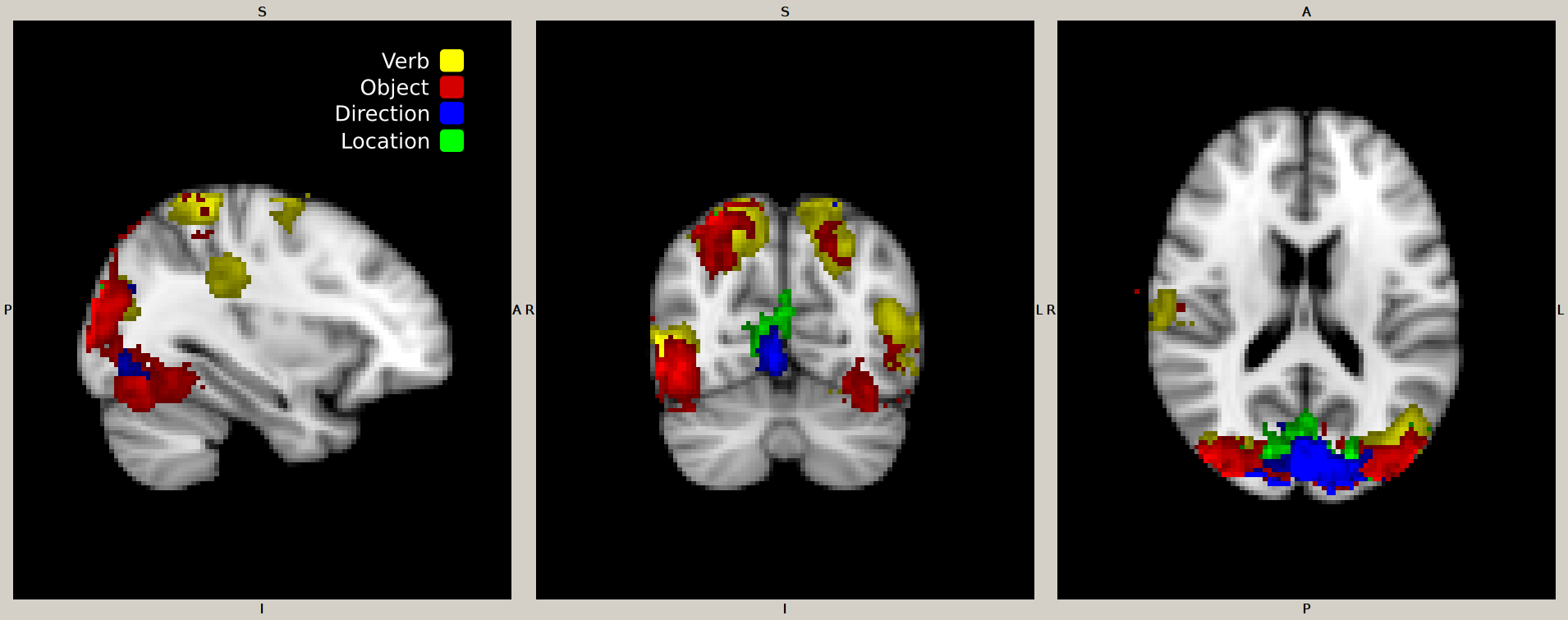}&
    \includegraphics[height=0.125\textheight]{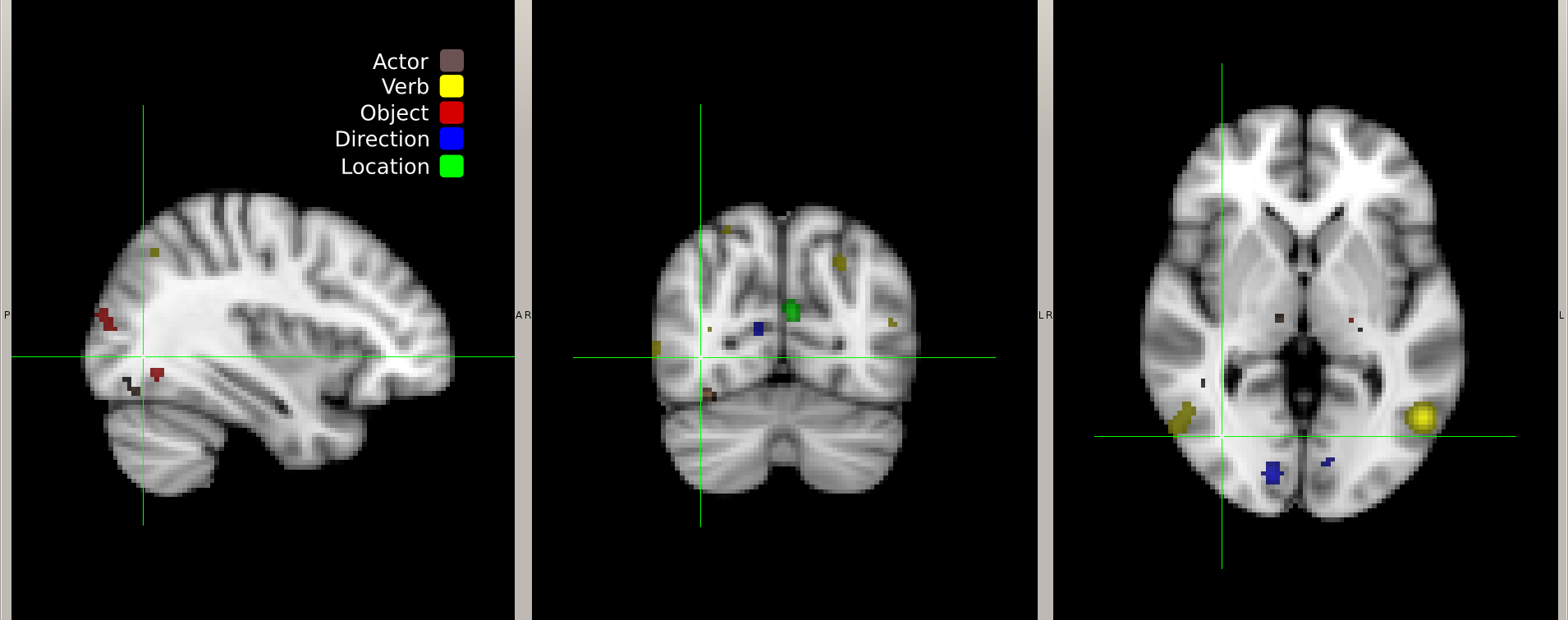}\\
    \includegraphics[height=0.125\textheight]{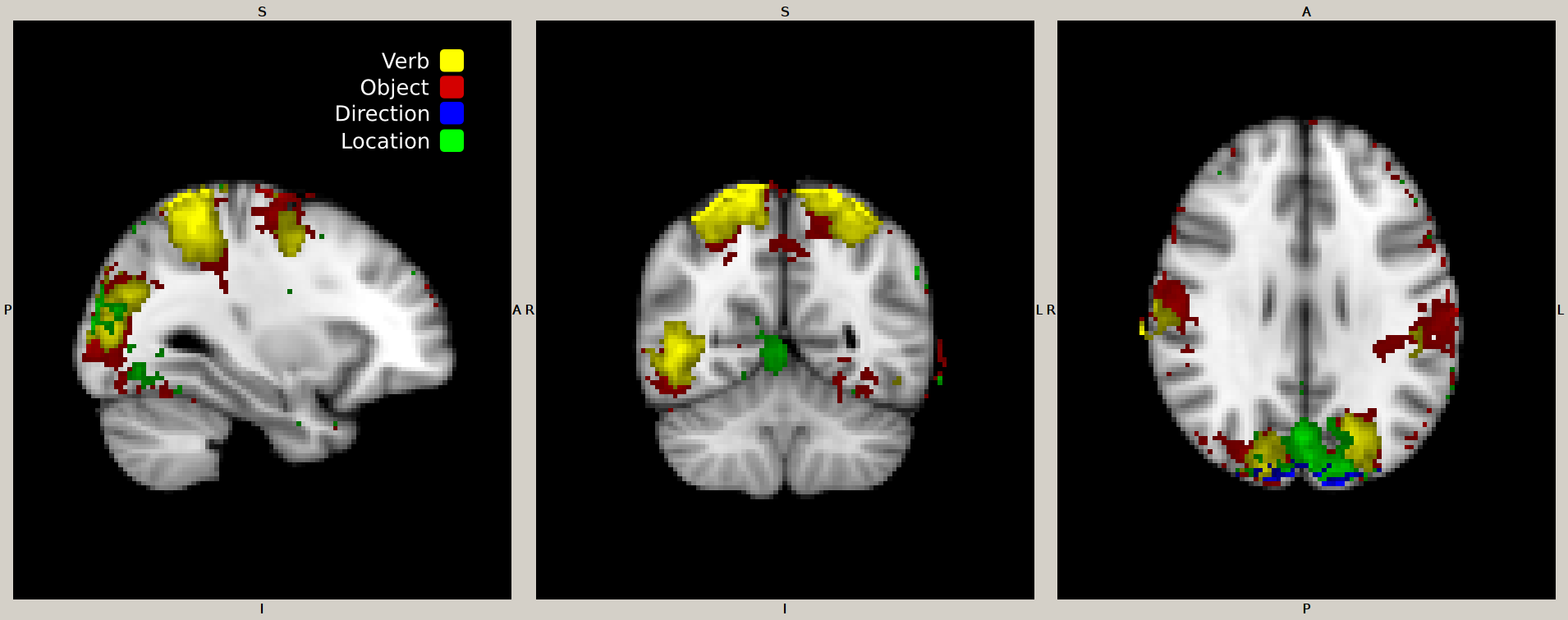}&
    \includegraphics[height=0.125\textheight]{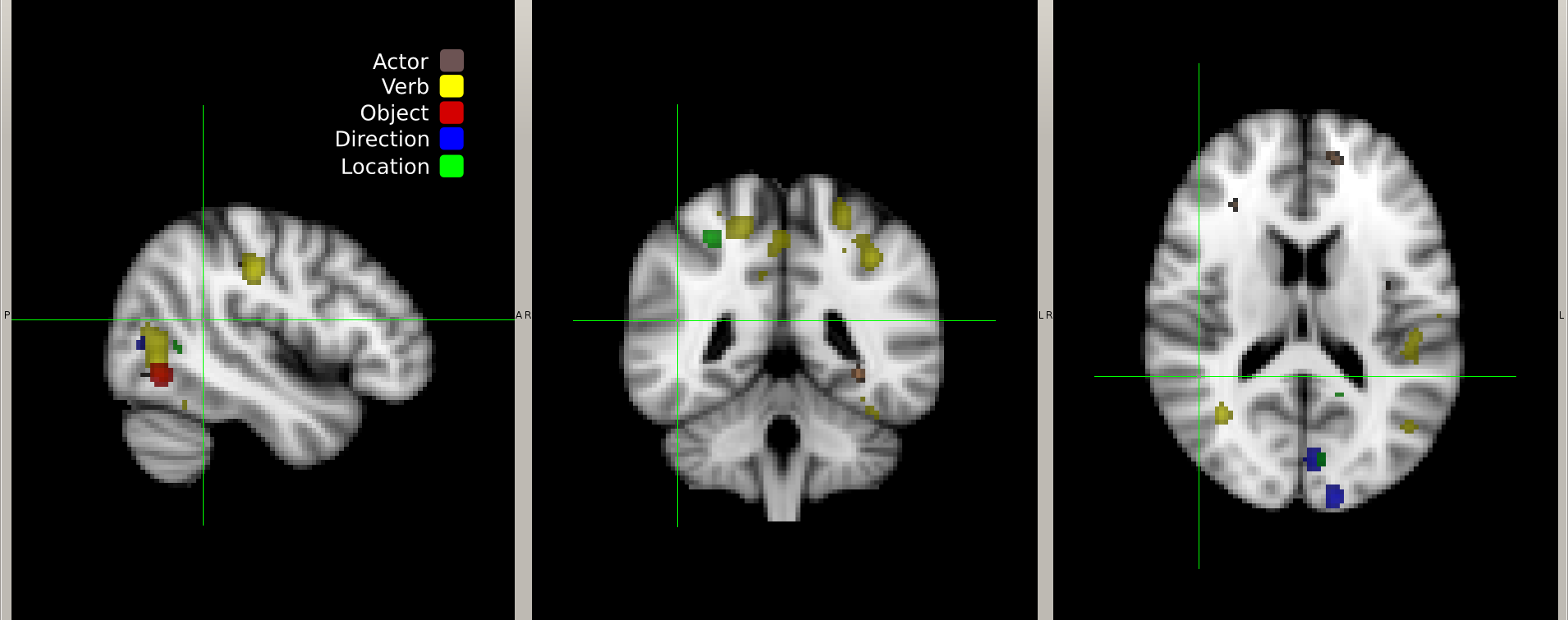}\\
    \includegraphics[height=0.125\textheight]{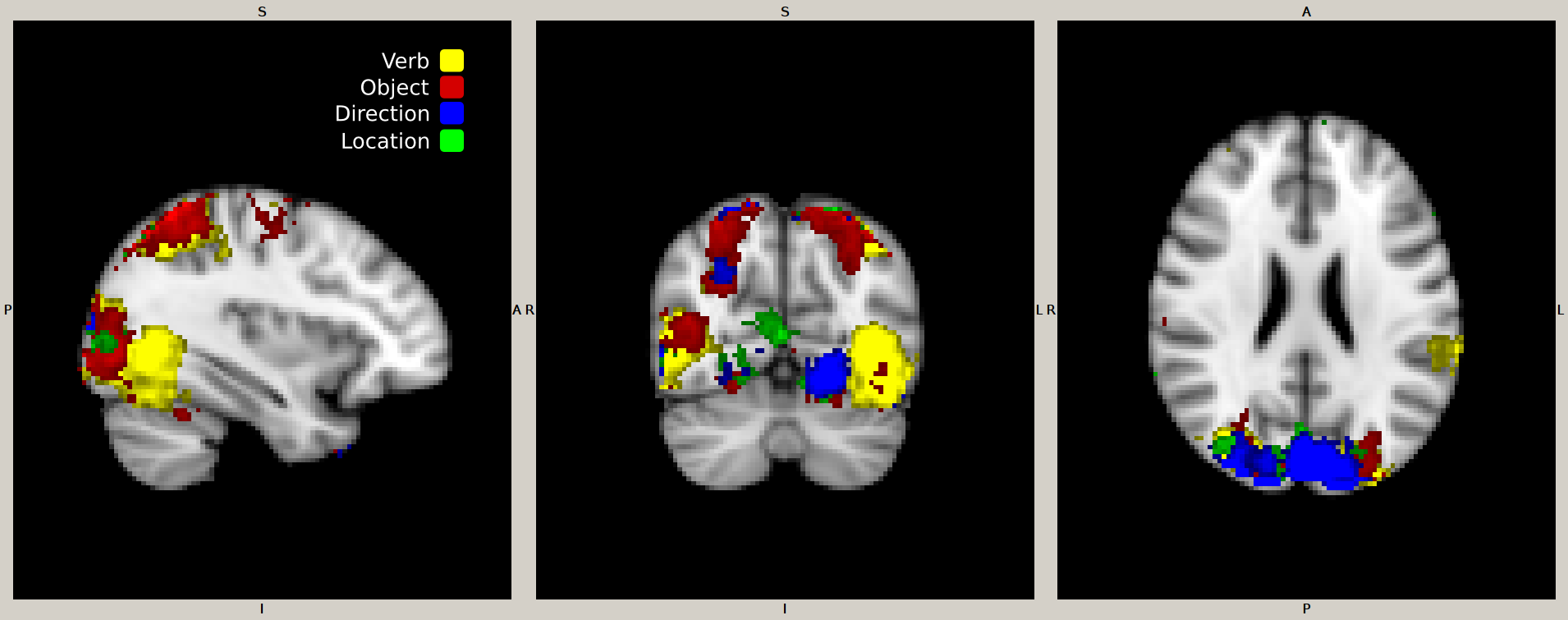}&
    \includegraphics[height=0.125\textheight]{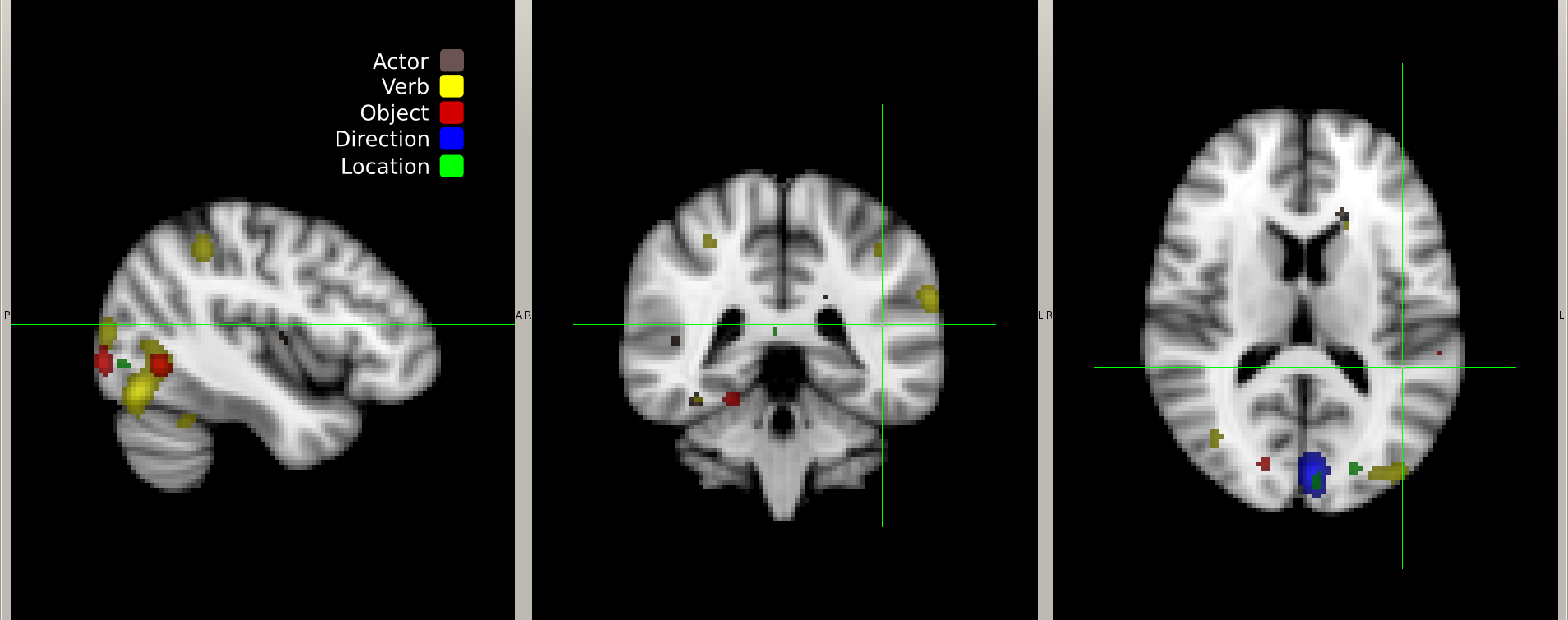}\\
    \includegraphics[height=0.125\textheight]{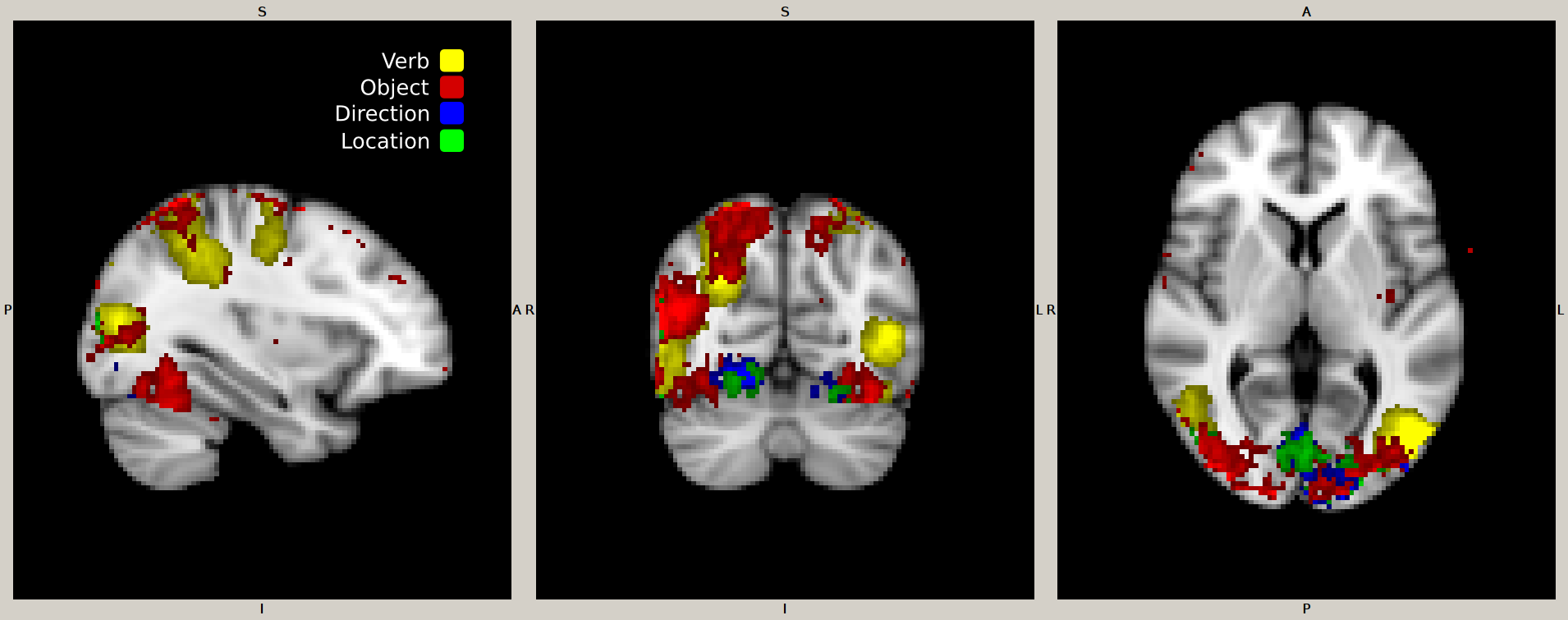}&
    \includegraphics[height=0.125\textheight]{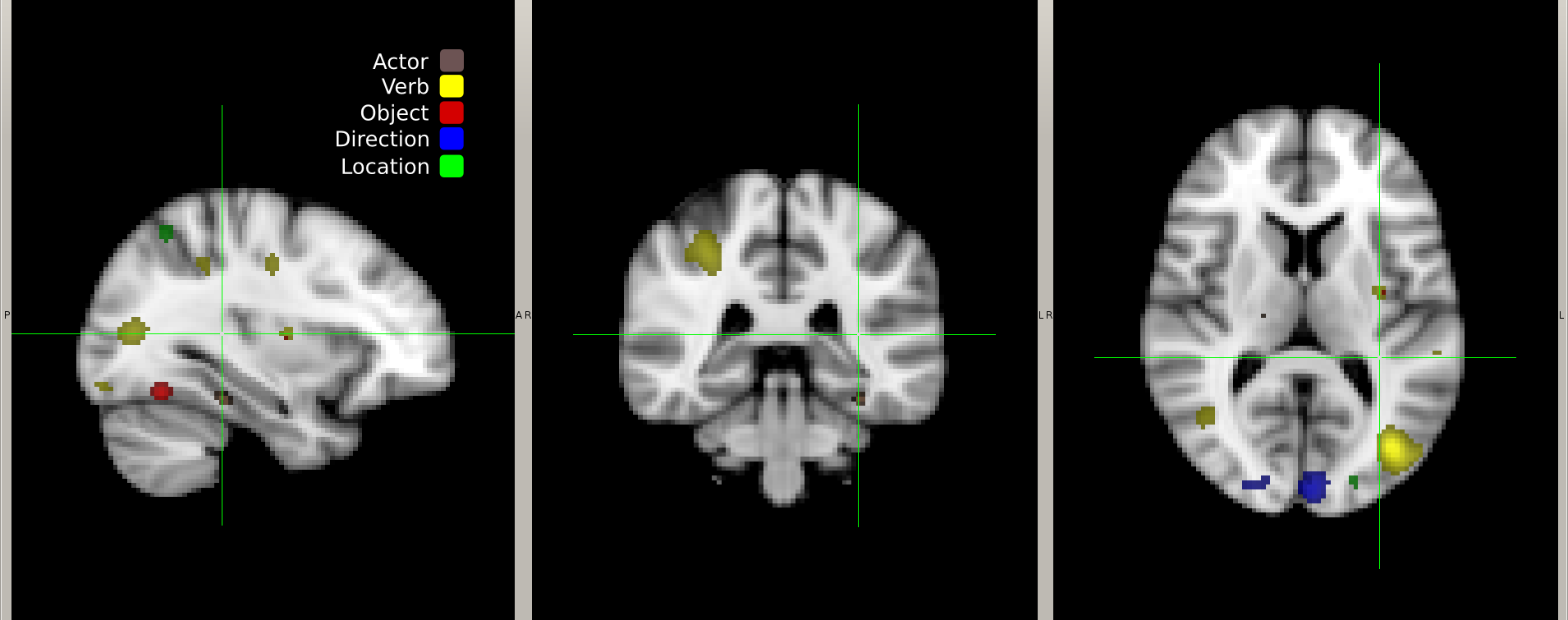}\\
    \includegraphics[height=0.125\textheight]{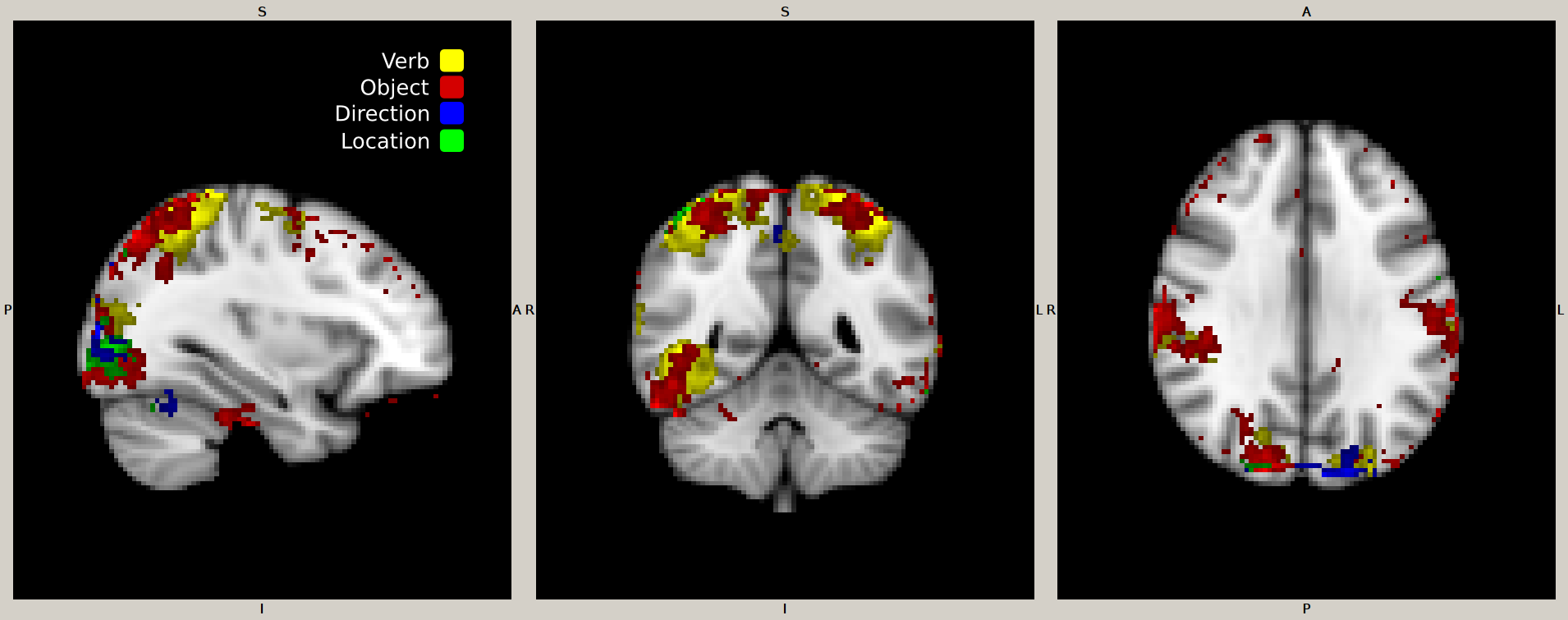}&
    \includegraphics[height=0.125\textheight]{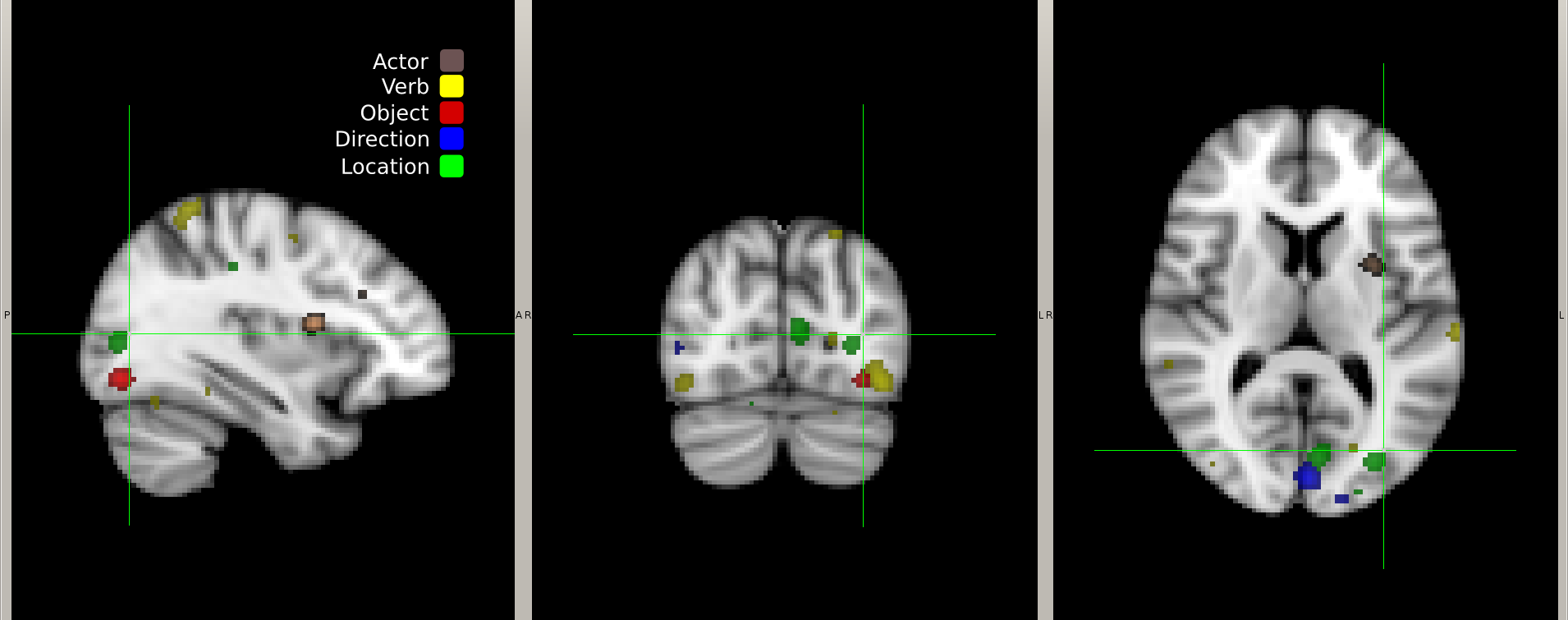}
  \end{tabular}
  \caption{%
    (left)~Searchlight analysis indicating the classification accuracy of
    different brain regions on the anatomical scans from subjects~1--7 averaged
    across stimulus, class, and run.
    (right)~Thresholded SVM coefficients for subjects~1--7, back-projected
    onto the anatomical scan, aggregated across run.}
  \vspace*{-3pt}
\end{figure}

\begin{table}
  \centering
  \begin{tabular}{@{}lrrrrrrr@{\hspace*{4ex}}r@{}}
    \multicolumn{9}{c}{$\displaystyle\frac{\displaystyle\left\lvert\bigcap_i\text{independent}_i\right\rvert}{\displaystyle\left\lvert\bigcup_i\text{independent}_i\right\rvert}$}\\
    \rule{0pt}{2ex}\\
    \toprule
    analysis &  1&  2&  3&  4&  5&  6&  7&mean\\
    \midrule
    \actor-\Verb             &    0.78\%&    5.75\%&    3.55\%&    3.39\%&    4.44\%&    2.57\%&    2.63\%&    3.30\%\\
    \actor-\object           &    3.72\%&    4.73\%&   14.46\%&    3.85\%&    3.61\%&    8.98\%&    7.85\%&    6.74\%\\
    \actor-\direction        &    1.52\%&    1.24\%&    4.67\%&    1.15\%&    0.87\%&    1.81\%&    0.90\%&    1.74\%\\
    \actor-\location         &    0.81\%&    0.56\%&    3.16\%&    1.05\%&    1.22\%&    0.65\%&    1.83\%&    1.32\%\\
    \Verb-\object            &    8.21\%&   25.32\%&    9.59\%&   18.34\%&   22.04\%&    9.60\%&   11.78\%&   14.98\%\\
    \Verb-\direction         &    1.11\%&    2.08\%&    0.18\%&    0.59\%&    3.74\%&    0.73\%&    0.00\%&    1.20\%\\
    \object-\direction       &   13.39\%&   11.72\%&    9.78\%&    8.85\%&    3.77\%&    8.65\%&    2.84\%&    8.43\%\\
    \object-\location        &    1.47\%&    7.15\%&    5.61\%&    5.14\%&    3.02\%&    1.47\%&    7.52\%&    4.48\%\\
    \cmidrule(r{4ex}){1-8}\cmidrule{9-9}
    \actor-\Verb-\object      &    0.28\%&    2.14\%&    1.25\%&    1.16\%&    1.47\%&    1.05\%&    1.45\%&    1.26\%\\
    \actor-\Verb-\direction   &    0.02\%&    0.15\%&    0.00\%&    0.01\%&    0.22\%&    0.04\%&    0.00\%&    0.06\%\\
    \actor-\object-\direction &    0.60\%&    0.57\%&    1.79\%&    0.27\%&    0.18\%&    0.89\%&    0.25\%&    0.65\%\\
    \Verb-\object-\direction  &    0.14\%&    0.96\%&    0.08\%&    0.20\%&    1.76\%&    0.27\%&    0.00\%&    0.49\%\\
    \addlinespace[2ex]
    mean    &    2.67\%&    5.20\%&    4.51\%&    3.67\%&    3.86\%&    3.06\%&    3.09\%&    3.72\%\\
    \bottomrule
    \rule{0pt}{5ex}\\
    \multicolumn{9}{c}{$\displaystyle\frac{\displaystyle\left\lvert\left(\bigcup_i\text{independent}_i\right)\cap\text{joint}\right\rvert}{\displaystyle\left\lvert\text{joint}\right\rvert}$}\\
    \rule{0pt}{2ex}\\
    \toprule
    analysis &  1&  2&  3&  4&  5&  6&  7&mean\\
    \midrule
    \actor-\Verb             &   73.84\%&   84.17\%&   29.37\%&   69.92\%&   84.69\%&   62.24\%&   67.68\%&   67.42\%\\
    \actor-\object           &   58.13\%&   70.01\%&   12.69\%&   48.66\%&   58.19\%&   50.56\%&   42.26\%&   48.64\%\\
    \actor-\direction        &   70.57\%&   76.98\%&   44.23\%&   53.25\%&   92.59\%&   81.29\%&   60.79\%&   68.53\%\\
    \actor-\location         &   77.87\%&   78.21\%&   43.04\%&   60.20\%&   85.62\%&   84.59\%&   56.04\%&   69.37\%\\
    \Verb-\object            &   87.64\%&   91.44\%&   47.81\%&   79.92\%&   95.24\%&   72.63\%&   82.00\%&   79.53\%\\
    \Verb-\direction         &   78.78\%&   91.95\%&   38.70\%&   57.35\%&   91.39\%&   76.85\%&   86.66\%&   74.53\%\\
    \object-\direction       &   70.03\%&   38.04\%&   71.44\%&   73.35\%&   63.40\%&   81.09\%&   64.41\%&   65.97\%\\
    \object-\location        &   72.66\%&   97.45\%&   95.52\%&   87.33\%&   75.24\%&   68.75\%&   57.12\%&   79.15\%\\
    \cmidrule(r{4ex}){1-8}\cmidrule{9-9}
    \actor-\Verb-\object      &   19.80\%&   21.10\%&    9.13\%&   12.40\%&   52.40\%&   42.90\%&   13.60\%&   24.48\%\\
    \actor-\Verb-\direction   &   67.37\%&   69.78\%&   28.90\%&   44.15\%&   95.21\%&   56.72\%&   53.88\%&   59.43\%\\
    \actor-\object-\direction &   57.39\%&   21.04\%&   10.21\%&   38.26\%&   52.20\%&   43.39\%&   41.94\%&   37.78\%\\
    \Verb-\object-\direction  &   57.12\%&   66.48\%&   47.87\%&   42.93\%&   82.05\%&   43.58\%&   45.87\%&   55.13\%\\
    \addlinespace[2ex]
    mean    &   65.93\%&   67.22\%&   39.91\%&   55.64\%&   77.35\%&   63.72\%&   56.02\%&   60.83\%\\
    \bottomrule
  \end{tabular}
  \caption{%
    Per-subject quantitative comparison of the brain regions indicated by
    searchlight of the independent classifiers to the joint classifiers, for
    all constituent pairs and triples, together with means across subject,
    means across analysis, and means across both.
    (top)~The percentage of voxels in the union of the constituents for
    the independent classifier also in the intersection.
    (bottom)~The percentage of voxels in the joint classifier that
    are shared with the independent classifier.}
\end{table}

\begin{table}
  \centering
  \begin{tabular}{@{}lrrrrrrr@{\hspace*{4ex}}r@{}}
    \multicolumn{9}{c}{$\displaystyle\frac{\displaystyle\left\lvert\bigcap_i\text{independent}_i\right\rvert}{\displaystyle\left\lvert\bigcup_i\text{independent}_i\right\rvert}$}\\
    \rule{0pt}{2ex}\\
    \toprule
    analysis &  1&  2&  3&  4&  5&  6&  7&mean\\
    \midrule
    \actor-\Verb             &    1.21\%&    6.10\%&    2.19\%&    2.66\%&    1.88\%&    1.88\%&    3.95\%&    2.84\%\\
    \actor-\object           &    1.11\%&    2.61\%&    1.67\%&    3.84\%&    4.32\%&    1.98\%&    2.24\%&    2.54\%\\
    \actor-\direction        &    0.50\%&    0.70\%&    1.78\%&    0.55\%&    0.50\%&    3.14\%&    0.95\%&    1.16\%\\
    \actor-\location         &    0.80\%&    1.31\%&    2.51\%&    1.93\%&    2.04\%&    3.09\%&    2.77\%&    2.06\%\\
    \Verb-\object            &    5.54\%&    5.26\%&    4.82\%&    6.72\%&   11.04\%&    5.09\%&    3.89\%&    6.05\%\\
    \Verb-\direction         &    3.95\%&    3.25\%&    2.35\%&    3.68\%&    4.27\%&    2.82\%&    4.98\%&    3.61\%\\
    \object-\direction       &    7.87\%&    1.93\%&    2.66\%&    3.14\%&    3.78\%&    5.42\%&    1.06\%&    3.70\%\\
    \object-\location        &    0.90\%&    1.72\%&    3.03\%&    3.09\%&    3.95\%&    2.14\%&    1.67\%&    2.36\%\\
    \cmidrule(r{4ex}){1-8}\cmidrule{9-9}
    \actor-\Verb-\object      &    0.00\%&    0.76\%&    0.10\%&    0.75\%&    0.88\%&    0.21\%&    0.28\%&    0.42\%\\
    \actor-\Verb-\direction   &    0.00\%&    0.03\%&    0.00\%&    0.03\%&    0.00\%&    0.03\%&    0.00\%&    0.01\%\\
    \actor-\object-\direction &    0.00\%&    0.00\%&    0.03\%&    0.03\%&    0.00\%&    0.00\%&    0.00\%&    0.00\%\\
    \Verb-\object-\direction  &    0.29\%&    0.21\%&    0.17\%&    0.10\%&    0.15\%&    0.36\%&    0.10\%&    0.20\%\\
    \addlinespace[2ex]
    mean                   &    1.85\%&    1.99\%&    1.78\%&    2.21\%&    2.73\%&    2.18\%&    1.82\%&    2.08\%\\
    \bottomrule
    \rule{0pt}{5ex}\\
    \multicolumn{9}{c}{$\displaystyle\frac{\displaystyle\left\lvert\left(\bigcup_i\text{independent}_i\right)\cap\text{joint}\right\rvert}{\displaystyle\left\lvert\text{joint}\right\rvert}$}\\
    \rule{0pt}{2ex}\\
    \toprule
    analysis &  1&  2&  3&  4&  5&  6&  7&mean\\
    \midrule
    \actor-\Verb             &   58.59\%&   71.79\%&   57.89\%&   58.69\%&   57.79\%&   57.19\%&   50.00\%&   58.85\%\\
    \actor-\object           &   47.89\%&   55.70\%&   52.20\%&   52.90\%&   47.39\%&   55.70\%&   46.80\%&   51.22\%\\
    \actor-\direction        &   42.89\%&   39.80\%&   39.00\%&   34.30\%&   43.10\%&   49.89\%&   48.00\%&   42.42\%\\
    \actor-\location         &   25.00\%&   28.59\%&   21.39\%&   22.50\%&   42.89\%&   28.10\%&   25.50\%&   27.71\%\\
    \Verb-\object            &   64.70\%&   67.90\%&   65.60\%&   68.10\%&   70.59\%&   69.09\%&   56.39\%&   66.05\%\\
    \Verb-\direction         &   67.50\%&   47.69\%&   55.00\%&   62.00\%&   72.09\%&   64.79\%&   67.60\%&   62.38\%\\
    \object-\direction       &   54.40\%&   37.10\%&   58.69\%&   51.80\%&   56.10\%&   57.39\%&   53.40\%&   52.70\%\\
    \object-\location        &   45.10\%&   30.19\%&   37.10\%&   43.29\%&   42.29\%&   44.00\%&   29.69\%&   38.81\%\\
    \cmidrule(r{4ex}){1-8}\cmidrule{9-9}
    \actor-\Verb-\object      &   61.79\%&   68.30\%&   54.40\%&   62.70\%&   62.50\%&   60.39\%&   54.70\%&   60.68\%\\
    \actor-\Verb-\direction   &   68.89\%&   52.60\%&   51.60\%&   49.70\%&   55.10\%&   61.70\%&   56.00\%&   56.51\%\\
    \actor-\object-\direction &   45.10\%&   37.39\%&   45.89\%&   27.00\%&   32.70\%&   42.29\%&   38.10\%&   38.35\%\\
    \Verb-\object-\direction  &   69.59\%&   38.29\%&   53.70\%&   58.59\%&   64.50\%&   60.69\%&   62.39\%&   58.25\%\\
    \addlinespace[2ex]
    mean                   &   54.29\%&   47.95\%&   49.37\%&   49.30\%&   53.92\%&   54.27\%&   49.04\%&    51.16\%\\
    \bottomrule
  \end{tabular}
  \caption{%
    Per-subject quantitative comparison of the brain regions indicated
    by the thresholded SVM coefficients of the independent classifiers to the
    joint classifiers, for all constituent pairs and triples, together with
    means across subject, means across analysis, and means across both.
    (top)~The percentage of voxels in the union of the constituents for
    the independent classifier also in the intersection.
    (bottom)~The percentage of voxels in the joint classifier that
    are shared with the independent classifier.}
\end{table}

\begin{table}
  \centering
  \resizebox{\textwidth}{!}{\begin{tabular}{@{}llll@{\hspace*{4ex}}lllllll@{}}
      \toprule
      analysis                &  chance&    mean&  stddev&       1&       2&       3&       4&       5&       6&       7\\
      \midrule
      \actor                        &  0.2500&  0.2904$^{***}$&   0.057&  0.259&  0.267&  0.300$^{**}$&  0.321$^{***}$&  0.313$^{***}$&  0.281$^{*}$&  0.292$^{*}$\\
      \Verb                         &  0.3333&  0.4802$^{***}$&   0.071&  0.514$^{***}$&  0.399$^{**}$&  0.431$^{***}$&  0.514$^{***}$&  0.517$^{***}$&  0.530$^{***}$&  0.457$^{***}$\\
      \object                       &  0.3333&  0.4199$^{***}$&   0.070&  0.377$^{*}$&  0.399$^{**}$&  0.401$^{***}$&  0.410$^{***}$&  0.490$^{***}$&  0.470$^{***}$&  0.392$^{**}$\\
      \direction                    &  0.5000&  0.6548$^{***}$&   0.084&  0.654$^{***}$&  0.591$^{***}$&  0.695$^{***}$&  0.706$^{***}$&  0.633$^{***}$&  0.630$^{***}$&  0.674$^{***}$\\
      \location                     &  0.5000&  0.6057$^{***}$&   0.116&  0.661$^{***}$&  0.536&  0.672$^{***}$&  0.661$^{***}$&  0.620$^{**}$&  0.547&  0.542\\
      \midrule
      \actor-\Verb                  &  0.0833&  0.1342$^{***}$&   0.045&  0.139$^{***}$&  0.118$^{**}$&  0.108$^{*}$&  0.156$^{***}$&  0.170$^{***}$&  0.132$^{***}$&  0.116$^{**}$\\
      \actor\&\Verb                 &  0.0833&  0.1451$^{***}$&   0.049&  0.137$^{***}$&  0.104$^{*}$&  0.132$^{***}$&  0.168$^{***}$&  0.191$^{***}$&  0.155$^{***}$&  0.128$^{***}$\\[2ex]
      \actor-\object                &  0.0833&  0.1161$^{***}$&   0.045&  0.111$^{*}$&  0.127$^{***}$&  0.094&  0.125$^{***}$&  0.142$^{***}$&  0.109$^{*}$&  0.104$^{*}$\\
      \actor\&\object               &  0.0833&  0.1235$^{***}$&   0.040&  0.102&  0.115$^{*}$&  0.106$^{*}$&  0.142$^{***}$&  0.156$^{***}$&  0.132$^{***}$&  0.111$^{*}$\\[2ex]
      \actor-\direction             &  0.1250&  0.1782$^{***}$&   0.060&  0.177$^{**}$&  0.146&  0.190$^{***}$&  0.182$^{**}$&  0.193$^{***}$&  0.172$^{**}$&  0.188$^{***}$\\
      \actor\&\direction            &  0.1250&  0.1968$^{***}$&   0.059&  0.161$^{*}$&  0.177$^{**}$&  0.224$^{***}$&  0.240$^{***}$&  0.187$^{***}$&  0.188$^{***}$&  0.201$^{***}$\\[2ex]
      \actor-\location              &  0.1250&  0.1793$^{***}$&   0.093&  0.172$^{*}$&  0.182$^{*}$&  0.130&  0.203$^{**}$&  0.219$^{***}$&  0.198$^{**}$&  0.151\\
      \actor\&\location             &  0.1250&  0.1711$^{***}$&   0.090&  0.135&  0.130&  0.198$^{**}$&  0.203$^{**}$&  0.224$^{***}$&  0.151&  0.156\\[2ex]
      \Verb-\object                 &  0.1111&  0.2004$^{***}$&   0.058&  0.189$^{***}$&  0.167$^{***}$&  0.167$^{***}$&  0.222$^{***}$&  0.234$^{***}$&  0.240$^{***}$&  0.184$^{***}$\\
      \Verb\&\object                &  0.1111&  0.2039$^{***}$&   0.056&  0.196$^{***}$&  0.155$^{**}$&  0.172$^{***}$&  0.215$^{***}$&  0.262$^{***}$&  0.243$^{***}$&  0.184$^{***}$\\[2ex]
      \Verb-\direction              &  0.2500&  0.4174$^{***}$&   0.074&  0.430$^{***}$&  0.375$^{***}$&  0.398$^{***}$&  0.461$^{***}$&  0.440$^{***}$&  0.396$^{***}$&  0.422$^{***}$\\
      \Verb\&\direction             &  0.2500&  0.3199$^{***}$&   0.069&  0.344$^{***}$&  0.247$^{***}$&  0.310$^{***}$&  0.357$^{***}$&  0.354$^{***}$&  0.305$^{***}$&  0.323$^{***}$\\[2ex]
      \object-\direction            &  0.1667&  0.2500$^{***}$&   0.059&  0.263$^{***}$&  0.211$^{*}$&  0.242$^{***}$&  0.271$^{***}$&  0.260$^{***}$&  0.250$^{***}$&  0.253$^{***}$\\
      \object\&\direction           &  0.1667&  0.2716$^{***}$&   0.070&  0.255$^{***}$&  0.227$^{**}$&  0.271$^{***}$&  0.307$^{***}$&  0.305$^{***}$&  0.281$^{***}$&  0.255$^{***}$\\[2ex]
      \object-\location             &  0.1667&  0.3162$^{***}$&   0.099&  0.333$^{***}$&  0.271$^{***}$&  0.307$^{***}$&  0.359$^{***}$&  0.391$^{***}$&  0.323$^{***}$&  0.229$^{*}$\\
      \object\&\location            &  0.1667&  0.2626$^{***}$&   0.092&  0.276$^{***}$&  0.193&  0.271$^{***}$&  0.292$^{***}$&  0.307$^{***}$&  0.292$^{***}$&  0.208\\
      \midrule
      \actor-\Verb-\object          &  0.0278&  0.0494$^{***}$&   0.025&  0.054$^{***}$&  0.054$^{***}$&  0.035&  0.050$^{**}$&  0.064$^{***}$&  0.045$^{*}$&  0.043$^{*}$\\
      \actor\&\Verb\&\object        &  0.0278&  0.0647$^{***}$&   0.030&  0.056$^{***}$&  0.054$^{***}$&  0.047$^{*}$&  0.078$^{***}$&  0.097$^{***}$&  0.075$^{***}$&  0.047$^{*}$\\[2ex]
      \actor-\Verb-\direction       &  0.0625&  0.1183$^{***}$&   0.049&  0.128$^{***}$&  0.112$^{***}$&  0.109$^{***}$&  0.125$^{***}$&  0.133$^{***}$&  0.115$^{***}$&  0.107$^{**}$\\
      \actor\&\Verb\&\direction     &  0.0625&  0.0986$^{***}$&   0.047&  0.102$^{***}$&  0.076$^{**}$&  0.096$^{***}$&  0.117$^{***}$&  0.122$^{***}$&  0.083$^{***}$&  0.094$^{***}$\\[2ex]
      \actor-\object-\direction     &  0.0417&  0.0692$^{***}$&   0.036&  0.057&  0.055&  0.070$^{*}$&  0.070$^{*}$&  0.076$^{**}$&  0.083$^{***}$&  0.073$^{**}$\\
      \actor\&\object\&\direction   &  0.0417&  0.0807$^{***}$&   0.040&  0.065$^{*}$&  0.068$^{*}$&  0.076$^{**}$&  0.115$^{***}$&  0.086$^{***}$&  0.094$^{***}$&  0.063$^{*}$\\[2ex]
      \Verb-\object-\direction      &  0.0833&  0.1533$^{***}$&   0.056&  0.185$^{***}$&  0.109$^{*}$&  0.146$^{***}$&  0.203$^{***}$&  0.146$^{***}$&  0.154$^{***}$&  0.130$^{**}$\\
      \Verb\&\object\&\direction    &  0.0833&  0.1373$^{***}$&   0.054&  0.146$^{***}$&  0.094$^{**}$&  0.120$^{***}$&  0.161$^{***}$&  0.169$^{***}$&  0.141$^{***}$&  0.130$^{***}$\\
      \midrule
      sentence\&                    &  0.0139&  0.0441$^{***}$&   0.025&  0.035$^{***}$&  0.038$^{***}$&  0.028$^{*}$&  0.063$^{***}$&  0.061$^{***}$&  0.052$^{***}$&  0.033$^{**}$\\
      \bottomrule
  \end{tabular}}
  \caption{Cross-subject variant of Table~\ref{xtab:results}.}
\end{table}

\begin{figure}
  \centering
  \begin{tabular}{@{}cc@{}}
  \includegraphics[height=0.134\textheight]
                  {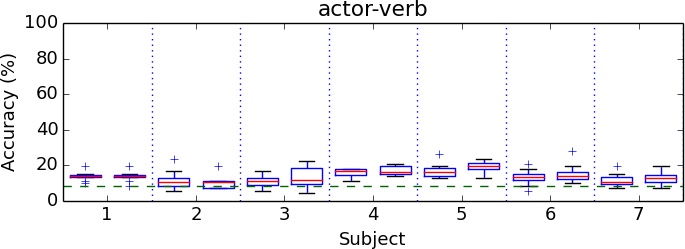}&
  \includegraphics[height=0.134\textheight]
                  {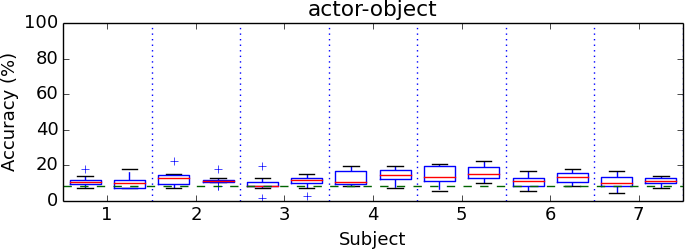}\\
  \includegraphics[height=0.134\textheight]
                  {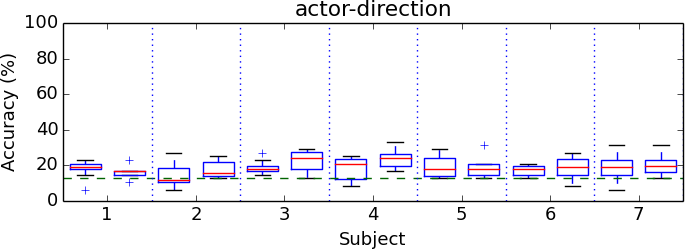}&
  \includegraphics[height=0.134\textheight]
                  {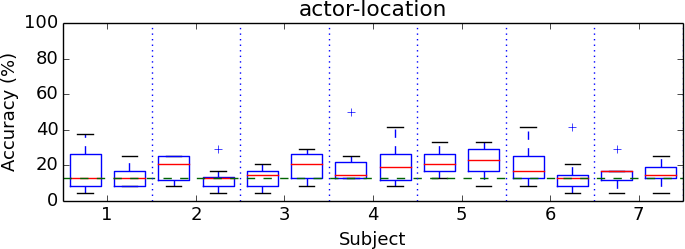}\\
  \includegraphics[height=0.134\textheight]
                  {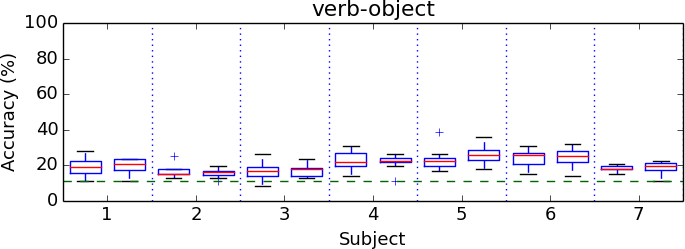}&
  \includegraphics[height=0.134\textheight]
                  {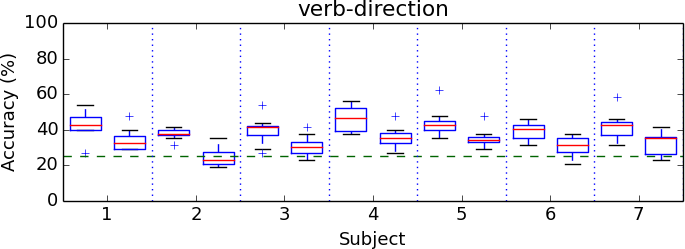}\\
  \includegraphics[height=0.134\textheight]
                  {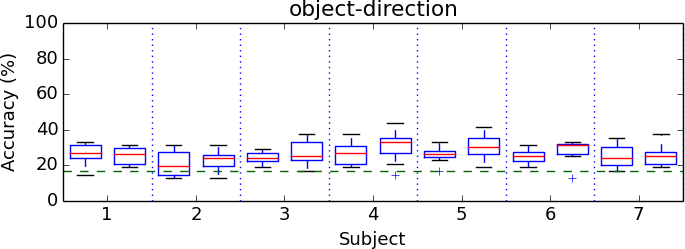}&
  \includegraphics[height=0.134\textheight]
                  {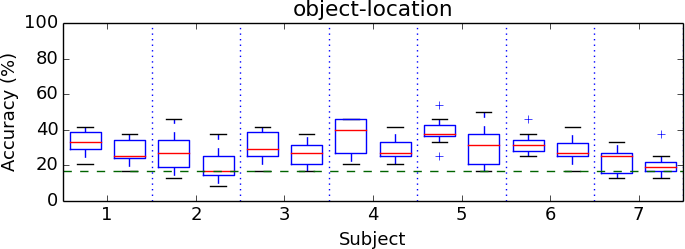}\\*[2ex]
  \includegraphics[height=0.134\textheight]
                  {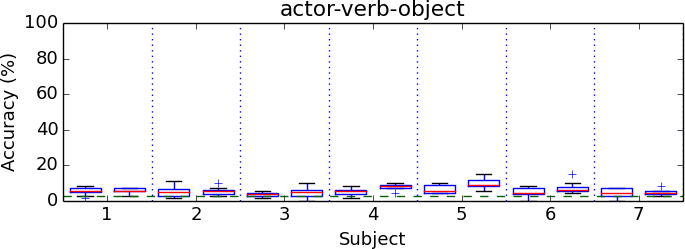}&
  \includegraphics[height=0.134\textheight]
                  {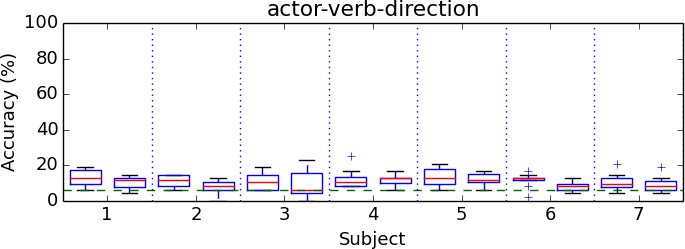}\\
  \includegraphics[height=0.134\textheight]
                  {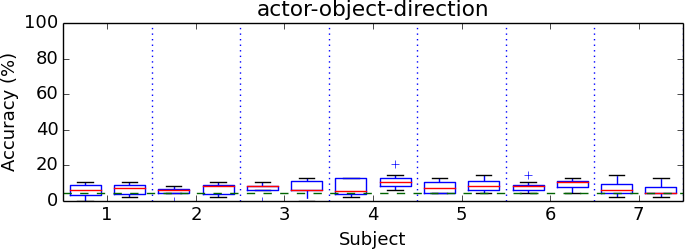}&
  \includegraphics[height=0.134\textheight]
                  {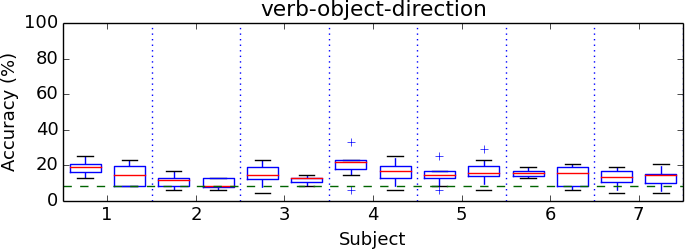}
  \end{tabular}
  \caption{Cross-subject variant of Fig.~\ref{xfig:pair-triple}.}
\end{figure}

\end{document}